\catcode`\@=11

\font\tenmsa=msam10
\font\sevenmsa=msam7
\font\fivemsa=msam5
\font\tenmsb=msbm10
\font\sevenmsb=msbm7
\font\fivemsb=msbm5
\newfam\msafam
\newfam\msbfam
\textfont\msafam=\tenmsa  \scriptfont\msafam=\sevenmsa
  \scriptscriptfont\msafam=\fivemsa
\textfont\msbfam=\tenmsb  \scriptfont\msbfam=\sevenmsb
  \scriptscriptfont\msbfam=\fivemsb

\def\hexnumber@#1{\ifnum#1<10 \number#1\else
 \ifnum#1=10 A\else\ifnum#1=11 B\else\ifnum#1=12 C\else
 \ifnum#1=13 D\else\ifnum#1=14 E\else\ifnum#1=15 F\fi\fi\fi\fi\fi\fi\fi}

\def\msa@{\hexnumber@\msafam}
\def\msb@{\hexnumber@\msbfam}
\mathchardef\boxdot="2\msa@00
\mathchardef\boxplus="2\msa@01
\mathchardef\boxtimes="2\msa@02
\mathchardef\square="0\msa@03
\mathchardef\blacksquare="0\msa@04
\mathchardef\centerdot="2\msa@05
\mathchardef\lozenge="0\msa@06
\mathchardef\blacklozenge="0\msa@07
\mathchardef\circlearrowright="3\msa@08
\mathchardef\circlearrowleft="3\msa@09
\mathchardef\rightleftharpoons="3\msa@0A
\mathchardef\leftrightharpoons="3\msa@0B
\mathchardef\boxminus="2\msa@0C
\mathchardef\Vdash="3\msa@0D
\mathchardef\Vvdash="3\msa@0E
\mathchardef\vDash="3\msa@0F
\mathchardef\twoheadrightarrow="3\msa@10
\mathchardef\twoheadleftarrow="3\msa@11
\mathchardef\leftleftarrows="3\msa@12
\mathchardef\rightrightarrows="3\msa@13
\mathchardef\upuparrows="3\msa@14
\mathchardef\downdownarrows="3\msa@15
\mathchardef\upharpoonright="3\msa@16

\mathchardef\downharpoonright="3\msa@17
\mathchardef\upharpoonleft="3\msa@18
\mathchardef\downharpoonleft="3\msa@19
\mathchardef\rightarrowtail="3\msa@1A
\mathchardef\leftarrowtail="3\msa@1B
\mathchardef\leftrightarrows="3\msa@1C
\mathchardef\rightleftarrows="3\msa@1D
\mathchardef\Lsh="3\msa@1E
\mathchardef\Rsh="3\msa@1F
\mathchardef\rightsquigarrow="3\msa@20
\mathchardef\leftrightsquigarrow="3\msa@21
\mathchardef\looparrowleft="3\msa@22
\mathchardef\looparrowright="3\msa@23
\mathchardef\circeq="3\msa@24
\mathchardef\succsim="3\msa@25
\mathchardef\gtrsim="3\msa@26
\mathchardef\gtrapprox="3\msa@27
\mathchardef\multimap="3\msa@28
\mathchardef\therefore="3\msa@29
\mathchardef\because="3\msa@2A
\mathchardef\doteqdot="3\msa@2B

\mathchardef\triangleq="3\msa@2C
\mathchardef\precsim="3\msa@2D
\mathchardef\lesssim="3\msa@2E
\mathchardef\lessapprox="3\msa@2F
\mathchardef\eqslantless="3\msa@30
\mathchardef\eqslantgtr="3\msa@31
\mathchardef\curlyeqprec="3\msa@32
\mathchardef\curlyeqsucc="3\msa@33
\mathchardef\preccurlyeq="3\msa@34
\mathchardef\leqq="3\msa@35
\mathchardef\leqslant="3\msa@36
\mathchardef\lessgtr="3\msa@37
\mathchardef\backprime="0\msa@38
\mathchardef\risingdotseq="3\msa@3A
\mathchardef\fallingdotseq="3\msa@3B
\mathchardef\succcurlyeq="3\msa@3C
\mathchardef\geqq="3\msa@3D
\mathchardef\geqslant="3\msa@3E
\mathchardef\gtrless="3\msa@3F
\mathchardef\sqsubset="3\msa@40
\mathchardef\sqsupset="3\msa@41
\mathchardef\trianglerighteq="3\msa@44
\mathchardef\trianglelefteq="3\msa@45
\mathchardef\bigstar="0\msa@46
\mathchardef\between="3\msa@47
\mathchardef\blacktriangledown="0\msa@48
\mathchardef\blacktriangleright="3\msa@49
\mathchardef\blacktriangleleft="3\msa@4A
\mathchardef\blacktriangle="0\msa@4E
\mathchardef\triangledown="0\msa@4F
\mathchardef\eqcirc="3\msa@50
\mathchardef\lesseqgtr="3\msa@51
\mathchardef\gtreqless="3\msa@52
\mathchardef\lesseqqgtr="3\msa@53
\mathchardef\gtreqqless="3\msa@54
\mathchardef\Rrightarrow="3\msa@56
\mathchardef\Lleftarrow="3\msa@57
\mathchardef\veebar="2\msa@59
\mathchardef\barwedge="2\msa@5A
\mathchardef\doublebarwedge="2\msa@5B
\mathchardef\angle="0\msa@5C
\mathchardef\measuredangle="0\msa@5D
\mathchardef\sphericalangle="0\msa@5E
\mathchardef\varpropto="3\msa@5F
\mathchardef\smallsmile="3\msa@60
\mathchardef\smallfrown="3\msa@61
\mathchardef\Subset="3\msa@62
\mathchardef\Supset="3\msa@63
\mathchardef\Cup="2\msa@64

\mathchardef\Cap="2\msa@65

\mathchardef\curlywedge="2\msa@66
\mathchardef\curlyvee="2\msa@67
\mathchardef\leftthreetimes="2\msa@68
\mathchardef\rightthreetimes="2\msa@69
\mathchardef\subseteqq="3\msa@6A
\mathchardef\supseteqq="3\msa@6B
\mathchardef\bumpeq="3\msa@6C
\mathchardef\Bumpeq="3\msa@6D
\mathchardef\lll="3\msa@6E

\mathchardef\ggg="3\msa@6F

\mathchardef\circledS="0\msa@73
\mathchardef\pitchfork="3\msa@74
\mathchardef\dotplus="2\msa@75
\mathchardef\backsim="3\msa@76
\mathchardef\backsimeq="3\msa@77
\mathchardef\complement="0\msa@7B
\mathchardef\intercal="2\msa@7C
\mathchardef\circledcirc="2\msa@7D
\mathchardef\circledast="2\msa@7E
\mathchardef\circleddash="2\msa@7F
\def\ulcorner{\delimiter"4\msa@70\msa@70 }
\def\urcorner{\delimiter"5\msa@71\msa@71 }
\def\llcorner{\delimiter"4\msa@78\msa@78 }
\def\lrcorner{\delimiter"5\msa@79\msa@79 }
\def\yen{\mathhexbox\msa@55 }
\def\checkmark{\mathhexbox\msa@58 }
\def\circledR{\mathhexbox\msa@72 }
\def\maltese{\mathhexbox\msa@7A }
\mathchardef\lvertneqq="3\msb@00
\mathchardef\gvertneqq="3\msb@01
\mathchardef\nleq="3\msb@02
\mathchardef\ngeq="3\msb@03
\mathchardef\nless="3\msb@04
\mathchardef\ngtr="3\msb@05
\mathchardef\nprec="3\msb@06
\mathchardef\nsucc="3\msb@07
\mathchardef\lneqq="3\msb@08
\mathchardef\gneqq="3\msb@09
\mathchardef\nleqslant="3\msb@0A
\mathchardef\ngeqslant="3\msb@0B
\mathchardef\lneq="3\msb@0C
\mathchardef\gneq="3\msb@0D
\mathchardef\npreceq="3\msb@0E
\mathchardef\nsucceq="3\msb@0F
\mathchardef\precnsim="3\msb@10
\mathchardef\succnsim="3\msb@11
\mathchardef\lnsim="3\msb@12
\mathchardef\gnsim="3\msb@13
\mathchardef\nleqq="3\msb@14
\mathchardef\ngeqq="3\msb@15
\mathchardef\precneqq="3\msb@16
\mathchardef\succneqq="3\msb@17
\mathchardef\precnapprox="3\msb@18
\mathchardef\succnapprox="3\msb@19
\mathchardef\lnapprox="3\msb@1A
\mathchardef\gnapprox="3\msb@1B
\mathchardef\nsim="3\msb@1C
\mathchardef\napprox="3\msb@1D
\mathchardef\nsubseteqq="3\msb@22
\mathchardef\nsupseteqq="3\msb@23
\mathchardef\subsetneqq="3\msb@24
\mathchardef\supsetneqq="3\msb@25
\mathchardef\subsetneq="3\msb@28
\mathchardef\supsetneq="3\msb@29
\mathchardef\nsubseteq="3\msb@2A
\mathchardef\nsupseteq="3\msb@2B
\mathchardef\nparallel="3\msb@2C
\mathchardef\nmid="3\msb@2D
\mathchardef\nshortmid="3\msb@2E
\mathchardef\nshortparallel="3\msb@2F
\mathchardef\nvdash="3\msb@30
\mathchardef\nVdash="3\msb@31
\mathchardef\nvDash="3\msb@32
\mathchardef\nVDash="3\msb@33
\mathchardef\ntrianglerighteq="3\msb@34
\mathchardef\ntrianglelefteq="3\msb@35
\mathchardef\ntriangleleft="3\msb@36
\mathchardef\ntriangleright="3\msb@37
\mathchardef\nleftarrow="3\msb@38
\mathchardef\nrightarrow="3\msb@39
\mathchardef\nLeftarrow="3\msb@3A
\mathchardef\nRightarrow="3\msb@3B
\mathchardef\nLeftrightarrow="3\msb@3C
\mathchardef\nleftrightarrow="3\msb@3D
\mathchardef\divideontimes="2\msb@3E
\mathchardef\varnothing="0\msb@3F
\mathchardef\nexists="0\msb@40
\mathchardef\mho="0\msb@66
\mathchardef\thorn="0\msb@67
\mathchardef\beth="0\msb@69
\mathchardef\gimel="0\msb@6A
\mathchardef\daleth="0\msb@6B
\mathchardef\lessdot="3\msb@6C
\mathchardef\gtrdot="3\msb@6D
\mathchardef\ltimes="2\msb@6E
\mathchardef\rtimes="2\msb@6F
\mathchardef\shortmid="3\msb@70
\mathchardef\shortparallel="3\msb@71
\mathchardef\smallsetminus="2\msb@72
\mathchardef\thicksim="3\msb@73
\mathchardef\thickapprox="3\msb@74
\mathchardef\approxeq="3\msb@75
\mathchardef\succapprox="3\msb@76
\mathchardef\precapprox="3\msb@77
\mathchardef\curvearrowleft="3\msb@78
\mathchardef\curvearrowright="3\msb@79
\mathchardef\digamma="0\msb@7A
\mathchardef\varkappa="0\msb@7B
\mathchardef\hslash="0\msb@7D
\mathchardef\hbar="0\msb@7E
\mathchardef\backepsilon="3\msb@7F
\def\Bbb{\ifmmode\let\next\Bbb@\else
 \def\next{\errmessage{Use \string\Bbb\space only in math mode}}\fi\next}
\def\Bbb@#1{{\Bbb@@{#1}}}
\def\Bbb@@#1{\fam\msbfam#1}

\catcode`\@=\active

\def\inv{^{\raise.15ex\hbox{${
  \scriptscriptstyle -}$}\kern-.05em 1}}

\def\Dsl{\,\raise.15ex\hbox{$/$}\mkern-13.5mu D}
\def\dsl{\raise.15ex\hbox{$/$}\kern-.57em\hbox{$\partial$}}

\def\lspace{\ifx\answ\bigans{}\else\qquad\fi}
\def\del{\partial}
\def\CF{\hbox{{$\cal F$}}} 
\def\CL{\hbox{{$\cal L$}}}

\def\CR{\hbox{{$\cal R$}}}

 \def\CZ{\hbox{{$\cal Z$}}}

\def\cv{{\vartheta}} % used for special element v

\def\cg{\hbox{{\sl g}}} % used for Lie algebra 'gothic g'

\def\lform{\hbox{$\sqcup$}\llap{\hbox{$\sqcap$}}}
\def\darr#1{\raise1.5ex\hbox{$\leftrightarrow$}
\mkern-16.5mu #1}
\def\h{{{1\over2}}}

\def\INT{{\textstyle \int\kern-.642em\int}}

\def\R{{\Bbb R}}
\def\C{{\Bbb C}}
\def\Z{{\Bbb Z}}

\def\Rrel{{{\bf R'}}}
\def\Rmul{{{{\bf R}_\cdot}}}
\def\Radd{{{\bf R}}}

\def\eps{{\epsilon}}

\def\trace{{\rm Tr\, }}

\def\dcross{{\bowtie}}

\def\rbiprod{{\cdot\kern-.33em\triangleright\!\!\!<}}
\def\lbiprod{{>\!\!\!\triangleleft\kern-.33em\cdot}}

\def\tens{\mathop{\otimes}}

\def\la{{\triangleright}}

\def\extd{{{\rm d}}}

\def\isom{{\cong}}

\def\ev{{\rm ev}}
\def\coev{{\rm coev}}

\def\id{{\rm id}}

\def\<{\langle}
\def\>{\rangle}
\def\dila{{\varsigma}}

\def\haj#1{{\mathaccent20 {#1}}}
\def\Vhaj{{V\haj{\ }}}

\def\veca{{\bf a}}
\def\vecb{{\bf b}}
\def\vecc{{\bf c}}
\def\vecd{{\bf d}}
\def\vect{{\bf t}}\def\vecs{{\bf s}}\def\vecv{{\bf v}}
\def\vecu{{\bf u}}\def\vecx{{\bf x}}\def\vecp{{\bf p}}
\def\vecl{{\bf l}}\def\vecy{{\bf y}}\def\veca{{\bf a}}
\def\vecw{{\bf w}}
\def\vecf{{\bf f}}

\def\<{\langle}
\def\>{\rangle}

\def\equad{\kern -1.7em}

\def\nquad{{\!\!\!\!\!\!}}

\def\o{{}_{\scriptscriptstyle(1)}}
\def\t{{}_{\scriptscriptstyle(2)}}
\def\th{{}_{\scriptscriptstyle(3)}}
\def\fo{{}_{\scriptscriptstyle(4)}}
\def\fiv{{}_{\scriptscriptstyle(5)}}

\def\und#1{{\underline {#1}}}

\def\uo{{{}^{\scriptscriptstyle(1)}}}
\def\ut{{{}^{\scriptscriptstyle(2)}}}

\def\Bo{{{}_{\und{\scriptscriptstyle(1)}}}}
\def\Bt{{{}_{\und{\scriptscriptstyle(2)}}}}

\def\text#1{\mbox{\rm #1}}
\def\note#1{}

\def\blacksquare{{\lform}}%AMS Tex Fakes
\def\frac#1#2{{{#1\over#2}}}

\def\proof{\goodbreak\noindent{\bf Proof\quad}}

\def\endproof{{\ $\lform$}\bigskip }

\def\eqn#1#2{\begin{equation}#2\label{#1}\end{equation}}
\def\align#1{\begin{eqnarray*}#1\end{eqnarray*}}
\def\alignn#1#2{\begin{eqnarray}\label{#1}#2
\end{eqnarray}}
\def\cmath#1{\[\begin{array}{c} #1 \end{array}\]}
\def\ceqn#1#2{\begin{equation}\label{#1}\begin{array}{c}#2\end{array}
\end{equation}}

\documentstyle[epsf]{article}
\newtheorem{lemma}{Lemma}[section]

\newtheorem{example}[lemma]{Example}

\begin{document}
\section*{\Large Introduction to Braided Geometry and $q$-Minkowski Space}
\medskip
\smallskip

\noindent S. Majid\footnote{Royal Society University Research Fellow and Fellow
of
Pembroke College, Cambridge. This paper is in final form and no version of it
will be submitted for publication elsewhere}

\medskip

\noindent  Department of Applied Mathematics \& Theoretical Physics, University
of Cambridge, Cambridge CB3 9EW, UK
\bigskip
\medskip

\noindent{\bf Abstract} We present a systematic introduction to the geometry of
linear braided spaces. These are versions of $\R^n$ in which the coordinates
$x_i$ have braid-statistics described by an R-matrix. From this starting point
we survey the author's braided-approach to $q$-deformation: braided
differentiation, exponentials, Gaussians, integration and forms, i.e. the basic
ingredients for $q$-deformed physics are covered. The braided approach includes
natural $q$-Euclidean and $q$-Minkowski spaces in R-matrix form.

\bigskip
\noindent Keywords: quantum groups -- noncommutative
geometry -- braided geometry -- $q$-Minkowski -- $q$-Euclidean

\tableofcontents

% Leonardo/Editor: obviously the page numbers will change (if you keep the
% contents)
% better, I'd recommend an overall table of contents for the entire
% volume. That, for all the contributions, would be really useful
% and unique if they are lectures. Shahn

\section{Introduction}

It is often thought that quantum groups provide the key to $q$-deforming the
basic structures of physics from the point of view of non-commutative geometry.
If one considered a classical algebra of observables and quantised it relative
to some Poisson bracket, one might obtain a quantum group. The underlying
semiclassical theory is the theory of Poisson-Lie groups -- see Reyman's
lectures and Reshetikhin's lectures on classical inverse scattering. But this
is only part of the story. Our goal in these lectures is to explain  that the
fundamental concept needed for the full structure of even the simplest
$q$-deformed spaces, such as the quantum plane, is not so much a quantum group
as
one of the more exotic objects called {\em braided groups}. These were
introduced by the author in 1989\cite{Ma:rec} and subsequently developed in the
course of 40
or so papers into a systematic theory of {\em braided geometry}. Quantum groups
play a background role in this theory as the {\em  quantum symmetry} or
covariance of the geometry, but the spaces themselves tend to be braided ones.

 My intention is to provide here a pedagogical introduction to this theory of
braided geometry. Braided groups provide a new beginning for the theory of
$q$-deformation and can be developed along-side quantum groups without
requiring much experience of them. Instead, some experience with Grassmann
algebras or supersymmetry will be quite helpful although not essential. We try
to cover here only $q$-deformed or braided versions of $\R^n$, where the theory
is fairly complete. This includes important examples such as $q$-Euclidean and
$q$-Minkowski space. Only when this linear theory is thoroughly understood
could one reasonably expect to move on to define $q$-manifolds etc. For some
first steps in quantum geometry, see \cite{BrzMa:gau}. Braided Yang-Mills
theory on a general braided manifold is not yet understood.

We begin with the concept itself of a braided
group. This is a new concept. On the one hand we replace old ideas form the
theory of superspaces by similar ones with braid statistics in place of
Bose-Fermi ones. This makes it easy for the reader to get the idea of braided
groups. On the other hand the true meaning and abstract definition of braided
groups involves writing its algebraic structure  diagrammatically as a joining
of strings (the product) or a splitting of strings (the coproduct or
coaddition). All information flows along these strings which can form braids
and knots. Each braid crossing $\Psi$ corresponds to a $q$-factor or more
generally, to an R-matrix. This is much more fun and more systematic than
trying to introduce $q$ or an R-matrix by guesswork or by other ad-hoc means,
which is the usual approach to $q$-deforming physics. We will not see too much
of
this diagrammatic side here, since we will try for a more hands-on and less
abstract treatment. One can see \cite{Ma:introp} for the diagrammatic theory as
well as for a review of braided groups up to about mid 1992. Section~2 below
provides the briefest of introductions. In addition, there are two introductory
papers \cite{Ma:carib}\cite{Ma:mex} in conference proceedings, which cover the
braided-groups programme since then. The present work is based in part on
Chapter~10 of my forthcoming book\cite{Ma:book}.

\subsection{Why $q$-deform?}

There are several reasons to want to $q$-deform the basic structures of physics
in the first place. We outline some of them here.

\begin{itemize}
\item To begin with it is simply a fact that many of our usual concepts of
geometry are a special $q=1$ case of something more general which works just as
well, i.e. mathematically we can $q$-deform and have no particular reason to
limit ourselves to $q=1$ in every physical situation.

\item The $q\ne 1$ world seems to be less singular than the $q=1$ world:
perhaps some of the infinities we encounter in quantum field theory are really
poles in $1\over q-1$ and appear singular because we used $q=1$ geometry in
the bare theory. This has two points of view.

\subitem (a) It may be that  the real world is only $q=1$ and that expressing
infinities in this way as poles is a mathematical tool of
$q$-regularisation\cite{Ma:reg}. Even so it is useful because $q$-deformation
is elegant and (in the braided approach) systematic. We will see that one of
the themes of the $q$-deformed world is that $q$-deformed quantities bear the
same
mathematical relationships with each other as in the undeformed case. So we do
not do serious damage to the mathematical structure as is done in more physical
but brutal regularisation methods such momentum cut-off. Also, we do not have
ad-hoc problems like what to do with the $\eps$ tensor as in dimensional
regularisation. In this context it is fitting that $q$ is dimensionless and
`orthogonal to physics'.

\subitem (b) It may be that really $q\ne 1$ as a crude model of quantum or
other corrections to our usual concept of geometry. In this case $q$ could be
an exponential of the ratio of masses in our system to the Planck mass, for
example. Quantum groups do have explicit connections with Planck-scale physics.
We do not cover this here, but see \cite{Ma:pla} where this connection was
introduced  for the first time.

\item Some physical models are harder to $q$-deform than others. The principle
of
$q$-deformisability or continuity of physics at $q=1$ may help to single out
some
Lagrangians as more natural than others. Some physical Lagrangians may be based
for example on accidental isomorphisms at $q=1$ in the various families of Lie
groups: such degeneracies tend to be removed by $q$-deformation.

\item $q$-deformation and quantum or braided geometry in general unifies
concepts. Thus ideas which at $q=1$ are quite different, may in fact be
isomorphic as soon as $q\ne 1$. In particular, the concept in physics of
covariance or symmetry is one and the same as the concept of statistics or
grading (as in supersymmetry) when both are expressed in the language of Hopf
algebras\cite{Ma:introp}.

\item Related to this, there are possible some very spectacular `self-duality'
unifications of particular algebras. Thus the enveloping algebra of
$SU(2)\times U(1)$ becomes isomorphic to the co-ordinates $x_\mu$ of
$q$-Minkowski space when both are $q$-deformed in a natural way within braided
geometry\cite{Ma:lie}\cite{Ma:mex}.

\end{itemize}

The reader should bear in mind all of these ideas as well as any others she or
he can think of. We will see the ones above realised to some extent below.

\subsection{What is braided geometry?}

Keeping in mind the above ideas, how can we develop a systematic and universal
approach to $q$-deforming structures in physics? Braided geometry claims to do
this. Here we explain the key idea behind it and where it may be that more
fashionable ideas such as non-commutative geometry went wrong.

The point is that in our experience in quantum physics there are in fact {\em
two} kinds of non-commutativity which we encounter. The first of these I
propose to call {\em  inner noncommutativity} or {\em  noncommutativity of the
first kind} because it is a property within a quantum system or algebra. It is
the kind that we encounter when we start with a classical algebra of
observables and quantise it by making it non-commutative. It is customary to
make an analogy with this process of quantisation by considering algebras in
which there is a parameter $q\ne 1$ analogous to $\hbar\ne 0$. In mathematical
terms, an algebra is regarded as like functions on a manifold, but all
geometrical constructions are developed in such a way that the algebra need not
be commutative and hence need not really be the algebra of functions on any
space. It could, for example, be an algebra arising by quantisation, but this
is not a prerequisite.
In this usual formulation of {\em noncommutative geometry} the tensor product
of algebras (corresponding to direct product of the manifolds) is the usual one
in which the factors commute. It is the algebras themselves which become
noncommutative.

The idea of {\em braided geometry} is to associate $q$ not with quantisation
but rather with a different {\em  outer noncommutativity} that can exist
between independent systems. This is {\em  noncommutativity of the second kind}
and is encountered in physics when we consider fermions: independent fermionic
systems anticommute rather than commute. So the idea is to consider $q$ as a
generalisation of the $-1$ factor for fermions. In mathematical terms it is the
notion of $\tens$ product between algebras which we will $q$-deform and not
directly the algebras themselves. These, as far as we are concerned, can remain
classical or `commutative' albeit in a deformed sense appropriate to the
noncommutative tensor product.

This is conceptually quite a different role for $q$ than its usual picture as
quantisation. It turns out to be the key if one wants to $q$-deform not one
algebra but an entire universe of structures: lines, planes, matrices,
differentials etc in a systematic and mutually consistent way. The reason is
that we can use the systematic machinery of braided categories to deform the
entire category of vector spaces with its usual $\tens$ to a {\em braided
category} with tensor product $\tens_q$. Most constructions in physics and many
in mathematics take place in the category of vector spaces, so by deforming the
category itself we carry over all our favourite constructions without any
further effort. The reader should not be afraid of the term `category' here. It
just means a collection of objects of some specified type. The outer
non-commutativity is manifested in the construction (due to the author) of the
{\em  braided tensor product algebra structure} $B\und\tens C$ of two braided
algebras $B,C$. The tensor product is physically the joint system and contains
$B,C$ as subalgebras. But the notion of braided-independence or {\em braid
statistics} means that the two factors do not mutually commute as they would in
a usual tensor product. The concept here is obviously quite general and is not
tied to a single parameter $q$: its role can be played by a general matrix or
collection of matrices $R$ obeying suitable braid relations. So we develop in
fact a braided theory of $R$-deformation. The standard R-matrices depend on a
single parameter $q$ but the reader can just as easily put in multi-parameter
or non-standard R-matrices into our formalism.

It should be clear by now that this new approach to deformation is quite
independent of, or orthogonal to, the usual role of quantum groups and
non-commutative geometry. Quantum groups play no very direct role in braided
geometry and moreover, the fundamental concepts here did {\em not} arise in the
context of Quantum Inverse Scattering where quantum groups arose. The point of
contact is covariance, which we come to at the end of our studies. Our starting
point, which is that of a braided group, is due to the
author\cite{Ma:eul}\cite{Ma:exa}\cite{Ma:bra} and came out of experience with
fermionic systems and supersymmetry. As well as being important to keep the
history straight (now that these ideas have become popular for physicists) it
is also important mathematically because the two kinds of non-commutativity
here are not at all mutually exclusive. They are orthogonal in the sense that
one can just as well have {\em quantum braided groups} in which both ideas are
present. We will not emphasise this here, but see
\cite{Ma:tra}\cite{Ma:any}\cite{Hla:bra}\cite{MaPla:any} and the appendix.

Let us note also that our point of view on $q$ does not preclude the
possibility that other physical effects may induce these braid statistics. We
have discussed various physical reasons to consider $q\ne 1$ in the previous
subsection. The fact is that any of these lead us to $q$-deform geometry and in
this $q$-deformed world the usual spin-statistics theorem fails. Braid
statistics
are allowed and indeed are a general feature of $q$-deformation.

\section{Diagrammatic definition of a braided group}

I would like to  begin with a lightening sketch of the abstract definition of a
braided group. This is not essential for the later sections, so the reader who
wants to learn the definition by experience should proceed directly to the next
section where we see lots of examples. Even so, it is useful to know that there
is a firm mathematical foundation to this concept\cite{Ma:bg}\cite{Ma:tra} and
this is what we outline here. For much more detail on this topic, see
\cite{Ma:introp}.

\begin{figure}
\[
\epsfbox{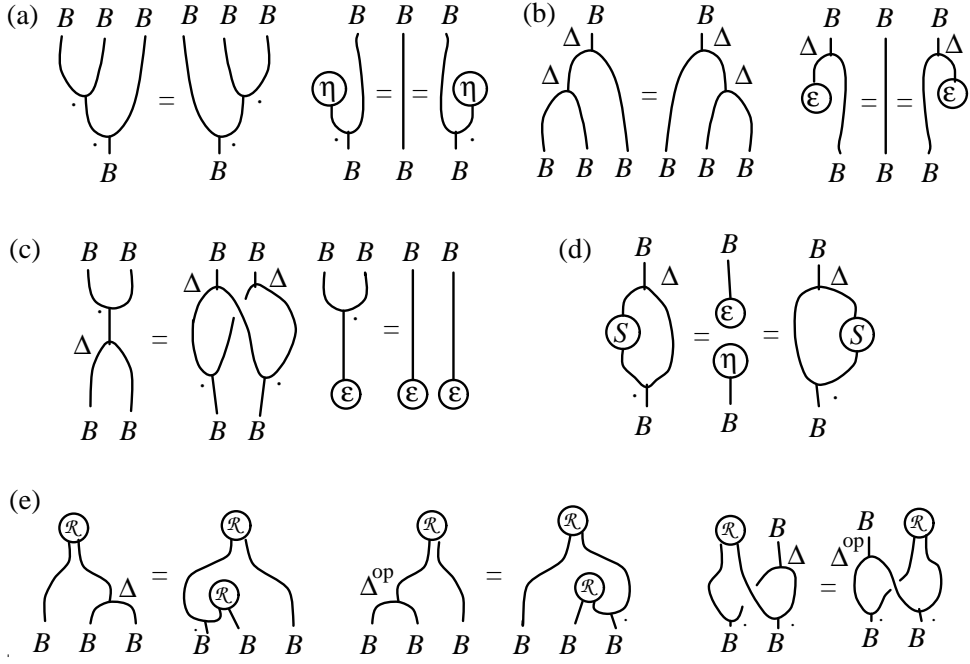}
\]
\caption{Axioms of a braided group showing (a) associativity and unit (b)
coassociativity and counit (c) braided homomorphism property (d) antipode (e)
quasitriangular structure}
\end{figure}

The axioms of a braided group $B$ are summarised in parts (a) -- (d) of
Figure~1 in a diagrammatic notation. We write morphisms or maps pointing
downwards. There is a product $\cdot=\epsfbox{prodfrag.eps}$ which should be
associative and have a unit $\eta$ as we see in part (a). This is a braided
algebra. The axiom for the unit says that grafting it on via the product map
does not change anything. In addition we should have a {\em coproduct}
$\Delta=\epsfbox{deltafrag.eps}$ which should be coassociative, and a counit
$\eps$. This is shown in part (b), which is just part (a) up-side-down. This is
a braided coalgebra. These two structures should be compatible in the sense
that $\Delta,\eps$ are braided-multiplicative as shown in part (c). In concrete
terms this means
\eqn{brahom}{ \Delta(ab)=(\Delta a)(\Delta b),\quad (a\tens c)(b\tens
d)=a\Psi(c\tens b)d}
which says that $\Delta$ is a homomorphism from $B$ to the braided tensor
product algebra $B\und\tens B$. The braid crossing here corresponds to an
operator $\Psi=\epsfbox{braid.eps}$ obeying the braid relations. We can pull
nodes through such braid crossings as if they are on strings in a
three-dimensional space. This space is not physical space but an abstract space
in which braided mathematics is written. Sometimes we also have an antipode or
`inversion map' obeying the axioms in part (d). It turns out that all the
elementary group theory that the reader is familiar with can be developed in
this diagrammatic setting, including representations or modules, adjoint
actions, cross products etc. For example, Figure~2 shows the proof of the
property
\eqn{braSanti}{ S\circ\cdot=\cdot\circ\Psi_{B,B}\circ (S\tens S),\quad
\Delta\circ S=(S\tens S)\circ\Psi_{B,B}\circ \Delta}
\begin{figure}
\[
\epsfbox{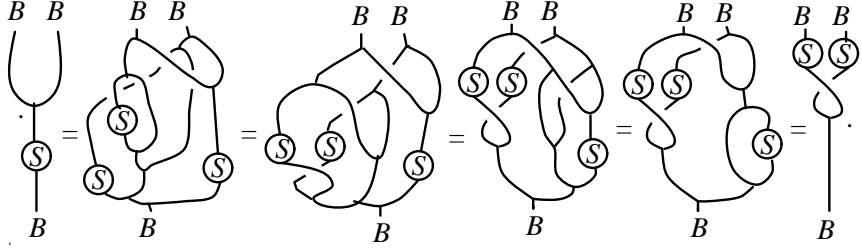}
\]
\caption{Diagrammatic proof of braided antihomomorphism property of $S$}
\end{figure}
which we will need  later. The proof grafts on two loops involving $S$, knowing
that they are trivial from  Figure~1(d). After some reorganisation using parts
(a)--(b), we use~(c) and then~(d) again for the final result. For the second
half
of (\ref{braSanti}) just turn this volume up-side-down and read Figure~2 again.
On the more esoteric side, we sometimes also have a braided universal-R-matrix
or
quasitriangular structure shown in Figure~1(e).

The simplest example of such a braided group is the {\em braided line}.  This
is just the polynomials in a single variable $x$ endowed with \[  \Delta
x=x\tens 1+1\tens x,\quad \eps(x)=0,\quad Sx=-x,\quad \Psi(x^m\tens
x^n)=q^{mn}x^n\tens x^m.\]
The first three look on the generator $x$ the same as the usual definitions for
functions in one variable. The coproduct corresponds in this usual case to
addition on the underlying copy of $\R$ for which $x$ is the linear co-ordinate
function. The new ingredient is the braiding $\Psi$ and means for example that
\[ \Delta x^m=\sum_{r=0}^m [{m\atop r};q]  x^r\tens  x^{m-r},\quad
Sx^m=q^{m(m+1)\over 2}(-x)^m.\]
We see here the origin in braided geometry of the $q$-integers  and
$q$-binomials
 \[  [m;q]={1-q^m\over 1-q},\quad [{m\atop r};q]={[m;q]!\over [r;q]![m-r;q]!}\]
which are familiar when working with $q$-deformations. It turns out that many
formulae in $q$-deformed analysis, such as differentiation, integration etc.
are
immediately recovered once one takes the braided point of view. In this example
$q$ is arbitrary but non-zero. If we take $q^2=1$ we can consistently add the
relation $x^2=0$ which gives us the usual Grassmann algebra in one variable,
i.e. the super-line. If we take $q^n=1$ we can consistently add the relation
$x^n=0$ and arrive at the {\em anyonic
line}\cite{Ma:any}\cite{MaPla:ran}\cite{MaPla:any}.

The next simplest example is the braided plane $B$ generated by $x,y$
with\cite{Ma:poi}
\cmath{ yx=qxy,\quad \Delta x=x\tens 1+1\tens x,\quad \Delta y=y\tens 1+1\tens
y\\
\eps x=\eps y=0,\quad Sx=-x,\quad  Sy=-y\\
 \Psi(x\tens x)=q^2 x\tens x,\quad \Psi(x\tens y)=q y\tens x,\quad
\Psi(y\tens y)=q^2 y\tens y \\
\Psi(y\tens x)=q x\tens y+(q^2-1)y\tens x}
The algebra here is sometimes called the `quantum plane'; the new part is the
coproduct $\Delta$ and the braiding $\Psi$. The latter is the same one that
leads to the Jones knot polynomial or the quantum group $SU_q(2)$ in another
context. It is a nice exercise for the reader to verify that $\Delta$ is indeed
an algebra homomorphism using the braided tensor product (\ref{brahom}). Again,
this seems innocent enough but has the result that we generalise to
2-dimensions all the familiar ideas from one-dimensional $q$-analysis. We will
see this quite generally in the next section for $n$-dimensions and general
braidings. This is one of the successes of the theory of braided groups.

The coproducts $\Delta$ in these examples are linear on the generators. They
could better be called {\em coaddition}. All the interesting coadditions I know
are braided ones. If they were not braided, they would have to be cocommutative
and hence correspond essentially to ordinary Lie algebras and not quantum
groups. This is why we need braided groups as the foundation of braided
geometry. There are also plenty of other more complicated braided groups,
including a canonical one for every strict quantum group by a transmutation
construction\cite{Ma:bra}. In this way the theory of braided groups contains
braided versions of the quantum groups $U_q(\cg)$ for example,  and is a good
way of getting to grips with their geometry as well\cite{Ma:lie}. One can also
make partial transmutations to obtain any number of other (quantum) braided
groups which lie in between quantum groups and their completely transmuted
braided group versions. The theory of transmutation is covered in the Appendix.

\section{Braided coaddition}

We describe in this section the basic braided groups which will be the object
of our study. We begin with deformations of  co-ordinates $x_i$ or vectors
$v^i$, i.e. versions of $\R^n$. In the braided world there are many such
versions depending on the precise commutation relations of the algebra and the
precise braid statistics, which we encode  by matrices $R',R$ respectively. In
Sections~3.2 and~3.3 we give braided versions of $\R^{n^2}$ using the same
formalism on a matrix of generators.

\subsection{Braided coaddition on vectors and covectors}

Let $R,R'$ be invertible matrices in $M_n\tens M_n$. We suppose that they
obey\cite{Ma:poi}
\eqn{QYBE}{ R_{12}R_{13}R_{23}=R_{23}R_{13}R_{12}}
\eqn{coveca}{ R_{12}R_{13}R'_{23}=R'_{23}R_{13}R_{12},\quad
R_{23}R_{13}R'_{12}=R'_{12}R_{13}R_{23}}
\eqn{covecb}{(PR+1)(PR'-1)=0}
\eqn{covecc}{R_{21}R' =R'_{21}R }
where $P$ is the permutation matrix. The suffices refer to the position in
tensor power of $M_n$. Thus in (\ref{QYBE}), which is called the {\em Quantum
Yang-Baxter Equations (QYBE)}, we have $R_{12}=R\tens \id$ and $R_{23}=\id\tens
R$ etc.

It is pretty easy to solve these equations. Just start for example with a
matrix $R$ solving the QYBE. Any matrix $PR$ necessarily obeys some minimal
polynomial $\prod_i(PR-\lambda_i)=0$ and for each nonzero eigenvalue
$\lambda_i$ we can just normalise $R$ so that $\lambda_i=-1$ and take
\eqn{R'(R)}{ R'=P+P\prod_{j\ne i} (PR-\lambda_j).}
This clearly solves (\ref{coveca})--(\ref{covecc}) and gives us at least one
braided covector space for each nonzero eigenvalue of $PR$. The simplest case
is when there are just two eigenvalues, which is called the {\em Hecke} case.

Given a solution of (\ref{QYBE})--(\ref{covecc}) we have the {\em
braided-covector algebra} $\Vhaj(R',R)$ defined by generators $1,x_i$ and
relations and braided group structure\cite{Ma:poi}
\ceqn{xhopfa}{ x_ix_j=x_bx_aR'{}^a{}_i{}^b{}_j,\quad {\it i.e.,}\quad
\vecx_1\vecx_2=\vecx_2\vecx_1R' \\
 \Delta x_i=x_i\tens 1+1\tens x_i,\quad \eps x_i=0,\quad Sx_i=-x_i\\
\Psi(x_i\tens x_j)=x_b\tens x_a R^a{}_i{}^b{}_j,\quad {\it i.e.,}\quad
\Psi(\vecx_1\tens \vecx_2)=\vecx_2\tens \vecx_1R}
extended multiplicatively with braid statistics. We use the compact notation
shown on the right were bold $\vecx$ refers to the entire covector and its
numerical suffices to the position in a tensor product of indices.

Next we introduce a notation for this map $\Delta$. It is a homomorphism from
the algebra to two copies of the algebra. If we denote the
generators of the first copy by $x_i\equiv x_i\tens 1$ and the generators of
the
second copy by $x_i'\equiv 1\tens x_i$ then the assertion that $\Delta$ of the
above linear form is a homomorphism
is just that\cite{Ma:poi}\cite{Ma:fre}
\eqn{xhopfb}{ x''_i=x_i+x'_i,\quad {\it i.e.},\quad \vecx''=\vecx+\vecx'}
obey the same relations of $\Vhaj(R',R)$. In other words, we can treat our
noncommuting generators $x_i$ like row vector coordinates and add them,
provided we remember that in the braided tensor product they do not commute but
rather obey the
{\em  braid statistics}
\eqn{xhopfc}{ x'_ix_j=x_bx_aR{}^a{}_i{}^b{}_j,\quad {\it i.e.,}\quad
\vecx'_1\vecx_2=\vecx_2\vecx'_1R .}
This is the most compact way of working with our braided groups. We can really
add them and treat them like
covectors provided we have the appropriate braid statistics between independent
copies. In this notation, the essential fact that the coproduct extends to
products as a well-defined braided Hopf algebra is checked as
\align{ \vecx''_1\vecx''_2\equad
&&=(\vecx_1+\vecx'_1)(\vecx_2+\vecx'_2)=\vecx_1\vecx_2+\vecx'_1\vecx'_2
+\vecx_1\vecx'_2+\vecx_2\vecx'_1R\\
\vecx''_2\vecx''_1R'\equad&&=(\vecx_2+\vecx'_2)(\vecx_1+\vecx'_1)R'
=\vecx_2\vecx_1R'+\vecx'_2\vecx'_1R'+\vecx_2\vecx'_1R'
+\vecx_1\vecx'_2R_{21}R'}
which indeed coincide by (\ref{covecb}). Note that there is a lot more to be
checked for a braided-Hopf algebra. For example,  we also have to check that
$\Psi$ likewise extends consistently to products in such a way as
to be functorial with respect to the product map. Details are in \cite{Ma:poi}.
But the homomorphism property is the most characteristic and the one which we
stress here.

The simplest example is provided by the 1-dimensional matrices $R=(q)$, $R'=1$,
where $q$ is arbitrary but non-zero. This is the braided line which was given
more explicitly in Section~2. The braided plane also given there is likewise an
example of the above:

\begin{example}\cite{Ma:poi} The standard quantum plane algebra $\C_q^{2|0}$
with relations $yx=qxy$ is
a braided-covector algebra with
\[ x' x=q^2 x x',\quad x' y=q yx',\quad y' y=q^2 y y',\quad y'x=q x y'+(q^2-1)y
x'\]
i.e.,
\[ (x'',y'')=(x,y)+(x',y')\]
obeys the same relations provided we remember these braid statistics.
\end{example}
\proof We use the standard solution of the QYBE associated to the Jones knot
invariant and the quantum group $SU_q(2)$ in another context, namely
\[ R=\pmatrix{q^2&0&0&0\cr0&q&q^2-1&0 \cr 0&0&q&0\cr0&0&0&q^2},\qquad
R'=q^{-2}R\]
which we put into the above. The algebra $\C_q^{2|0}$ here is a well-known and
much-studied one: the new features are the addition law and the
braid-statistics. \endproof

\begin{example} The mixed quantum plane $\C_q^{1|1}$ with relations
$\theta^2=0$, $\theta x=q x\theta$ is a
braided covector algebra with
\[ x' x=q^2 x x',\quad x' \theta=q \theta x',\quad \theta' \theta=- \theta
\theta',\quad \theta'x=q x \theta'+(q^2-1)\theta x'\]
i.e.,
\[ (x'',\theta'')=(x,\theta)+(x',\theta')\]
obeys the same relations provided we remember these braid statistics.
\end{example}
\proof We use the solution of the QYBE associated to the Alexander-Conway knot
invariant in another context\cite{CLS:con}, namely
\[ R=\pmatrix{q^2&0&0&0\cr0&q&q^2-1&0 \cr 0&0&q&0\cr0&0&0&-1},\qquad
R'=q^{-2}R\]
which we put into the above.
\endproof

\begin{example} The usual mixed $1|1$-superplane with relations $x\theta=\theta
x$, $\theta^2=0$ is a braided covector algebra with
\[ x' x=q^2 x x',\quad x' \theta=q^2\theta x',\quad \theta' \theta=- \theta
\theta',\quad \theta'x=x \theta'+(q^2-1)\theta x'\]
i.e.,
\[ (x'',\theta'')=(x,\theta)+(x',\theta')\]
obeys the same relations provided we remember these braid statistics.
\end{example}
\proof We use the close cousin of the preceding R-matrix,
\[ R=\pmatrix{q^2&0&0&0\cr0&q^2&q^2-1&0 \cr 0&0&1&0\cr0&0&0&-1},\qquad
R'=q^{-2}R\]
which we put into the above. This example is interesting because like the
braided line in Section~2, the $q$-deformation enters
only into the braid statistics while the algebra is the usual one. \endproof

\begin{example} The fermionic quantum plane $\C_q^{0|2}$ with relations
$\theta^2=0$, $\vartheta^2=0$ and $\vartheta\theta=-q^{-1}\theta\vartheta$ is a
braided covector algebra with
\[ \theta' \theta=- \theta \theta',\quad \theta' \vartheta=-q^{-1} \vartheta
\theta',\quad \vartheta' \vartheta=- \vartheta \vartheta',\quad
\vartheta'\theta=-q^{-1} \theta \vartheta'+(q^{-2}-1)\vartheta \theta'\]
i.e.,
\[ (\theta'',\vartheta'')=(\theta,\vartheta)+(\theta',\vartheta')\]
obeys the same relations provided we remember these braid statistics.
\end{example}
\proof We use
\[ R=-q^{-2}\pmatrix{q^2&0&0&0\cr0&q&q^2-1&0 \cr 0&0&q&0\cr0&0&0&q^2},\qquad
R'=q^2R.\]
These are the same R-matrix as in Example~3.1 but with different
normalisations. In fact, we use now for $R$ the matrix which was $-R'$ in
Example~3.1 and vice-versa. We return to this symmetry in Section~5.7.
\endproof

These ideas work just as well for vector algebras with generators $1, v^i$ with
indices up.
So for the same data (\ref{QYBE})--(\ref{covecc}) we have
also a {\em  braided vector algebra} $V(R',R)$ defined with generators $1,v^i$
and relations
\eqn{vhopfa}{ v^iv^j=R'{}^i{}_a{}^j{}_b v^b v^a ,\quad {\it i.e.,}\quad
\vecv_1\vecv_2=R' \vecv_2\vecv_1.}
This has a braided addition law whereby $\vecv''=\vecv+\vecv'$ obeys the same
relations if $\vecv'$ is a second copy with
braid statistics\cite{Ma:poi}
\eqn{vhopfb}{ v'{}^iv^j=R{}^i{}_a{}^j{}_b v^b v'{}^a ,\quad {\it i.e.,}\quad
\vecv'_1\vecv_2=R \vecv_2\vecv'_1.}
More formally, it forms a braided-Hopf algebra with
\ceqn{vhopfc}{ \Delta v^i=v^i\tens 1+1\tens v^i,\quad \eps v^i=0,\quad
Sv^i=-v^i\\
 \Psi(v^i\tens v^j)=R^i{}_a{}^j{}_bv^b\tens v^a,\quad {\it i.e.,}\quad
\Psi(\vecv_1\tens \vecv_2)=R \vecv_2\tens \vecv_1}
extended multiplicatively with braid statistics.  The proof is similar to the
covector case. In the shorthand notation the key braided-homomorphism or
additivity property is checked as
\align{
\vecv''_1\vecv''_2\equad&&=(\vecv_1+\vecv'_1)(\vecv_2+\vecv'_2)
=\vecv_1\vecv_2+\vecv'_1\vecv'_2+\vecv_1\vecv'_2+R\vecv_2\vecv'_1\\
R'\vecv''_2\vecv''_1\equad&&=R'(\vecv_2+\vecv'_2)(\vecv_1+\vecv'_1)
=R'\vecv_2\vecv_1+R'\vecv'_2\vecv'_1+R'\vecv_2\vecv'_1+R'R_{21}
\vecv_1\vecv'_2}
which coincide by (\ref{covecb}). As before, one also has to check other
properties too, such as the fact that $\Psi$ also
extends consistently to products in a natural manner.

\begin{example} The quantum plane $\C_{q^{-1}}^{2|0}$ with relations
$wv=q^{-1}vw$ is a
braided-vector algebra with braid statistics
\[ v'v=q^2vv',\quad v'w=qwv'+(q^2-1)vw',\quad w'v=qvw',\quad w'w=q^2ww'\]
i.e.
\[\pmatrix{v''\cr w''}=\pmatrix{v\cr w}+\pmatrix{v'\cr w'}\]
obeys the same relations.
\end{example}
\proof We take the standard R-matrix as in Example~3.1. Again, the resulting
algebra is standard. To this we now add the braiding and coaddition. \endproof

Similarly for the other standard examples $\C_q^{0|2}$, $\C_q^{1|1}$ etc. The
possibilities are the same as for the covector case. Note that it is a mistake
to think that the vectors are correlated with the fermionic normalisation and
the covectors with the bosonic one: in the braided approach to such algebras we
(a) have more than two types of algebra if $PR$ has more than two eigenvalues
(we will see such examples below) and (b) we have both vectors and covectors
for each choice of eigenvalue or more generally for each pair $R,R'$ obeying
our matrix conditions (\ref{QYBE})--(\ref{covecc}).

A typical application of fermionic co-ordinates in differential geometry is as
describing the properties of forms $\theta=\extd x$. The braided vector version
of Example~3.3 could be viewed for example as the exterior algebra in
1-dimension. It comes out as

\begin{example} The 1-dimensional exterior algebra $\Omega(\C_q)$ with
relations $\extd x^2=0$, $\extd x\, x=q^{-2} x\extd x$ is a braided vector
algebra with
braid statistics
\[ x'x=q^2xx',\quad x'\extd x=q^2\extd x\, x'+(q^2-1)x\extd x',\quad \extd
x'\, x=x\extd x',\quad \extd x'\extd x=-\extd x\extd x'\]
i.e.,
\[\pmatrix{x''\cr \extd x''}=\pmatrix{x\cr \extd x}+\pmatrix{x'\cr \extd x'}\]
obeys the same relations.
\end{example}
\proof We take the same R-matrix as in Example~3.3 but compute the
corresponding vector rather than covector algebra.
\endproof

This example transforms covariantly as a vector under a $q$-deformed
supersymmetry quantum group which mixes $x,\extd x$, here from the right. We
will use a left-covariant version of it later in Example~4.4.

\subsection{Braided coaddition on matrices $A(R)$ and $\bar A(R)$}

We have seen how to coadd vectors and covectors, an idea that was missing
without the braided approach. The same problem occurs for the familiar {\em
quantum matrices} $A(R)$ studied in \cite{FRT:lie} and elsewhere. These are
defined with generators $1,t^i{}_j$ and relations
\eqn{thopfa}{ R^i{}_a{}^k{}_b t^a{}_j t^b{}_l=t^k{}_b t^i{}_a
R^a{}_j{}^b{}_l,\quad {\it i.e.},\quad R\vect_1\vect_2=\vect_2\vect_1 R.}
It is well-known that any algebra of this type (without any condition at all on
$R$) forms an ordinary quantum group with
\eqn{thopfb}{ \Delta_\cdot t^i{}_j=t^i{}_a\tens t^a{}_j,\quad\eps
t^i{}_j=\delta^i{}_j,\quad {\it i.e.},\quad
\Delta_\cdot\vect=\vect\tens\vect,\quad \eps\vect=\id}
and (usually) without an antipode. An ordinary quantum group means we just use
the trivial braiding when extending $\Delta_\cdot$ to products. It means that
\eqn{thopfc}{t''{}^i{}_j=t^i{}_at'{}^a{}_j,\quad [t^i{}_j,t'{}^k{}_l]=0,\quad
{\it i.e.,}\quad \vect''=\vect\vect',\quad [\vect_1,\vect'_2]=0}
is also a realisation of the same algebra if $\vect,\vect'$ are. This
well-known coproduct $\Delta_\cdot$ corresponds to multiplication of matrices
in terms of the possibly non-commuting co-ordinate functions $t^i{}_j$.

But classical matrices can also be added. So what about a corresponding
coaddition law for $A(R)$? Again, this can be handled with braided geometry, at
least when $R$ solves the QYBE and obeys the Hecke condition
\eqn{q-Hecke}{(PR-q)(PR+q^{-1})=0.}
In this case the quantum matrices form in fact a braided covector algebra with
addition law\cite{Ma:add}
\eqn{thopfd}{ t''{}^i{}_j=t^i{}_j+t'{}^i{}_j,\quad {\it i.e.}\quad
\vect''=\vect+\vect'}
obeying the same relations of $A(R)$ provided $\vect'$ is a second copy with
braid statistics
\eqn{thopfe}{ t'{}^i{}_jt^k{}_l=R^k{}_b{}^i{}_a t^b{}_d t'{}^a{}_c
R^c{}_j{}^d{}_l,\quad {\it i.e.},\quad \vect'_1\vect_2=R_{21}\vect_2\vect'_1
R.}
More formally, $A(R)$ is a braided-Hopf algebra with
\eqn{thopff}{ \Psi(\vect_1\tens\vect_2)=R_{21}\vect_2\tens\vect_1R,\quad \Delta
\vect=\vect\tens 1+1\tens \vect,\quad \eps\vect=0,\quad S\vect=-\vect.}
Moreover, the coaddition $\Delta$ is compatible with the usual matrix
comultiplication $\Delta_\cdot$ in the sense\cite{Ma:add}
\ceqn{thopfg}{
(\id\tens\cdot)\circ(\id\tens\tau\tens\id)(\Delta_\cdot\tens\Delta_\cdot)
\circ\Delta=(\Delta\tens\id)\circ\Delta_\cdot\\
(\cdot\tens\id)\circ(\id\tens\tau\tens\id)(\Delta_\cdot\tens\Delta_\cdot)
\circ\Delta=(\id\tens\Delta)\circ\Delta_\cdot}
where $\tau$ the usual transposition map. To see this, we have to show that
$\vect''$ in our short-hand notation obeys the same algebra relations. This is
\align{R(\vect_1+\vect'_1)(\vect_2+\vect'_2)\equad&&=R\vect_1\vect_2+
R\vect'_1\vect'_2+RR_{21}\vect_2\vect'_1R+R\vect_1\vect'_2\\
(\vect_2+\vect'_2)(\vect_1+\vect'_1)R\equad
&&=\vect_2\vect_1R+\vect'_2\vect'_1R+ R\vect_1\vect'_2
R_{21}R+\vect_2\vect'_1 R}
which are equal because $R_{21}R=1+(q-q^{-1})PR$ and $RR_{21}=1+(q-q^{-1})RP$
from the $q$-Hecke assumption. One can also check that $\Psi$  extends
consistently to products in such a way as
to be functorial. Details are in \cite{Ma:add}.

We have given here a direct proof of the coaddition structure on $A(R)$.
Alternatively, we can put it  more explicitly in the braided covector algebra
form (\ref{xhopfa})--(\ref{xhopfc}) by working with the
covector notation $t_I=t^{i_0}{}_{i_1}$ for the generators where $I=(i_0,i_1)$,
$J=(j_0,j_1)$ etc are
multi-indices. Then\cite{Ma:add}
\ceqn{tRR'}{ R\vect_1\vect_2=\vect_2\vect_1R\quad\Leftrightarrow\quad
t_It_J=t_Bt_A{\bf
R'}^A{}_I{}^B{}_J;\quad {\bf
R'}^I{}_J{}^K{}_L=R^{-1}{}^{j_0}{}_{i_0}{}^{l_0}{}_{k_0}
R^{i_1}{}_{j_1}{}^{k_1}{}_{l_1}\\
\vect'_1\vect_2=R_{21}\vect_2\vect'_1R\quad\Leftrightarrow\quad
t'_It_J=t_Bt'_A{\bf R}^A{}_I{}^B{}_J;\quad {\bf
R}^I{}_J{}^K{}_L=R{}^{l_0}{}_{k_0}{}^{j_0}{}_{i_0}
R^{i_1}{}_{j_1}{}^{k_1}{}_{l_1}}
puts $A(R)$ explicitly into the form of a braided covector algebra with $n^2$
generators. We use the bold multi-index $\bf R, R'$ matrices built from our
original $R$. They obey the conditions (\ref{QYBE})--(\ref{covecc}) and also
the additional (\ref{covecd}) just in virtue of the QYBE and $q$-Hecke
condition
on $R$. The corresponding braided vector algebra (\ref{vhopfa})--(\ref{vhopfc})
in matrix form is
\eqn{A(R)vec}{v^I\equiv v^{i_1}{}_{i_0};\qquad
R_{21}\vecv_1\vecv_2=\vecv_2\vecv_1R_{21},\quad
\vecv'_1\vecv_2=R\vecv_2\vecv'_1R_{21}.}

\begin{example} The standard quantum matrices $M_q(2)$ with generators
$\vect=\pmatrix{a&b\cr c&d}$ and relations
\cmath{ab=q^{-1}ba,\quad ac=q^{-1}ca,\quad bd=q^{-1}db,\quad cd=q^{-1}dc\\
bc=cb,\quad ad-da=(q^{-1}-q)bc}
have the usual multiplication law whereby\cite{Man:non}
\[ \pmatrix{a''&b''\cr c''&d''}=\pmatrix{a&b\cr c&d}\pmatrix{a'&b'\cr
c'&d'}\]
obey the same $M_q(2)$ relations provided the second primed copy commutes with
the first copy. They also have a braided addition law whereby\cite{Ma:add}
\[ \pmatrix{a''&b''\cr c''& d''}=\pmatrix{a&b\cr c&d}+\pmatrix{a'&b'\cr
c'&d'}\]
also obeys the relations of $M_q(2)$ provided the second primed copy has the
braid
statistics
\cmath{ a'a=q^2 aa',\quad b'b=q^2bb',\quad c'c=q^2 cc',\quad d'd=q^2 dd'\\
 a'b=qba',\quad a'c=qca',\quad a'd=da',\quad b'd=qdb',\quad c'd=qdc' \\
 b'a=qab'+(q^2-1)ba',\quad b'c=cb'+(q-q^{-1})da'\\
c'a=qac'+(q^2-1)ca',\quad c'b=bc'+(q-q^{-1})da'\\
d'b=qbd'+(q^2-1)db',\quad d'c=qcd'+(q^2-1)dc'\\
d'a=ad'+(q-q^{-1})(cb'+bc')+(q-q^{-1})^2 da'.}
Moreover, this addition law distributes in the expected way over the
multiplication.
\end{example}
\proof We take the standard $R$ as in Example~3.1  but in the
normalisation required for the $q$-Hecke condition (\ref{q-Hecke}), which is
$q^{-1}$ times the one shown in Example~3.1. This is not relevant to the
algebra but is needed for the correct braiding. We then compute from the
formulae (\ref{thopfa})--(\ref{thopff}). \endproof

We have begun with the above quantum matrices $A(R)$ because they are
well-known
quantum groups and probably the reader has seen then somewhere before. But they
are not really the example we need for braided geometry. A more interesting
algebra, which we will need in Section~7.1, is the variant
$\bar A(R)$ studied by the author in \cite{Ma:euc}. It is defined with
generators $1, x^i{}_j$ and
relations
\eqn{ehopfa}{ R^k{}_b{}^i{}_a x^a{}_j x^b{}_l=x^k{}_b x^i{}_a
R^a{}_j{}^b{}_l,\quad {\it i.e.},\quad R_{21}\vecx_2\vecx_2=\vecx_2\vecx_1R}
and forms a braided covector algebra if $R$ is a Hecke solution of the QYBE,
with addition law and braid statistics\cite{Ma:euc}
\eqn{ehopfb}{ \vecx''=\vecx+\vecx';\quad x'{}^i{}_jx^k{}_l=R^i{}_a{}^k{}_b
x^b{}_d x'{}^a{}_c R^c{}_j{}^d{}_l,\quad {\it i.e.},\quad
\vecx'_1\vecx_2=R\vecx_2\vecx'_1R.}
More formally, it forms a braided-Hopf algebra with
\eqn{ehopfc}{ \Psi(\vecx_1\tens\vecx_2)=R\vecx_2\tens\vecx_1R,\quad
\Delta\vecx=\vecx\tens 1+1\tens\vecx,\quad \eps\vecx=0,\quad S\vecx=-\vecx}
To see this we check that $\Delta$ extends to products as an algebra
homomorphism to the braided tensor product algebra, i.e. that $\vecx''$ obeys
the same relations. This is checked as
\align{R_{21}(\vecx_1+\vecx'_1)(\vecx_2+\vecx'_2)\equad&&=R_{21}\vecx_1\vecx_2+
R_{21}\vecx'_1\vecx'_2+R_{21}R\vecx_2\vecx'_1R+R_{21}\vecx_1\vecx'_2\\
(\vecx_2+\vecx'_2)(\vecx_1+\vecx'_1)R\equad
&&=\vecx_2\vecx_1R+\vecx'_2\vecx'_1R+ R_{21}\vecx_1\vecx'_2
R_{21}R+\vecx_2\vecx'_1 R}
which are equal by the $q$-Hecke assumption much as before. We also have to
check
that $\Psi$  extends consistently to products of the generators in such a way
as
to be functorial. This reduces to the QYBE for $R$ along the lines for $A(R)$
in \cite{Ma:add}.

The usual matrix coproduct of $\vecx$ forms neither a quantum group nor a
braided one but something in between. On the other hand, as  before, we can put
the coaddition explicitly into our usual braided covector form by working with
the multi-index notation $x_I=x^{i_0}{}_{i_1}$. Then\cite{Ma:euc}
\ceqn{xRR'}{ R_{21}\vecx_1\vecx_2=\vecx_2\vecx_1R\quad \Leftrightarrow\quad
x_Ix_J=x_Bx_A{\bf
R'}^A{}_I{}^B{}_J;\quad {\bf
R'}^I{}_J{}^K{}_L=R^{-1}{}^{l_0}{}_{k_0}{}^{j_0}{}_{i_0}
R^{i_1}{}_{j_1}{}^{k_1}{}_{l_1}\\
 \vecx'_1\vecx_2=R\vecx_2\vecx'_1R\quad \Leftrightarrow\quad
x'_Ix_J=x_Bx'_A{\bf
R}^A{}_I{}^B{}_J;\quad {\bf R}^I{}_J{}^K{}_L=R{}^{j_0}{}_{i_0}{}^{l_0}{}_{k_0}
R^{i_1}{}_{j_1}{}^{k_1}{}_{l_1}}
puts $\bar A(R)$ into the form of a braided covector algebra with $n^2$
generators. Its corresponding braided vector algebra
(\ref{vhopfa})--(\ref{vhopfc}) in matrix form is $\bar A(R)$ again,
\eqn{barA(R)vec}{v^I\equiv v^{i_1}{}_{i_0};\qquad
R_{21}\vecv_1\vecv_2=\vecv_2\vecv_1R,\quad \vecv'_1\vecv_2=R\vecv_2\vecv'_1R.}

\begin{example}\cite{Ma:euc} The $q$-Euclidean space algebra $\bar M_q(2)$ with
generators
$\vecx=\pmatrix{a&b\cr c&d}$ and relations
\cmath{ba=qab,\quad ca=q^{-1}ac,\quad da=ad,\quad db=q^{-1}bd\quad dc=qcd\\
bc=cb+(q-q^{-1})ad}
has a braided addition law whereby
\[ \pmatrix{a''&b''\cr c''& d''}=\pmatrix{a&b\cr c&d}+\pmatrix{a'&b'\cr
c'&d'}\]
also obeys the relations of $\bar M_q(2)$ provided the second primed copy has
the braid
statistics
\cmath{ c'c=q^2 cc',\quad d'd=q^2dd',\quad a'a=q^2 aa',\quad b'b=q^2 bb'\\
c'd=qdc',\quad c'a=qac',\quad c'b=bc',\quad d'b=qbd',\quad a'b=qba'\\
 d'c=qcd'+(q^2-1)dc',\quad d'a=ad'+(q-q^{-1})bc'\\
a'c=qca'+(q^2-1)ac',\quad a'd=da'+(q-q^{-1})bc'\\
b'd=qdb'+(q^2-1)bd',\quad b'a=qab'+(q^2-1)ba'\\
b'c=cb'+(q-q^{-1})(ad'+da')+(q-q^{-1})^2 bc'.}
\end{example}
\proof We take the standard Jones invariant or $SU_q(2)$  $R$-matrix as in
Example~3.1  but in the
normalisation required for the $q$-Hecke condition (\ref{q-Hecke}), which is
$q^{-1}$ times the one shown there. This is needed for the correct braiding. We
then compute from the formulae (\ref{ehopfa})--(\ref{ehopfc}). \endproof

The interpretation of this standard example $\bar M_q(2)$  as $q$-Euclidean
space will be covered in Section~7.1. The general $\bar A(R)$ construction is
however, more general. A less standard example is:

\begin{example} The algebra $\bar M_q(1|1)$ with generators
$\vecx=\pmatrix{a&b\cr c&d}$ and relations
\cmath{ b^2=0,\quad c^2=0,\quad  ba= a bq,\quad c a= q^{-1}a c ,\quad  db = -q
bd,\quad
 dc=- c dq^{-1}\\
da= a d,\quad bc= cb +(q -q^{-1}) ad}
has a braided addition law whereby
\[ \pmatrix{a''&b''\cr c''& d''}=\pmatrix{a&b\cr c&d}+\pmatrix{a'&b'\cr
c'&d'}\]
also obeys the relations of $\bar M_q(1|1)$ provided the second primed copy has
the braid  statistics
\cmath{
a'a=q^2aa', \quad b'b= - b b',\quad c' c= - cc',\quad d' d=d d'q^{-2}\\
a'b=qb a' ,\quad a'c=ca' q + (q^2 -1) ac',\quad a'd=d a' + (q-q^{-1})b c'\\
b'a= ab' q+(q^2-1)b a', \quad b'c=cb'+ (q-q^{-1})^2bc'+ (q-q^{-1})(da'+ ad')\\
b'd=- q^{-1}d b' + (q^{-2}-1)b d',\quad c' a=qac',\quad c' b=b c',\quad c' d= -
q^{-1}d c'\\
d' a=ad'+ (q-q^{-1})b c',\quad d' b= - q^{-1}b d',\quad d' c=-q^{-1}cd'
+(q^{-2}-1)d c' }
\end{example}
\proof We take the Alexander-Conway $R$-matrix as in Example~3.2  but in the
normalisation required for the q-Hecke condition (\ref{q-Hecke}), which is
$q^{-1}$ times the one shown. This is needed for the correct braiding. We then
compute from the formulae (\ref{ehopfa})--(\ref{ehopfc}). \endproof

\subsection{Braided coaddition on matrices $B(R)$}

Next we consider the {\em braided matrices} $B(R)$ introduced and
studied as a braided group by the author in
\cite{Ma:exa}\cite{Ma:skl}\cite{Ma:lin}.
These are defined with generators $1,u^i{}_j$ and relations
\eqn{uhopfa}{ R^k{}_b{}^i{}_a u^a{}_c R^c{}_j{}^b{}_d u^d{}_l=u^k{}_b
R^b{}_c{}^i{}_a u^a{}_d R^d{}_j{}^c{}_l,\quad{\it i.e.},\quad
R_{21}\vecu_1R\vecu_2=\vecu_2R_{21}\vecu_1R.}
Such relations are perhaps more familiar as among the relations
obeyed by the matrix generators $l^+Sl^-$ of the quantum groups
$U_q(\cg)$ in \cite{FRT:lie} but these
have many other relations too beyond (\ref{uhopfa}) and are not relevant
for us now. They have been used by Zumino and others to describe the
differential calculus on quantum groups; see \cite{Ma:skl} for the full story
here. We are interested instead in (\ref{uhopfa}) purely as a quadratic algebra
with generators $u^i{}_j$ and these relations, which is not in general a
quantum group at all.

The main property of these braided matrices in \cite{Ma:exa}, from which they
take their name, is their multiplicative braided group structure. We have
\cite{Ma:exa}
\ceqn{uhopfb}{ \Delta_\cdot u^i{}_j= u^i{}_a\tens u^a{}_j,\ \eps
u^i{}_j=\delta^i{}_j,\ {\it i.e.},\ \Delta_\cdot\vecu=\vecu\tens\vecu,\
\eps\vecu=\id\\
\Psi_\cdot(u^i{}_j\tens u^k{}_l)=u^p{}_q\tens u^m{}_n  R^{i}{}_a{}^d{}_{p}
R^{-1}{}^a{}_{m}{}^{q}{}_b
R^{n}{}_c{}^b{}_{l} {\widetilde R}^c{}_j{}^k{}_d\\
{\it i.e.},\quad
\Psi_\cdot(R^{-1}\vecu_1 \tens R\vecu_2)=\vecu_2 R^{-1}\tens \vecu_1 R.}
It means that if $\vecu'$ is another copy of $B(R)$ then the matrix product
\eqn{uhopfc}{u''{}^i{}_j=u^i{}_a u'{}^a{}_j,\quad {\it i.e.},\quad
\vecu''=\vecu\vecu'}
obeys the relations of $B(R)$ also provided $\vecu'$ has the {\em
multiplicative braid statistics}
\eqn{uhopfd}{ R^{-1}{}^i{}_a{}^k{}_b u'{}^a{}_c R^c{}_j{}^b{}_d u^d{}_l=u^k{}_b
R^{-1}{}^i{}_a{}^b{}_c u'{}^a{}_d R^d{}_j{}^c{}_l,\quad {\it i.e.},\quad
R^{-1}\vecu'_1 R\vecu_2=\vecu_2 R^{-1}\vecu'_1 R.}
To see this, we check
\align{&&\nquad R_{21}\vecu_1\vecu'_1
R\vecu_2\vecu'_2=R_{21}\vecu_1 R(R^{-1}\vecu'_1
R\vecu_2)\vecu'_2=(R_{21}\vecu_1 R\vecu_2) R^{-1}R_{21}^{-1}(R_{21}\vecu'_1
R\vecu'_2)\\
&&= \vecu_2 R_{21}(\vecu_1 R_{21}^{-1}\vecu'_2 R_{21})\vecu'_1
R=\vecu_2 R_{21}R_{21}^{-1}\vecu'_2 R_{21}\vecu_1\vecu'_1 R
= \vecu_2\vecu'_2 R_{21}\vecu_1\vecu'_1 R}
as required for $\Delta_\cdot$ to extend to $B(R)$ as a braided-Hopf algebra.
The other details such as functoriality of $\Psi_\cdot$  can also be checked in
the
same explicit way\cite{Ma:exa}. This is was the first braided group
construction known.

Note that we have stated $\Psi_\cdot$ implicitly. To give it explicitly (for a
proper braided-group structure) we need that $R$ is {\em bi-invertible} in the
sense that both $R^{-1}$ and the {\em second inverse} $\widetilde R$ exist. The
latter is characterised by
\eqn{Rtilde}{ \widetilde{R}^i{}_a{}^b{}_l
R^a{}_j{}^k{}_b=\delta^i{}_j\delta^k{}_l=R^i{}_a{}^b{}_l
\widetilde{R}^a{}_j{}^k{}_b.}

If we have also that $R$ obeys the $q$-Hecke condition (\ref{q-Hecke}) then
there is also a braided-covector algebra structure, discovered by U. Meyer,
with addition law $\vecu''=\vecu+\vecu'$ and braid statistics\cite{Mey:new}
\eqn{uhopfe}{   R^{-1}{}^i{}_a{}^k{}_b u'{}^a{}_c R^c{}_j{}^b{}_d
u^d{}_l=u^k{}_b R^b{}_c{}^i{}_a u'{}^a{}_d R^d{}_j{}^c{}_l,\quad{\it
i.e.},\quad  R^{-1}\vecu_1'R\vecu_2= \vecu_2 R_{21}\vecu_1'R.}
More formally, $B(R)$ is a braided-Hopf algebra with
\eqn{uhopff}{ \Psi(R^{-1}\vecu_1\tens
R\vecu_2)=\vecu_2R_{21}\tens\vecu_1R,\quad \Delta
\vecu=\vecu\tens 1+1\tens \vecu,\  \eps\vecu=0,\ S\vecu=-\vecu.}
To see this we show as usual that $\Delta$ extends to products
as an algebra homomorphism to the braided tensor product algebra, i.e. that
$\vecu''$ obeys the same relations. This is checked as
\align{R_{21}(\vecu_1+\vecu'_1)R(\vecu_2+\vecu'_2)\equad
&&=R_{21}\vecu_1R\vecu_2+R_{21}\vecu'_1R\vecu'_2
+R_{21}R\vecu_2R_{21}\vecu'_1R+R_{21}\vecu_1R\vecu'_2\\
(\vecu_2+\vecu'_2)R_{21}(\vecu_1+\vecu'_1)R\equad
&&=\vecu_2R_{21}\vecu_1R+\vecu'_2R_{21}\vecu'_1R +
R_{21}\vecu_1R\vecu'_2R_{21}R+\vecu_2R_{21}\vecu'_1 R}
which are equal by the $q$-Hecke assumption (\ref{q-Hecke}). Functoriality of
$\Psi$ under the product map can also be checked explicitly by these
techniques, as well as the antipode and other properties needed for a
braided-Hopf algebra.

This gives a direct proof of the (braided) comultiplication and coaddition
structures on $B(R)$. We can put the latter explicitly into the braided
covector form  (\ref{xhopfa})--(\ref{xhopfc}) by working with the multi-index
notation $u_I=u^{i_0}{}_{i_1}$ and\cite{Ma:exa}\cite{Mey:new}
\ceqn{uRR'a}{\Rrel^I{}_J{}^K{}_L=R^{-1}{}^{d}{}_{k_0}{}^{j_0}{}_{a}
R^{k_1}{}_{b}{}^{a}{}_{i_0}R^{i_1}{}_c{}^b{}_{l_1} {\widetilde
R}^c{}_{j_1}{}^{l_0}{}_d\\
\Rmul^I{}_J{}^K{}_L=R^{j_0}{}_a{}^d{}_{k_0}
R^{-1}{}^a{}_{i_0}{}^{k_1}{}_b
R^{i_1}{}_c{}^b{}_{l_1} {\widetilde R}^c{}_{j_1}{}^{l_0}{}_d\nonumber\\
\Radd^I{}_J{}^K{}_L=R^{j_0}{}_a{}^d{}_{k_0}
R^{k_1}{}_b{}^a{}_{i_0}
R^{i_1}{}_c{}^b{}_{l_1} {\widetilde R}^c{}_{j_1}{}^{l_0}{}_d.}
Then we have
\ceqn{uRR'b}{R_{21}\vecu_1R\vecu_2= \vecu_2 R_{21} \vecu_1 R\quad
\Leftrightarrow\quad
 u_Iu_J= u_Bu_A \Rrel ^A{}_I{}^B{}_J\\
\vecu''=\vecu\vecu';\quad R^{-1}\vecu_1'R\vecu_2=\vecu_2R^{-1}\vecu_1'R\quad
\Leftrightarrow\ u_I'u_J= u_Bu_A' \Rmul ^A{}_I{}^B{}_J \nonumber\\
\vecu''=\vecu+\vecu';\quad  R^{-1}\vecu_1'R\vecu_2= \vecu_2
R_{21}\vecu_1'R\quad
\Leftrightarrow\ \quad u_I'u_J= u_Bu_A' \Radd ^A{}_I{}^B{}_J.}
It is easy to see that $\Rrel,\Radd$ obey the conditions
(\ref{QYBE})--(\ref{covecc}) needed for our braided covector space as well as
the supplementary ones (\ref{covecd}) needed later for the coaddition of forms.
The corresponding braided vector algebra (\ref{vhopfa})--(\ref{vhopfc}) in
matrix form for the relations and additive braid statistics is
\eqn{B(R)vec}{v^I\equiv v^{i_1}{}_{i_0};\quad \vecv_1\widetilde{R}_{21}\vecv_2
R_{21}=R\vecv_2\widetilde{R}\vecv_1,\quad
\vecv''=\vecv+\vecv';\quad \vecv'_1\widetilde{R}_{21}\vecv_2
R^{-1}=R\vecv_2\widetilde{R}\vecv'_1.}
A braided coaddition on the following example of a braided matrix covector
space was obtained by the author
in \cite{Ma:poi} but not in such a nice R-matrix form, which is due to
\cite{Mey:new}.

\begin{example} The $q$-Minkowski space algebra $BM_q(2)$ with generators
$\vecu=\pmatrix{a&b\cr c&d}$ and relations\cite{CWSSW:ten}\cite{Ma:exa}
\cmath{ba=q^2ab,\quad ca=q^{-2}ac,\quad d a=ad,\qquad
bc=cb+(1-q^{-2})a(d-a)\\
d b=bd+(1-q^{-2})ab,\quad cd=d c+(1-q^{-2})ca}
has a braided multiplication law whereby\cite{Ma:exa}
\[ \pmatrix{a''&b''\cr c''&d''}=\pmatrix{a&b\cr c&d}\pmatrix{a'&b'\cr c'&d'}\]
obey the same relations of $BM_q(2)$ if the primed copy does and if we use the
multiplicative braid statistics
\cmath{a'  a=a a'+(1-q^2)b c',\quad a'  b=b  a',\quad  a'  c=c a'+(1-q^2)(d-a)
c'\\
a'  d=d  a'+(1-q^{-2})b  c',\quad b'  a=a  b'+(1-q^2)b  (d'- a'),\quad b'b=q^2b
 b' \\
b'  c=q^{-2}c  b'+(1+q^2)(1-q^{-2})^2b  c'-(1-q^{-2})(d-a) (d'-a')\\
b'  d=d  b'+(1-q^{-2})b (d'-a'),\quad c'  a=a  c',\quad c'  b=q^{-2}b  c'\\
c'  c=q^2c  c',\quad c'  d=d  c',\quad d'  a=ad'+(1-q^{-2})b  c'\\
d'  b=b  d',\quad d'  c=c  d'+(1-q^{-2})(d-a)c',\quad d'  d=d
d'-q^{-2}(1-q^{-2})b  c'.}
Here $q^{-1}a+qd$ is central and bosonic\cite{Ma:exa}. At the same time we have
a braided addition law whereby\cite{Mey:new}
\[ \pmatrix{a''&b''\cr c''&d''}=\pmatrix{a&b\cr c&d}+\pmatrix{a'&b'\cr c'&d'}\]
obey the relations again if the primed copy does and has the additive braid
statistics
\cmath{
a'a=q^2aa',\quad a'b=ba',\quad b'b=q^2bb',\quad c'a=ac',\quad c'c=q^2cc' \\
a'c=ca' q^2 + (q^2-1)ac',\quad a'd=da' +  (q^2-1) bc'  + (q-q^{-1})^2aa'\\
b'a=(q^2-1)ba'+ ab' q^2,\quad  b'c=cb'+(1-q^{-2})(da'+ad')  +
(q-q^{-1})^2bc'-(2  - 3 q^{-2} + q^{-4})aa'\\
b'd=db'+ (q^2-1)bd' +(q^{-2}-1)ba' + (q-q^{-1})^2ab',\quad c'b=bc' +
(1-q^{-2})aa'\\
c'd=dc' q^2 + (q^2-1)ca',\quad  d'a=ad'+(q^2-1)bc' + (q-q^{-1})^2aa'\\
d'b=bd' q^2 + (q^2 -1) ab',\quad d'c=cd' +(q^2-1)dc'  +
(q-q^{-1})^2ca'+(q^{-2}-1)ac' \\
d'd=dd' q^2 + (q^2-1)cb' +(q^{-2}-1)bc' - (1-q^{-2})^2aa'}
So we have both  multiplication and addition of these braided matrices.
\end{example}
\proof  We use the R-matrix from Example~3.1 in the $q$-Hecke normalisation (as
in Example~3.7), which we put into (\ref{uhopfa})--(\ref{uhopff}). The
normalisation and Hecke condition do not enter at all into the multiplicative
braided group structure, but are needed for the additive one. \endproof

The interpretation of this standard example $BM_q(2)$  as $q$-Minkowski space
will be covered in Section~7.2. Its specific six relations were first proposed
as $q$-Minkowski space by Carow-Watamura et. al. \cite{CWSSW:ten} in another
context as a tensor product of two quantum planes. We will see the connection
later in Section~4.1.
The braided matrix $B(R)$ construction is however, more general. A less
standard example is:

\begin{example} The algebra $BM_q(1|1)$ with generators $\vecu=\pmatrix{a&b\cr
c&d}$ and relations\cite{Ma:exa}
\cmath{b^2=0,\qquad c^2=0,\qquad d-a\  {\rm central}, \\
ab=q^{-2}ba,\qquad ac=q^2 ca,\qquad bc=-q^{2}cb+(1-q^2)(d-a)a}
has a braided multiplication law whereby\cite{Ma:exa}
\[ \pmatrix{a''&b''\cr c''&d''}=\pmatrix{a&b\cr c&d}\pmatrix{a'&b'\cr c'&d'}\]
obey the same relations of $BM_q(1|1)$ if the primed copy does and if we use
the multiplicative braid statistics consisting of $d-a$
bosonic and
\cmath{a'a=a a' + (1-q^2) bc',\quad b'b=-bb',\quad c'c=-cc',\quad a'b=ba'\\
b'c=-c b'- (1-q^2)(d-a)(d'-a'),\quad b'a=ab'+(1-q^2)b(d'-a')\\
c'b=-bc',\quad a'c=ca'+ (1-q^2)(d-a)c',\quad c'a=ac'.}
At the same time we have a braided addition law whereby
\[ \pmatrix{a''&b''\cr c''&d''}=\pmatrix{a&b\cr c&d}+\pmatrix{a'&b'\cr c'&d'}\]
obey the relations again if the primed copy does and if we use the additive
braid statistics
\cmath{
a' a =q^2 a a' ,\quad a' b =b a' ,\quad
a' c =q^2c  a' + (q^2-1)a c'\\
a' d =d  a' + (q^{-2}-1)b c' + (q-q^{-1})^2a a' ,\quad
b' a = q^2 a b' + (q^2-1)b a' ,\quad
b' b = - b b' \\
b' c =-q^2 cb'+ + (q-q^{-1})^2 bc'+ (1-q^2)( d  a'+ a d') + (2 q^2 - 3  +
q^{-2})aa'\\
b' d =d  b'+(q^{-2}-1)b d'+ (q^2-1)b a' + (q-q^{-1})^2a b',\quad
c' a =a c'\\
c' b =- q^{-2}b c' + (1-q^{-2})a a',\quad
c' c = - c  c' ,\quad
c' d =q^{-2}d  c' + (q^{-2}-1)c  a'\\
d' a =ad'+ (q^{-2}-1)bc'+ (q-q^{-1})^2a a',\quad
d' b =q^{-2}b d' + (q^{-2}-1)a b' \\
d' c =cd'+ (q^{-2}-1)dc'+ (q-q^{-1})^2c  a'  + (q^2-1)a c' \\
d' d =q^{-2}d  d' +(q^{-2}-1)(c  b' +b c')  + (q-q^{-1})^2a a'.}
So we have both  multiplication and addition of these non-standard braided
matrices.
\end{example}
\proof  We use the R-matrix from Example~3.2 in the $q$-Hecke normalisation (as
in Example~3.9), which we put into (\ref{uhopfa})--(\ref{uhopff}). The
normalisation and Hecke condition do not enter at all into the multiplicative
braided group structure,  only the additive one. \endproof

This example is supercommutative in the limit $q\to 1$ with $b,c$ odd and $a,d$
even. The braid statistics also become $\pm1 $ according to the degree. Thus we
recover exactly the superbialgebra $M(1|1)$ consisting of these generators and
their appropriate supercommutation relations. Thus the notion of braided
matrices really generalises both ordinary and supermatrices\cite{Ma:exa}.

For completeness we note also that there is a similarity of the braided matrix
algebra (\ref{uhopfa}) with the `reflection equation' of Cherednik
\cite{Che:ref} whose constant form is called `RE' in \cite{KulSkl:alg}. This
paper then went on to repeat some of the algebraic constructions in
\cite{Ma:exa}\cite{Ma:skl}\cite{Ma:lin} in terms of RE, with (\ref{uhopfa}) as
a variant RE. The braided matrix algebra $B(R)$ is quite interesting from a
purely homological point of view too, see\cite{LeB:hom}. Further results such
as covariance of this braided matrix quadratic algebra under a background
quantum group\cite{Ma:exa}\cite{Ma:lin}, its quantum trace central
elements\cite{Ma:lin} and a Hermitian $*$-structure\cite{Ma:mec} were likewise
introduced as far as I know for the first time in the braided approach. We will
cover the latter two properties in the next section and covariance in
Section~6.

\section{Braided linear algebra}

So far we have introduced braided covectors $x_i$, vectors $v^i$ and matrices
$u^i{}_j$, and linear structures on the quantum matrices $t^i{}_j$ as well. Now
we begin to explore the relationships between these
objects in analogy with the usual formulae in linear algebra. This justifies
our terminology further and shows that they all fit together into a systematic
generalisation of our usual concepts.

\subsection{Braided linear transformations}

In braided geometry all independent objects enjoy mutual braid statistics with
respect to each other. We therefore have to extend the braiding $\Psi$ to work
between objects and not only for a given braided group as we have done until
now. This depends in fact on the applications being made, as we have seen
already in Examples~3.10 and~3.11.  For the present set of applications we
take\cite{Ma:lin}
\ceqn{xvustata}{ \Psi(x_i\tens x_j)=x_n\tens x_m R^m{}_i{}^n{}_j,\quad
\Psi(v^i\tens v^j)=R^i{}_m{}^j{}_n v^n\tens v^m\\
 \Psi(x_i\tens v^j)=\widetilde R^m{}_i{}^j{}_n v^n\tens x_m,\quad \Psi(v^i\tens
x_j)=x_n\tens v^m R^{-1}{}^i{}_m{}^n{}_j\\
 \Psi(u^i{}_j\tens x_k)=x_m\tens u^a{}_b R^{-1}{}^i{}_a{}^m{}_n
R^b{}_j{}^n{}_k,\quad \Psi(x_k\tens u^i{}_j)=u^a{}_b\tens x_m\widetilde
R^n{}_k{}^i{}_a R^m{}_n{}^b{}_j\\
 \Psi(u^i{}_j\tens v^k)=v^m\tens u^a{}_b R^i{}_a{}^n{}_m\widetilde
R^b{}_j{}^k{}_n,\quad \Psi(v^k\tens u^i{}_j)=u^a{}_b\tens v^m R^k{}_n{}^i{}_a
R^{-1}{}^n{}_m{}^b{}_j\\
\Psi(u^i{}_j\tens u^k{}_l)=u^p{}_q\tens u^m{}_n  R^{i}{}_a{}^d{}_{p}
R^{-1}{}^a{}_{m}{}^{q}{}_b
R^{n}{}_c{}^b{}_{l} {\widetilde R}^c{}_{j}{}^{k}{}_d}
or equivalently the braid statistics\cite{Ma:lin}
\ceqn{xvustatb}{ \vecx'_1\vecx_2=\vecx_2\vecx'_1R,\quad
\vecv'_1\vecv_2=R\vecv_2\vecv'_1\\
 \vecx'_1R\vecv_2=\vecv_2\vecx'_1,\quad \vecv'_1\vecx_2=\vecx_2R^{-1}\vecv'_1\\
 \vecu'_1\vecx_2=\vecx_2 R^{-1}\vecu'_1 R,\quad
\vecx'_1R\vecu_2R^{-1}=\vecu_2\vecx'_1 \\
 R^{-1}\vecu'_1 R\vecv_2=\vecv_2\vecu'_1,\quad \vecv'_1\vecu_2=R\vecu_2
R^{-1}\vecv'_1 \\
R^{-1}\vecu'_1 R\vecu_2=\vecu_2 R^{-1}\vecu'_1 R}
where the primes denote the second algebra in the relevant braided tensor
product algebra as in (\ref{brahom}).
The systematic way to derive these braid statistics will be covered in
Section~6.1 using the background quantum group covariance.  For the moment we
verify directly that they are suitable. They are quite natural for, but not
uniquely determined by, these applications alone.

With the help of such braid statistics we show now that our braided matrices
act as (braided) transformations on braided vectors and covectors, as well as
on themselves. The fundamental notion here is that of a braided comodule
algebra of $B(R)$. On covectors for example it should be an algebra
homomorphism $\Vhaj\to \Vhaj\und\tens B(R)$  to the braided tensor product
algebra structure as in (\ref{brahom}), determined now by the mutual braid
statistics between braided matrices and braided covectors. This indeed works
with the transformation
\eqn{coactxu}{x_i\mapsto x_a\tens u^a{}_i,\quad  {\it i.e .},\quad
\vecx''=\vecx\vecu'}
obeys the same relations as the $\vecx$. To see this, it is  convenient to
assume that $PR'$ is given as some function of $PR$, as for example in
(\ref{R'(R)}). Since nothing depends on the precise form of the function, it is
clear that this is not a strong restriction. Then we compute
$\vecx_1\vecu'_1\vecx_2\vecu'_2= \vecx_1\vecx_2R^{-1}\vecu'_1
R\vecu'_2=\vecx_2 \vecx_1 R'R^{-1}\vecu'_1 R\vecu'_2=\vecx_2\vecx_1
R_{21}^{-1}\vecu_2' R_{21}\vecu_1' R'=\vecx_2\vecu_2'\vecx_1\vecu_1'R'$ as
required. Here we used the braid statistics from (\ref{xvustatb}) between
$\vecx,\vecu$ for the first and last equalities, and the relations for
$\Vhaj(R',R)$ and $B(R)$ in the form $(PR)R^{-1}\vecu_1 R\vecu_2=R^{-1}\vecu_1
R\vecu_2 (PR)$
for the middle equalities. The $B(R)$ relations imply therefore that
$R'R^{-1}\vecu_1R\vecu_2=R_{21}^{-1}\vecu_2R_{21}\vecu_1R'$, which could be
verified directly if $PR'$ is not given explicitly as a function of $PR$.

\begin{example} The braided matrices $BM_q(2)$ coact from the right on the
braided plane $\C_q^{2|0}$ in the sense that
\[ \pmatrix{x''&y''}=\pmatrix{x&y}\pmatrix{a'&b'\cr c'&d'}\]
obeys the quantum plane relations if $x,y$ do and $a',b',c',d'$ are a copy of
the braided-matrices, and provided we remember the braid-statistics
\[ \pmatrix{a'&b'\cr c'&d'}x=\pmatrix{ xa'+(1-q^2)yc'&
q^{-1}xb'+(q-q^{-1})y(a'-d')\cr qxc'& xd'+ (1-q^{-2})yc'}\]
\[\pmatrix{a'&b'\cr c'&d'}y=y\pmatrix{a'&qb'\cr q^{-1}c'&d'}\]
\end{example}
\proof We use (\ref{coactxu}) with the standard R-matrix, computing the
relevant braid statistics from (\ref{xvustatb}) and deducing the coaction of
$BM_q(2)$ in Example~3.10 on the usual quantum plane in Example~3.1. One can
easily confirm that the transformed $x'',y''$ obey the same relations
$y''x''=qx''y''$. \endproof

\begin{example} The braided matrices $BM_q(1|1)$ coact from the right on the
braided plane $\C_q^{1|1}$ in the sense that
\[ \pmatrix{x''&\theta''}=\pmatrix{x&\theta}\pmatrix{a'&b'\cr c'&d'}\]
obeys the $\C_q^{1|1}$ relations if $x,\theta$ do and $a',b',c',d'$ are a copy
of
$BM_q(1|1)$, and provided we remember the braid-statistics
\[ \pmatrix{a'&b'\cr c'&d'}x=\pmatrix{ xa'+(1-q^2)yc'&
q^{-1}xb'+(q-q^{-1})y(a'-d')\cr qxc'& xd'+ (1-q^2)yc'}\]
\[\pmatrix{a'&b'\cr c'&d'}\theta=\theta\pmatrix{a'&-q^{-1}b'\cr- qc'&d'}\]
\end{example}
\proof We use (\ref{coactxu}) and (\ref{xvustatb}) with the non-standard
R-matrix in Example~3.2, and deducing the coaction of $BM_q(1|1)$ in
Example~3.11
on it. One can easily confirm that the transformed $x'',\theta''$ obey the same
relations $\theta''x''=qx''\theta''$ and $\theta''{}^2=0$. \endproof

Similarly for $BM_q(2)$ acting on the quantum superplane $\C_q^{0|2}$ in
Example~3.4. It has the same $R$-matrix in a different normalisation to that in
Example~3.1, but this does not affect the braid statistics which therefore
comes out analogous to Example~4.1. Note that the reader has seen quantum
matrices characterised as transformations of the two types of quantum plane in
Manin's lectures\cite{Man:non}. In our case we have non-trivial braid
statistics and because of this we obtain the braided matrices $BM_q(2)$
instead. This is the general principle behind the process of transmutation as
explained in the Appendix. Note also that it is only in the Hecke case that two
quantum planes such as $\C_q^{2|0}$ and $\C_q^{0|2}$ are enough for this. More
generally one needs to consider all quantum planes for each choice of
eigenvalue of $PR$ in (\ref{R'(R)}). Braided covectors alone are enough to do
the job of characterising the braided matrices as (braided) transformations in
this way provided we consider the various possibilities for them.

In the same way, we have braided coactions of $B$ on the braided vectors $v^i$
and on the braided matrices $u^i{}_j$ itself. Here $B$ is a braided group with
antipode which we must make from $B(R)$ in a way compatible with the braidings.
This is possible for regular $R$ in the sense explained in Section~6.1. The
braided antipode $S$ can be found either by hand or by the systematic
transmutation technique in the Appendix. Assuming it, the relevant
transformations are\cite{Ma:lin}
\eqn{bracoactvu}{\nquad\ v^i\mapsto (1\tens Su^i{}_a)(v^a\tens 1),\
u^i{}_j\mapsto
(1\tens Su^i{}_a)(u^a{}_b\tens u^b{}_j)\quad {\it i.e.},\quad \vecv\to
\vecu'{}^{-1}\vecv,\ \vecu\to \vecu'^{-1}\vecu\vecu'}
where  the expressions are written in the braided tensor product algebra
$V\und\tens B$ or $B(R)\und\tens B$ respectively. In the compact notation
$\vecu'$ denotes the generator of the $B$ factor. The second of the
transformations here is the {\em braided adjoint action} and can be used as a
foundation for a theory of braided-Lie algebras\cite{Ma:lie}.

Next we show that the tensor product of a quantum covector with a quantum
vector, when treated with the correct braid statistics, is a braided matrix,
i.e. the map $B(R)\to V(R',R)\und\tens \Vhaj(R',R)$ given by\cite{Ma:lin}
\eqn{rank1}{u^i{}_j\mapsto v^i\tens x_j,\quad {\it i.e.},\quad \vecu\mapsto
\vecv\vecx'=\pmatrix{v^1x'_1&\cdots &v^1x'_n\cr \vdots&&\vdots\cr
v^nx'_1&\cdots &v^nx'_n}}
is an algebra homomorphism provided we use the braid statistics from
(\ref{xvustatb}). Again, this result from \cite{Ma:lin} is most easily checked
under the mild assumption that $PR$ used for the braided covectors and vectors
in Section~3 is given as some function of $PR'$. Then we have
$R_{21}\vecv_1\vecx'_1 R
\vecv_2\vecx'_2=R_{21}\vecv_1\vecv_2\vecx'_1\vecx'_2
=f(PR')P\vecv_1\vecv_2\vecx'_1\vecx'_2 =\vecv_2\vecv_1 f(1)\vecx'_1\vecx'_2
= \vecv_2\vecv_1\vecx'_1\vecx'_2 f(PR')=\vecv_2\vecv_1\vecx'_2
\vecx'_1R=\vecv_2\vecx'_2R_{21}\vecv_1\vecx'_1R $. The first and last
equalities use the  braid statistics relations. The middle equalities use the
defining relations in the algebras $V(R',R), \Vhaj(R',R)$. Hence $\vecv\vecx'$
is a realisation of the braided matrices $B(R)$. If $PR$ is not given
explicitly as some function of $PR'$ then the proposition typically still holds
but has to be verified directly according to the form in which $PR',PR$ are
given. An example is\cite{Ma:lin}
\eqn{bm2dec}{ BM_q(2) \to \C_{q^{-1}}^{2|0}\und\tens \C_q^{2|0}}
which relates the braided-matrix point of view which we will later adopt as a
definition of $q$-Minkowski space (see Example~3.10) with an independent
approach pioneered in \cite{CWSSW:lor}\cite{CWSSW:ten}\cite{OSWZ:def}.

It is also easy to see that the `inner product' element
\eqn{xv}{x'_iv^i=v^bx'_a\cv^a{}_b=\trace\vecv\vecx'\cv;\quad
\cv^i{}_j=\widetilde{R}^i{}_a{}^a{}_j}
in the braided tensor product algebra $V(R',R)\und\tens \Vhaj(R',R)$ is central
and bosonic with respect to the multiplicative braid statistics\cite{Ma:lin}.
This applies also to $c_1=\trace\vecu\cv$ as one would expect in view of
(\ref{rank1}). The trace elements in Examples~3.10 and~3.11 are of just this
form. More generally one has that all the powers $c_k=\trace \vecu^k\cv$ are
bosonic and central in $B(R)$\cite{Ma:lin}. This means of course that
\eqn{ck}{ c_k u^i{}_j=u^i{}_j c_k,\quad u'{}^i{}_j c_k=c_k u'^i{}_j,\quad c'_k
u^i{}_j=u^i{}_jc'_k}
in the algebras $B(R)$ and $B(R)\und\tens B(R)$ respectively. The first
equation is similar to the construction of Casimirs of $U_q(\cg)$  in
\cite{FRT:lie} in another context, while the latter two equations about the
bosonic nature are new features of the braided theory in \cite{Ma:lin}.

One can show (again as an application of transmutation) that the bosonic
central elements in the braided matrices $B(R)$ generate a subalgebra (are
closed under addition and multiplication). In this subalgebra one can expect to
find interesting bosonic central elements such as the braided-determinant
$\und\det(\vecu)$. This should be group-like in the sense
\eqn{bdeta}{ \Delta \und\det(\vecu)=\und\det(\vecu)\und\tens\det(\vecu),\quad
\eps\und\det(\vecu)=1}
with respect to the braided comultiplication. For example, in the 2-dimensional
case it means
\eqn{bdetb}{ \und\det\left(\pmatrix{a&b\cr c&d}\pmatrix{a'&b'\cr c'&d'}\right)=
\und\det\pmatrix{a&b\cr c&d} \und\det\pmatrix{a'&b'\cr c'&d'}}
where we use the multiplicative braid statistics as in (\ref{xvustatb}). One
can  also expect
\eqn{bdetc}{ \und\det(\vecv\vecx')=0}
from our above picture of $\vecv\vecx'$ as a rank-1 braided matrix. There is
also a general formula for $\und\det$ using the R-matrix formula for the
epsilon
tensor in Section~5.7. All of this suggests a fairly complete picture of our
braided vectors, covectors and matrices in terms of braided linear
algebra\cite{Ma:lin}.

\subsection{Gluing or direct sum of braided vectors}

Next we consider some finer points of linear algebra in our braided framework.
The first concerns how to tensor product braided groups. The braided tensor
product algebra and braided tensor product coalgebra (defined in an obvious way
again using the braiding) do not in general fit together to form a braided
group: one must also `glue together' the braidings of the two braided groups.

Such a gluing construction has been found by the author and M. Markl in
\cite{MaMar:glu} for the Hecke case. Firstly, if $R,S$ are two solutions of the
QYBE obeying the Hecke condition (\ref{q-Hecke}) then so is\cite{MaMar:glu}
\eqn{RS}{ R\oplus_q S=\pmatrix{R&0&0\cr 0&1 & (q-q^{-1})P&0\cr 0&0&1&0\cr
0&0&0&S}.}
Here $P$ is the permutation matric. The dimensions of $R,S$ need not be the
same. Such a phenomenon is encountered from time to time in the R-matrix
literature, see
\cite{Gur:alg} where Hecke R-matrices were extensively studied, although not as
far as
I know as a systematic gluing operation.

Many consequences of this associative gluing operation $\oplus_q$ are then
explored
in \cite{MaMar:glu}. Among them, it is shown that
\eqn{gluecoveca}{ \Vhaj_\lambda(R\oplus_q S)=\Vhaj_\lambda(R)\und\tens_\lambda
\Vhaj_\lambda(S);\quad \lambda=q \ {\rm or}\ -q^{-1}.}
Recall that in the Hecke case there are two natural choices for our braided
covector data, which we label by $\lambda$: we  keep $R,S$ fixed in the
Hecke normalisation and take $(R\lambda,\lambda^{-1}R)$ for $(R,R')$ in
Section~3.1. Likewise $(S\lambda,\lambda^{-1}S)$ for $\Vhaj_\lambda(S)$. The
$\und\tens_\lambda$ is the braided tensor product with respect to a braiding
given by the power of the braided covector generators (the scaling dimension)
much as in the braided-line example in Section~2. Explicitly, the isomorphism
(\ref{gluecoveca}) is given by writing the generators of
$\Vhaj_\lambda(R\oplus_qS)$ as
$(x_i,y_I)$ say for appropriate ranges of indices according to the dimensions
of $R,S$. Then its relations are
\eqn{gluevecb}{ \lambda\vecx_1\vecx_2=\vecx_2\vecx_1R,\quad
\lambda\vecy_1\vecy_2=\vecy_2\vecy_1S,\quad y_I x_j=\lambda x_j y_I,\quad {\it
i.e.},\quad \vecy_1\vecx_2=\lambda\vecx_2\vecy_1}
which is the right hand side of (\ref{gluecoveca}).

Note that our algebras are like those generated by the co-ordinate functions on
row vectors. Their tensor product is therefore like the direct sum of the
underlying braided vector spaces. For example,
\[ \C_q^{n|0}\und\tens_q\C_q^{m|0}=\C_q^{n+m|0},\quad
\C_q^{0|n}\und\tens_{(-q^{-1})}\C_q^{0|m}=\C_q^{0|n+m}\]
as is clear from the well-known form of these spaces. The above gluing however,
works quite generally for arbitrary $R,S$ of Hecke type.

This gluing is an example of a more general construction of {\em rectangular
quantum matrices} $A(R:S)$ introduced in \cite{MaMar:glu}. These are defined
with generators $1,x^i{}_I$ say and relations
\eqn{rectFRT}{ R^i{}_a{}^k{}_b x^a{}_J x^b{}_L=x^k{}_B x^i{}_A
S^A{}_J{}^B{}_L,\quad {\it i.e.},\quad R\vecx_1\vecx_2=\vecx_2\vecx_1 S.}
 and a braided  addition law $\vecx''=\vecx+\vecx'$ with braid
statistics\cite{MaMar:glu}\cite{Ma:add}
\eqn{rectstat}{ x'{}^i{}_Jx^k{}_L=R^k{}_b{}^i{}_a x^b{}_D x'{}^a{}_C
S^C{}_J{}^D{}_L,\quad {\it i.e.},\quad \vecx'_1\vecx_2=R_{21}\vecx_2\vecx'_1
S.}
They are not in general quantum groups but instead we have partial
comultiplication maps
\eqn{rectcomult}{ \Delta_{R,S,T}:A(R:T)\to A(R:S)\tens A(S:T),\quad \Delta
x^i{}_\alpha=x^i{}_I\tens x^I{}_\alpha,\quad {\it i.e.},\quad \Delta
\vecx=\vecx\tens\vecx}
for any three Hecke solutions $R,S,T$ of the QYBE, corresponding to matrix
multiplication of rectangular matrices. We associate the R-matrices $R,S$ to
the rows and columns of $A(R:S)$ and there is a comultiplication when the rows
match the columns to be contracted with.
There is also distributivity generalising (\ref{thopfg}) and expressing
linearity of $\Delta_{R,S,T}$ with respect to the coaddition. Finally, we can
regard
the rectangular quantum matrices $A(R:S)$ as a braided covector space as in
Section~3.1 with multi-index $\bf R',R$
now built from $R,S$, see \cite{Ma:add}. The corresponding braided vector space
is $A(S_{21}:R_{21})$.

It is clear that setting $R=(q)$ or $(-q^{-1})$ (the 1-dimensional Hecke
R-matrices) recovers the braided covectors as $1\times n$ rectangular quantum
matrices $\Vhaj_\lambda(R)=A(\lambda:R)$. The other case $A(R:\lambda)$ of
$n\times 1$ rectangular quantum matrices recovers a left handed version
$R\vecv_1\vecv_2=\vecv_2\vecv_1\lambda$ of the braided vectors. We have given
conventions in Section~3.1 in which everything is right-covariant under a
background quantum group, while these left-handed vectors are left-covariant by
contrast. The general $A(R:S)$ is bicovariant using the maps (\ref{rectcomult})
for comultiplication from the left or right by $A(R:R)=A(R)$ and $A(S:S)=A(S)$.
The diagonal case of the rectangular quantum matrices are of course the usual
quantum matrices of \cite{FRT:lie}. Another example is $A(R_{21}:R)=\bar A(R)$
which is the variant from
\cite{Ma:euc}. Finally, we have good behaviour of this construction under
gluing as\cite{MaMar:glu}
\eqn{rectgluea}{ A(R:S\oplus_qT)=A(R:S){}_R\und\otimes A(R:T),\quad
y^k{}_\alpha x^i{}_J=R^i{}_a{}^k{}_b x^a{}_J y^b{}_\alpha,\quad{\it i.e.},\quad
\vecy_2\vecx_1=R\vecx_1\vecy_2}
\eqn{rectglueb}{
A(R\oplus_qS:T)=A(R:T)\und\otimes_T A(S:T),\quad w^J{}_\delta
v^i{}_\gamma=v^i{}_\alpha w^J{}_\beta T^\beta{}_\delta{}^\alpha{}_\gamma,
\  {\it i.e.}, \ \vecw_1\vecv_2=\vecv_2\vecw_1 T}
which generalises the above. Here $\vecx\in A(R:S),\ \vecy\in A(R:T)$ which we
view as $(\vecx\ \vecy)$ and $\vecv\in A(R:T),\ \vecw\in A(S:T)$ which we view
as $\pmatrix{\vecv\cr\vecw}$. The braided tensor products are defined with
braidings given by $R,T$ respectively, which we have shown as the braid
statistics between the two factors in each case.

\begin{example} We can horizontally glue two quantum plane column vectors to
give a quantum matrix, or vertically glue two quantum plane row vectors to
achieve the same:
\[\begin{array}{rll}
M_q(2)\nquad\
&=\C_q^{2|0}{}_{R_{gl_2}}\und\tens\C_q^{2|0}\qquad\qquad\qquad\<\pmatrix{a&b\cr
c&d}\>\nquad\ &=\<\pmatrix{a\cr c}\>\und\tens\<\pmatrix{b\cr d}\>\\
&=\C_q^{2|0}\und\tens_{R_{gl_2}}\C_q^{2|0}\qquad{\displaystyle{\it
i.e.},\atop\strut}\qquad &=\<\pmatrix{a& b}\>\und\tens\<\pmatrix{c& d}\>
\end{array}\]
where $R_{gl_2}=(q)\oplus_q(q)$ is the standard R-matrix as in Example~3.1 but
in the Hecke normalisation. This is one way to derive its relations as quoted
in Example~3.7.
\end{example}
\proof For the first case we use (\ref{rectgluea}) with $R$ the standard
R-matrix as in Example~3.1 and~3.7. Then $A(R:q)=\C_q^{2|0}$ as a
left-covariant column vector and $A(R:q){}_R\und\otimes
A(R:q)=A(R:(q)\oplus_q(q))$. But $R=(q)\oplus_q(q)$ so the latter is just
$A(R)=M_q(2)$. Likewise for the second example we have $A(q:R)=\C_q^{2|0}$ as a
right-covariant row vector and $A(q:R)\und\tens_R
A(q:R)=A((q)\oplus_q(q):R)=A(R)=M_q(2)$ by (\ref{rectglueb}). The explicit
identification of the generators for the two cases is shown on the right, where
$\<\ \>$ denotes the algebra generated. \endproof

We could equally well glue $n$ rows of the quantum planes $\C_q^{n|0}$ or $n$
columns to arrive at the $n\times n$ quantum matrices $M_q(n)$. If we glue $m$
rows or columns then we arrive at rectangular quantum matrices
$A(R_{gl_m}:R_{gl_n})$ or $A(R_{gl_n}:R_{gl_m})$ etc. Such decomposition
properties of the standard quantum matrices $M_q(n)$ or $SL_q(n)$ are to some
extent known in other contexts, but recovered here systematically as part of a
general framework. For a less standard example we could just as easily glue a
row
vector quantum plane $\C_q^{2|0}$ and a `fermionic' row vector $\C_q^{0|2}$
say, i.e. glue Examples~3.1 and~3.4 vertically. Or we could glue horizontally
two copies of the mixed quantum plane $\C_q^{1|1}$ in Example~3.2. Either way,
the result by the gluing theorems (\ref{rectgluea})--(\ref{rectglueb}) will be
the `rectangular' quantum matrices $A(R_{gl_{1|1}}:R_{gl_2})$ where
$R_{gl_{1|1}}=(q)\oplus_q(-q^{-1})$ is the Alexander-Conway R-matrix as in
Example~3.2, in the Hecke normalisation. Clearly we have some powerful gluing
technology which we can use any number of ways. We content ourselves here with
a concrete application of such ideas to quantum differential geometry.

\begin{example} We can horizontally glue two copies of the (left-handed version
of) the 1-dimensional quantum exterior algebra in Example~3.6 to obtain the
left-handed quantum exterior algebra on the quantum plane as a rectangular
quantum matrix
\[\Omega(\C_q^{2|0})=\Omega(\C_q){}_{R_\Omega}\und\tens\Omega(\C_q),\qquad{\it
i.e.},\qquad \<\pmatrix{x&y\cr \extd x&\extd y}\>
=\<\pmatrix{x\cr \extd x}\>\und\tens\<\pmatrix{y\cr \extd y}\>\]
where $R_\Omega$ is the  R-matrix as in Examples~3.3 and~3.6, in the Hecke
normalisation.
The rectangular quantum matrix $\Omega(\C_q^{2|0})=A(R_\Omega:R_{gl_2})$
transforms covariantly under a quantum group $M_q^{\Omega}(1|1)=A(R_\Omega)$
for each column and under $M_q(2)$ for each row. It has relations
\cmath{yx = qxy,\quad (\extd x)^2=0,\quad (\extd y)^2=0,\quad \extd y\extd
x=-q^{-1}\extd x\extd y\\
\extd x\, x = q^2x \extd x,\quad \extd x\, y = qy\extd x,\quad \extd y\, y =
q^2y\extd y,\quad \extd y\, x = (q^2-1)\extd x\, y+ qx\extd y}
from (\ref{rectFRT}) and a braided addition law from (\ref{rectstat}) whereby
\[ \pmatrix{x''&y''\cr \extd x''&\extd y''}=\pmatrix{x&y\cr \extd x&\extd
y}+\pmatrix{x'&y'\cr \extd x'&\extd y'}\]
 obeys the same relations of $\Omega(\C_q^{2|0})$ provided the primed copy has
the braid statistics
\cmath{ x' x=q^2xx',\quad x' y=qy x',\quad y' x=(q^2-1)yx' + qxy',\quad
y' y=q^2yy'\\
x' \extd x=\extd x\, x',\quad
x' \extd y=q^{-1}\extd y\, x',\quad
y' \extd x=(1-q^{-2})\extd y\, x'+ q^{-1}\extd x\, y',\quad
y' \extd y=\extd y\, y\\
\extd x'\, x=(q^2-1)\extd x\, x' + q^2x\extd x',\quad
\extd x'\, y=(q-q^{-1})\extd y\, x' + qy\extd x'\\
\extd y'\, x=(q-q^{-1})^2\extd y\, x' + (q-q^{-1})(\extd x\, y'+ qy\extd x') +
qx\extd y',\quad
\extd y'\, y=(q^2-1)\extd y\, y' + q^2y\extd y'\\
\extd x' \extd x= - \extd x\extd x',\quad
\extd x' \extd y= - q^{-1}\extd y\extd x'\\
\extd y' \extd x=(q^{-2}-1)\extd y\extd x' - q^{-1} \extd x\extd y',\quad
\extd y' \extd y= - \extd y\extd y'}
\end{example}
\proof We take $\Omega(\C_q)=A(R_\Omega:q)$ which is a left-covariant vector
version of Example~3.6 (it has the opposite algebra). Then
$\Omega(\C_q){}_{R_\Omega}\und\tens\Omega(\C_q)=A(R_\Omega:q){}_{R_\Omega}
\und\tens A(R_\Omega:q)=A(R_\Omega:R_{gl_2})$ as an example of
(\ref{rectgluea}). We compute its relations from (\ref{rectFRT}) and obtain
the algebra generated by $x,y,\extd x,\extd y$ previously proposed as a
natural covariant differential calculus for the quantum plane in
\cite{PusWor:twi}\cite{WesZum:cov} and for which a coaddition was recently
proposed
in \cite{Vla:coa}. By obtaining it as a rectangular quantum
matrix, we know from the general theory above (without any work)
not only that it is covariant under the usual $M_q(2)$ in Example~3.7 whereby
\[\pmatrix{x&y\cr \extd x&\extd y}\pmatrix{a&b\cr c&d};\quad \pmatrix{a&b\cr
c&d}\ {\rm bosonic}\]
obeys the $\Omega(\C_q^{2|0})$ relations, which is the usual picture (cf.
Manin's lectures), but also that it is covariant under a different quantum
group $M_q^{\Omega}(1|1)$ with relations
\cmath{ba = ab,\quad ca = acq^2,\quad  db = -bd,\quad dc=-q^{-2} cd\\
cb = bcq^2,\quad ad - da= bc(1-q^2),\quad b^2=0,\quad c^2=0}
whereby
\[\pmatrix{a&b\cr c&d}\pmatrix{x&y\cr \extd x&\extd y};\quad \pmatrix{a&b\cr
c&d}\ {\rm bosonic}\]
obeys the same relations of $\Omega(\C_q^{2|0})$ as well. The quantum matrix
$A(R_\Omega:R_{gl_2})$ is `rectangular' in that its rows and columns transform
under different quantum groups (i.e., have a different flavour) even though
they have the same dimension. Moreover, we also know from \cite{MaMar:glu} that
every
rectangular quantum matrix in our setting has a braided addition law. We
compute the required braid-statistics at once from (\ref{rectstat}). They are
the same as Examples~3.1 and~3.4 for the two rows and (the opposite version of)
Example~3.6 for each column, but include cross statistics such as $\extd y'\,
x$
also. \endproof

The same remarks apply in $n$-dimensions by iterated gluing, or indeed for the
quantum plane associated to any Hecke R-matrix. We have taken so far the view
of horizontal gluing (of the exterior algebras for each dimension). We can also
take the vertical gluing point of view whereby we glue the position
co-ordinates and the basic forms as
\ceqn{ROmega}{
\Omega_q(R)=\Vhaj_q(R)\und\tens_{qR}\Vhaj_{-q^{-1}}(R)=A(R_\Omega:R),\quad {\it
i.e.},\quad \<\pmatrix{x_1\cdots x_n\cr \extd x_1\cdots \extd x_n}\>=\<
x_i\>\und\tens_{qR} \<\extd x_i\>\\
q\vecx_1\vecx_2=\vecx_2\vecx_1R,\quad \extd\vecx_1\extd\vecx_2=-q
\extd\vecx_2\extd\vecx_1 R,\quad \extd \vecx_1\, \vecx_2=q\vecx_2\extd\vecx_1
R}
which works for general Hecke $R$. Note that in \cite{MaMar:glu} we also gave
generalisations of the gluing procedure (\ref{RS}). One of these
generalisations allows operators in the inner diagonal of (\ref{RS}) and it is
this slightly generalised version which we use to obtain
$R_\Omega=(q)\oplus_q(-q^{-1})$. The corresponding version of (\ref{rectglueb})
involves $qR$ in the braided tensor product. This extra factor of $q$ is needed
to ensure that our notation is consistent with $\extd^2=0$ and a usual graded
Leibniz rule. We arrive at a version of the quantum exterior algebra
$\Omega_q(R)$ as in \cite{WesZum:cov} but obtained now as a rectangular quantum
matrix. As such, its $A(R)$ covariance from the right is automatic, while at
the same time we see automatically covariance from the left under
$M_q^{\Omega}(1|1)$. The quantum group $M_q^{\Omega}(1|1)$ clearly has some
super-like qualities and, indeed, there is a theory of superisation which
converts it strictly into a super-quantum group along the lines developed by
the author and M.J. Rodriguez-Plaza in \cite{MaPla:uni}. It is then a
$q$-deformation of a hidden supersymmetry in the exterior algebra of $\R^n$.
Also
automatic  is that $\Omega_q(R)$ has a braided addition law from
(\ref{rectstat}). We return to such exterior algebras from a more general and
more constructive point of view in Section~5.7. The recent papers
\cite{Vla:coa}\cite{Isa:int} on exterior algebras in the braided
setting likewise go beyond the Hecke case covered by the gluing theory
in\cite{MaMar:glu}.
In a somewhat different direction but also related to gluing, see
\cite{LyuSud:sup}.
One can envision many other applications
also of the gluing theory, such as formulating quantum path spaces and function
spaces as infinitely iterated  gluings, see \cite{Ma:add}.

Finally, we can consider the quantum groups $A(R\oplus_q S)$. From the above
they comes out as generated by four block matrices and block
relations\cite{MaMar:glu}
\ceqn{A(R)dec}{\vect=\pmatrix{\veca&\vecb\cr\vecc&\vecd},\quad \veca\in
A(R),\quad \vecb\in A(R:S),\quad \vecc\in A(S:R),\quad \vecd\in A(S)\\
R\veca_1\vecb_2=\vecb_2\veca_1,\quad
\vecc_1\veca_2=\veca_2\vecc_1R,\quad
S\vecc_1\vecd_2=\vecd_2\vecc_1,\quad
\vecd_1\vecb_2=\vecb_2\vecd_1S\\
\vecb_1\vecc_2=\vecc_2\vecc_1,\quad
\veca_1\vecd_2-\vecd_2\veca_1=(q^{-1}-q)P\vecc_1\vecb_2.}
This is a `blocked form' of the quantum matrices $M_q(2)$ in Example~3.7, where
the generators are themselves now rectangular quantum matrices as shown.

We see then that these general braided tensor product constructions provide
analogues of the usual constructions whereby matrices can be multiplied and
blocked into smaller ones. There is a further theory for gluing or
decomposition of braided matrices which remains to be studied, connecting with
the results (\ref{rank1})--(\ref{bm2dec}) in Section~4.1. Also, it would be
nice to go beyond the Hecke case as well as to use other possible `templates'
in (\ref{RS}) such as the 8-vertex
R-matrix in place of the standard 6-vertex one used in \cite{MaMar:glu}.

\subsection{Braided metric}

 Next we  consider the situation that our braided
vector and covector algebras are isomorphic. Recall that this is the true
meaning of the metric in differential geometry. So to complete our picture we
follow \cite{Mey:new} and define a {\em  braided metric} as a matrix
$\eta_{ij}$ such that
\eqn{metric}{x_i=\eta_{ia}v^a,\quad v^i=x_a\eta^{ai}}
is an isomorphism of braided-Hopf algebras. Here $\eta^{ij}$ is the inverse
transpose characterised by $\eta_{ja}\eta^{ia}=\delta^i_j=\eta_{aj}\eta^{ai}$.

There are two aspects to this definition, one for the algebra isomorphism and
one for an equivalence of the braiding so that the braided tensor product
algebras are also isomorphic. These immediately come out respectively
as\cite{Mey:new}
\eqn{etaRa}{
\eta_{ia}\eta_{jb}R'{}^a{}_k{}^b{}_l=R'{}^a{}_i{}^b{}_j
\eta_{ak}\eta_{bl},\quad
\eta_{ia}\eta_{jb}R^a{}_k{}^b{}_l=R^a{}_i{}^b{}_j
\eta_{ak}\eta_{bl}}
We will see later in Section~6.1 that the above constructions are generally
covariant under a background quantum group. It is natural to demand that $\eta$
preserves this (is an intertwiner for the coaction). This implies the second of
(\ref{etaRa}) and other relations too between $\eta$ and $R$ which we could
take along with (\ref{etaRa}) as axioms for a covariant metric. These
include\cite{KemMa:alg}
\eqn{etaRb}{\eta_{ka}R^i{}_j{}^a{}_l=\lambda^{-2}R^{-1}{}^i{}_j
{}^a{}_k\eta_{al},\quad
\eta_{ka}\widetilde{R}^i{}_j{}^a{}_l=\lambda^2 R^i{}_j{}^a{}_k\eta_{al}}
\eqn{etaRc}{R^a{}_j{}^k{}_l\eta_{ai}=\lambda^{-2}\eta_{ja}R^{-1}{}^a{}_i
{}^k{}_l,\quad \widetilde R^a{}_j{}^k{}_l\eta_{ai}
=\lambda^{2}\eta_{ja}R^a{}_i{}^k{}_l}
obtained by the methods in Section~6.1. The parameter $\lambda$ which shows up
here is called the {\em quantum group normalisation constant}\cite{Ma:lin} and
depends on the R-matrix. Finally, we  can also require our metric to be
symmetric in some sense. The natural condition is to use the same notion as the
sense in which the braided covector and vector algebras are commutative, i.e.,
\eqn{etaRd}{ \eta_{ba}R'{}^a{}_i{}^b{}_j=\eta_{ij}.}

The corresponding equations in terms of the inverse transposed metric
$\eta^{ij}$  are
\eqn{etainvRa}{
\eta^{ia}\eta^{jb}R'{}^k{}_a{}^l{}_b=R'{}^i{}_a{}^j{}_b
\eta^{ak}\eta^{bl},\quad \eta^{ia}\eta^{jb}R{}^k{}_a{}^l{}_b=R'{}^i{}_a{}^j{}_b
\eta^{ak}\eta^{bl}}
\eqn{etainvRb}{\eta^{la}R^i{}_j{}^k{}_a=\lambda^{-2
}R^{-1}{}^i{}_j{}^l{}_a\eta^{al},\quad
\eta^{la}\widetilde{R}^i{}_j{}^k{}_a=\lambda^2 R^i{}_j{}^l{}_a\eta^{ak}}
\eqn{etainvRc}{R^i{}_a{}^k{}_l\eta^{aj}=\lambda^{-2}\eta^{ia}R^{-1}
{}^j{}_a{}^k{}_l,\quad \widetilde{R}^i{}_a{}^k{}_l\eta^{aj}=\lambda^2
\eta^{ia} R^j{}_a{}^k{}_l}\eqn{etainvRd}{R'{}^i{}_a{}^j{}_b
\eta^{ba}=\eta^{ij}.}
We will see in Section~5.6 how to construct quantum metrics as an application
of braided-differentiation. It can also be obtained from knowledge of the
$*$-structure. We describe this next.

\subsection{Braided $*$-structures}

If we think of our algebras above as like the co-ordinate functions on a
manifold or (equally well)
as like a quantum system, we need to specify an operation $*$ from the algebra
to itself which is
like pointwise complex conjugation. It should be antilinear, square to 1 and be
an
antialgebra homomorphism, i.e., should make our algebras into $*$-algebras.
Such a structure is important in the non-commutative or quantum case
because it determines what it means for a representation of our algebras to be
`real'. Namely in the quantum case $*$ should map over to Hermitian
conjugation, which requirement generalises the notion of a unitary
representation of a
group. Since we do not have either points or groups, we should specify $*$
axiomatically by these and further properties.

For a braided group the most useful further axioms (as determined by experience
rather than by
abstract considerations) appear to be \cite{Ma:mec}
\eqn{bra*}{(*\tens *)\circ\Delta=\tau\circ\Delta\circ *,\quad \eps\circ
*=\bar{\ }\circ\eps,\quad *\circ S=S\circ *}
where $\tau$ denotes the usual transposition.

Whether or not our braided vectors, covectors and matrices etc have a natural
$*$-structure depends on the
particular algebras. These in turn depend on the chosen $R,R'$ matrices. We
discuss here what can be said at this
general level depending on the general properties of these matrices: there may
be other possibilities too
for individual algebras when one looks at them by hand. There are two useful
cases\cite{Ma:mec}
\eqn{Rreal}{ \overline{R^i{}_j{}^k{}_l}=R^l{}_k{}^j{}_i\quad {\rm (real\
type\ I)},\quad
\overline{R^i{}_j{}^k{}_l}=R^{-1}{}^j{}_i{}^l{}_k\quad {\rm (antireal\ type\
I)}.}
There are also type II cases needed to cover $q$-Minkowski space if we treat it
as a braided covector algebra\cite{Ma:star}. We use the same classification for
$R'$.

It is natural to ask our metric, when it exists, to respect $*$. We require
\eqn{metric*}{\overline{\eta^{ij}}=\eta_{ji}}
which is compatible with the various relations
(\ref{etainvRa})--(\ref{etainvRd}) when $\lambda$ is real, for either the
real or antireal type I case for $R,R'$. Then one can check that
\eqn{x*}{ x_i^*=x_a\eta^{ai}}
makes $\Vhaj(R',R)$ into a $*$-braided group obeying the above
axioms\cite{Ma:star}. The braided vector algebra is isomorphic when there is a
metric, so it has a corresponding operation $v^i{}^*=\eta_{ia}v^a$ by this
isomorphism. There is also a second $*$-structure given on the $v^i$ (say) by
$v^i{}^*=\eta_{ai}v^a$ which is needed for the duality theory in\cite{Ma:star}.

When there is no metric, it is more reasonable to think of our braided
covectors and vectors as holomorphic and antiholomorphic. This is because
rather than a $*$-structure as above, there is a natural map from one to the
other as $x_i^*=v^i$. The details and some non-trivial theorems in this
direction are in \cite{Ma:star}. This holomorphic situation is the one
that applies to the most simple examples such as the quantum plane
$\C_q^{2|0}$.

Next, we look at $*$-structures for our other objects. We concentrate on $R$ of
real type I, which holds for our standard $R_{gl_2}$ and similar R-matrices
when $q$ is real. In the case of the quantum matrices $A(R)$ it is known that
these often have a `unitary' $*$-structure\cite{Wor:com}
\eqn{t*}{ t^i{}_j{}^*=St^j{}_i}
at least when made into Hopf algebras with antipode $S$ for the multiplicative
coproduct. Usually the same formula can be used at the bialgebra level too. For
example, the usual quantum matrices $M_q(2)$ in Example~3.7 have a standard
$*$-structure
\eqn{m2*}{ \pmatrix{a^*&b^*\cr c^* & d^*}=\pmatrix{d&-q^{-1}c\cr -qb&a}}
when $q$ is real. This is the $*$-structure familiar for $SU_q(2)$ except that
the same formula works even when we do not demand the $q$-determinant relation.
Of course, we do not obey the antipode part of Woronowicz' axioms for a Hopf
$*$-algebra\cite{Wor:com}, but only the properties for a $*$-bialgebra with
respect to comultiplication.

At least in the Hecke case we also have the possibility of a second approach to
the $*$-structure on $A(R)$, namely to write it as a braided covector space as
in
Section~3.2. Then the bigger multi-index matrices $\bf R,\bf R'$ in
(\ref{tRR'}) will also be real type I, and hence when there is a quantum metric
$\eta^{IJ}$ we will have a $*$-structure by the above. This second approach
ensures that we obey the axioms (\ref{bra*}) for the additive braided group
structure.

\begin{example} The usual quantum matrices $M_q(2)$ in Example~3.7 have a
braided covector $*$-structure
\[ \pmatrix{a^*&b^*\cr c^* & d^*}=\pmatrix{-qd&c\cr b&-q^{-1}a}\]
when $q$ is real.
\end{example}
\proof We regard $M_q(2)$ as a 4-dimensional braided covector space and use the
quantum metric
\[\eta^{IJ}=\pmatrix{0&0&0&-q^{-1}\cr 0&0&1&0\cr 0&1&0&0\cr -q&0&0&0}.\]
One can check that it obeys the conditions in Section~4.3 with respect to $\bf
R, R'$ and $\lambda=q^{-1}$, as well as the condition (\ref{metric*}). Then we
get the braided covector $*$-structure from (\ref{x*}). \endproof

The braided covector approach works just as well on the variants $\bar A(R)$.
The multi-index matrices $\bf R,R'$ from (\ref{xRR'}) are again of real type I
and a quantum metric obeying (\ref{metric*}) then gives us a braided covector
$*$-structure. This includes our standard example of $q$-Euclidean space $\bar
M_q(2)$ from Example~3.8. We describe it further in Section~7.1.  Also, there
is a systematic process of twisting  which turns $A(R)$ into $\bar A(R)$. It
turns the multiplicative $*$-structure on the matrix generators $\vect$ of the
former into the braided covector $*$-structure on the generators $\vecx$ of the
latter. By this general theory one could begin with the usual (\ref{m2*}) on
$M_q(2)$, take the same form for the $*$-structure on $\bar M_q(2)$ and hence
by (\ref{x*}) deduce its quantum metric $\eta^{IJ}$.
This is the line taken by the author in \cite{Ma:euc}.

For $*$ structures on braided matrices $B(R)$ the discussion is simpler because
we do not need to consider an antipode. This topic was covered in \cite{Ma:mec}
where it was shown that for $R$ of real type I, we have a braided $*$-bialgebra
of Hermitian type
\eqn{u*}{ u^i{}_j{}^*=u^j{}_i}
with respect to the braided comultiplication in the axioms (\ref{bra*}).
Firstly, it is easy to see that $B(R)$ for real-type $R$ is a $*$-algebra. See
\cite{Ma:mec} or other places where the algebra relations have been studied.
Next note that $B(R)\und\tens B(R)$ has a natural $*$-structure defined not in
the obvious way but by\cite{Ma:mec}
\[ (b\tens c)^*=c^*\tens b^*,\quad\forall b,c\in B(R),\quad {\it i.e.}\quad
u^i{}_j{}^*=u'{}^j{}_i,\quad u'{}^i{}_j{}^*=u{}^j{}_i\]
on the generators. We just have to check that this is compatible with the
relations of the braided tensor product. Thus
\align{&&\equad (R^{-1}{}^i{}_a{}^k{}_b u'{}^a{}_c R^c{}_j{}^b{}_d
u^d{}_l)^*=u'{}^l{}_d R^d{}_b{}^j{}_c u^c{}_a R^{-1}{}^b{}_k{}^a{}_i \\
&&=R^l{}_a{}^j{}_b u^b{}_c R^{-1}{}^a{}_d{}^c{}_i
u'{}^d{}_k=(u^k{}_dR^{-1}{}^i{}_c{}^d{}_a u'{}^c{}_b R^b{}_j{}^a{}_l)^*}
as required. We used the multiplicative braid statistics in Section~3.3. After
this we just have to note that the multiplicative  $\Delta_\cdot$ is a
$*$-algebra map on the generators and hence extends to products as a
$*$-algebra map.

As before, a second way is to think of our braided matrices  as a braided
covector space, as explained in Section~3.3. We have $\bf R,R'$ from
(\ref{uRR'a}) obeying some different properties, called {\em type II} in
\cite{Ma:star} but again leading to a $*$-braided group structure on our
covectors, this time with (\ref{u*}). For a direct check, the computation is
like the above but this time with Meyer's additive braid statistics:
\align{&&\equad (R^{-1}{}^i{}_a{}^k{}_b u'{}^a{}_c R^c{}_j{}^b{}_d
u^d{}_l)^*=u'{}^l{}_d R^d{}_b{}^j{}_c u^c{}_a R^{-1}{}^b{}_k{}^a{}_i \\
&&=R^l{}_a{}^j{}_b u^b{}_c R^c{}_i{}^a{}_d u'{}^d{}_k=(u^k{}_dR^d{}_a{}^i{}_c
u'{}^c{}_b R^b{}_j{}^a{}_l)^*}
The braided coaddition extends to products as a $*$-algebra homomorphism to
this braided tensor product algebra. It is remarkable that the same Hermitian
form (\ref{u*}) works for both comultiplication and coaddition, making $B(R)$ a
$*$-braided group in two ways.

Note that under coaddition, the transposition $\tau$ in (\ref{bra*}) might be
unexpected -- it is not visible classically because the coaddition is then
cocommutative (symmetric in its output). But the
comultiplication on the other hand is not symmetric and here a $\tau$ really is
to be expected for Hermitian matrices: classically
the product of two Hermitian matrices $M,N$ is not Hermitian but rather obeys
$(MN)^*=NM$.

In addition, there is a strict relation of transmutation in \cite{Ma:mec} which
obtains this Hermitian $*$-structure on $B(R)$ from the multiplicative
unitary-like one on $A(R)$. There is also a theory of twisting in \cite{Ma:euc}
which turns the braided covector $*$-structure of $\bar A(R)$ into $B(R)$. So
all three algebras are connected by general theory. The Hermitian form
obviously means that the standard $2\times 2$ braided matrices $BM_q(2)$ in
Example~3.10 should be regarded as a natural $q$-Minkowski space. We can
deduce
its quantum metric $\eta^{IJ}$ from this $*$-structure and (\ref{x*}). It is
covered in Section~7.2. The twisting relation between $\bar M_q(2)$ and
$BM_q(2)$ becomes `quantum Wick rotation' in this context\cite{Ma:euc}.

Once one has a suitable $*$-structure on a braided space, it is natural to ask
about its implications for the various other constructions in braided geometry,
such as those to be described below. This is covered in \cite{Ma:star}, to
which we refer the interested reader. Not all problems in this direction are
yet solved.

\section{Braided analysis}

So far we have described how to $q$-deform $\R^n$ as a braided covector or
vector space and
developed the associated concepts of linear algebra. We now proceed to the
first steps of `braided analysis' on such spaces. We concentrate first on
understanding braided differentiation from \cite{Ma:fre}. This then determines
braided exponentials, braided-Gaussians, braided-integration and
braided-differential forms etc., i.e. some of the remaining basic concepts for
`analysis' on $\R^n$.

\subsection{Braided differentiation}

The next stage after braided addition and linear algebra is to make an
infinitesimal addition, which leads to the concept of differentiation. We can
think of the coaddition (\ref{xhopfb})--(\ref{xhopfc}) on our covectors $x_i$
equally well as a braided coaction of one copy of the covectors on another,
i.e. as  a global translation. If we denote the generators of the coacting copy
by $\veca$  then the content of Section~3.1
is that $\veca+\vecx$ also obeys the relations of a braided covector provided
we remember the braid statistics $\vecx_1\veca_2=\veca_2\vecx_1R$. We are now
ready to follow the ideas of I. Newton and define differentiation
$\del^i:\Vhaj(R',R)\to
\Vhaj(R',R)$ as an infinitesimal translation\cite{Ma:fre}
\eqn{diffx}{ \del^i f(\vecx)=
\left(a_i^{-1}(f(\veca+\vecx)-f(\vecx))\right)_{\veca=0}\equiv{\rm coeff\ of\
}a_i{\rm \ in\ } f(\veca+\vecx).}
We take the linear part in $a_i$, which is some function of $\vecx$ and does
not depend in fact on inverting $a_i$ or taking a limit. For example, on
monomials we have
\align{&&\equad{\rm coeff}_{a_i}
\left((\veca_1+\vecx_1)(\veca_2+\vecx_2)
\cdots(\veca_m+\vecx_m)\right)\\
&&={\rm coeff}_{a_i}
\left(\veca_1\vecx_2\cdots\vecx_m+\vecx_1\veca_2\vecx_3\cdots\vecx_m
+\cdots+\vecx_1\cdots\vecx_{m-1}\veca_m\right)\nonumber\\
&&= {\rm
coeff}_{a_i}(\veca_1\vecx_2\cdots\vecx_m(1+(PR)_{12}+(PR)_{12}(PR)_{23}
+\cdots+(PR)_{12}\cdots (PR)_{m-1, m}))}
giving us the result\cite{Ma:fre}
\eqn{dxxx}{\del^i(\vecx_1\cdots \vecx_m)= {\bf e}^i{}_1\vecx_2\cdots\vecx_m
\left[m;R\right]_{1\cdots m},\quad  \del^ix_{i_1}\cdots
x_{i_m}=\delta^i{}_{j_1}x_{j_2}\cdots
x_{j_m}\left[m;R\right]^{j_1\cdots j_m}_{i_1\cdots i_m}}
where ${\bf e}^i$ is a basis covector $({\bf e}^i){}_j=\delta^i{}_j$
and\cite{Ma:fre}
\eqn{branum}{\left[m;R\right]=1+(PR)_{12}+(PR)_{12}(PR)_{23}
+\cdots+(PR)_{12}\cdots (PR)_{m-1,m}}
is a certain {\em braided integer} matrix living in the $m$-fold matrix tensor
product of $M_n$.

The operators $\del^i$ have upper indices so one should hope that they obey the
relations of the braided vectors $v^i$ in Section~3.1. This is indeed true and
means that it defines an action of $V(R',R)$ on  $\Vhaj(R',R)$ given by this
differentiation. We have\cite{Ma:fre}
\align{\del^i\del^k\vecx_1\cdots\vecx_m\equad&&={\bf e}^k{}_1{\bf
e}^i{}_2\vecx_3\cdots\vecx_m\left[m-1;R\right]_{2\cdots
m}\left[m;R\right]_{1\cdots m}\\
R'{}^i{}_a{}^k{}_b\del^b\del^a\vecx_1\cdots\vecx_m\equad
&&=R'{}^i{}_a{}^k{}_b\del^b{\bf e}^a{}_1\vecx_2\cdots\vecx_m
\left[m;R\right]_{1\cdots m}\\
&&=R'{}^i{}_a{}^k{}_b{\bf e}^a{}_1{\bf
e}^b{}_2\vecx_3\cdots\vecx_m\left[m-1;R\right]_{2\cdots
m}\left[m;R\right]_{1\cdots m}\\
&&={\bf e}^k{}_1{\bf e}^i{}_2\vecx_3\cdots\vecx_m
(PR')_{12}\left[m-1;R\right]_{2\cdots m}\left[m;R\right]_{1\cdots m}}
which are equal due to the identity
\eqn{R'mm}{\left[m-1;R\right]_{2\cdots m}\left[m;R\right]_{1\cdots
m}=(PR')_{12}\left[m-1;R\right]_{2\cdots m}\left[m;R\right]_{1\cdots m}}
proven in \cite{Ma:fre}, to which we refer for further details.

Moreover, the braided covectors are braided-covariant under this action. The
concept is much like the covariance in Section~4.1: we use braid statistics. In
fact we need the inverse braiding $\Psi^{-1}$ as given between vectors and
covectors in (\ref{xvustata})--(\ref{xvustatb}). As usual, it extends to
products in such a way as to be compatible with the product map. This
consideration translates into the
{\em braided-Leibniz rule}\cite{Ma:fre}
\ceqn{braleiba}{ \del^i(ab)=(\del^i a)b+\cdot\Psi^{-1}(\del^i\tens a)b\\
\Psi^{-1}(\del^i\tens\vecx_1\cdots\vecx_r)={\bf
e}^i{}_1\vecx_2\cdots\vecx_r\vecx_{r+1}(PR)_{12}\cdots
(PR)_{r,r+1}\tens\del_{r+1}.}
There is an abstract or diagrammatic derivation for this based on
duality\cite{Ma:introp}. For
a direct proof one has\cite{Ma:fre}
\align{&&\equad
(\del^i\vecx_1\cdots\vecx_r)\vecx_{r+1}\cdots\vecx_{m}+\Psi^{-1}
(\del^i\tens\vecx_1\cdots\vecx_r)\vecx_{r+1}\cdots\vecx_m\\
&&={\bf e}^i{}_1\vecx_2\cdots\vecx_r\left[r;R\right]_{1\cdots
r}\vecx_{r+1}\cdots\vecx_m\\
&&\qquad+{\bf e}^i{}_1\vecx_2\cdots\vecx_r\vecx_{r'+1}(PR)_{12}\cdots
(PR)_{r,r'+1}\del_{r'+1}\vecx_{r+1}\cdots\vecx_m\\
&&={\bf
e}^i{}_1\vecx_2\cdots\vecx_r\vecx_{r+1}\cdots\vecx_m\left[r;R\right]_{1\cdots
r}
\\
&&\qquad +{\bf e}^i{}_1\vecx_2\cdots\vecx_r\vecx_{r+1}(PR)_{12}\cdots
(PR)_{r,r+1}\vecx_{r+2}\cdots\vecx_m  \left[m-r;R\right]_{r+1\cdots m}}
where we use (\ref{dxxx}) to evaluate the differentials. The primed $r'+1$
labels
a distinct matrix space from the existing $r+1$ index. These are then
identified by the ${\bf e}_{r+1}$ brought down by the action of $\del_{r'+1}$.
The resulting expression coincides with ${\bf
e}^i{}_1\vecx_2\cdots\vecx_r\vecx_{r+1}\cdots \vecx_m \left[m;R\right]_{1\cdots
m}$
due to the braided-integer identity\cite{Ma:fre}
\eqn{addbranum}{\left[r;R\right]_{1\cdots r}
+(PR)_{12}\cdots (PR)_{r,r+1}\left[m-r;R\right]_{r+1\cdots m}=
\left[m;R\right]_{1\cdots m}.}

Another way to express this braided-Leibniz rule is in terms of commutation
relations between differentiation and position operators
$\hat{x_i}:\Vhaj(R',R)\to \Vhaj(R',R)$ acting by multiplication from the left
by $x_i$. The braided-Leibniz rule applied to $\del^i(x_jf(\vecx))$ gives at
once that
\eqn{braleibb}{\del^i\hat{x_j}-\hat{x_a}R^a{}_j{}^i{}_b\del^b=\delta^i{}_j,\
{\it i.e.}\  \del_1\hat{\vecx_2}-\hat{\vecx_2}R_{21}\del_1=\id}
as operators on $\Vhaj(R',R)$. This is obviously the point of view that could
be called `braided quantum mechanics'. Indeed, we showed in \cite{Ma:fre} that
there is an abstract {\em braided Weyl algebra} defined by
\eqn{braweyl}{\vecx_1\vecx_2=\vecx_2\vecx_1R',\quad
\vecp_1\vecp_2=R'\vecp_2\vecp_1,\quad \vecp_1\vecx_2-\vecx_2 R_{21}\vecp_1=\id}
with (\ref{braleibb}) as an operator realisation. The construction was as a
braided cross product\cite{Ma:bos} and generalised the standard $GL_q(n)$ case
that had been considered before\cite{PusWor:twi}\cite{Kem:sym}.

\begin{example} The braided-line $\C_q$ at the end of Section~2 has braided
differentiation
\[\del x^m=x^{m-1}(1+q+\cdots+q^{m-1})=\left[m;q\right]
x^{m-1}\quad\Rightarrow\quad \del f(x)={f(qx)-f(x)\over (q-1)x}.\]
The braided-Leibniz rule is
\[ \del(x^{n}x^{m})=(\del x^n)x^m+q^{n}x^n(\del x^m),\ {\it i.e.}\
\del(fg)=(\del f)g+(L_qf)\del g\]
which is just as in the case of a superderivation, but with $q$ in the role of
$-1$ and a $\Z$-grading in the role of $\Z_2$-grading. The degree of $\del$
here is $-1$ and $L_q(f)(x)=f(qx)$.
\end{example}
\proof We use $R=(q)$ and $R'=(1)$ so that $\Vhaj(R',R)=\C[x]$ is the
braided-line. The braiding is $\Psi(\del\tens x)=q^{-1}x\tens\del$. We use the
inverse braiding when computing $\del x^m$ giving the $q$-integer as shown. Or
we
just use (\ref{dxxx}) with $R=(q)$ and hence $[m;R]=[m,q]$. It is easy enough
to verify the braided-Leibniz rule here explicitly. \endproof

This is where the familiar $q$-integers come from in braided geometry, and is
the
reason that we called
(\ref{branum}) braided integers. There is a $q$ each time $\del^i$ passes an
$x$ due to the braid statistics between them.
It is also the correct point of view on $q$-differentiation $\del$ and
reproduces easily well-known formulae in the 1-dimensional case. On the other
hand, the braided formalism works on any higher-dimensional braided space just
as well.

\begin{example} The quantum plane $\C_q^{2|0}$ as a braided covector space in
Example~3.1 has braided differentiation\cite{Ma:fre}
\[ {\del\over\del x} x^ny^m=\left[n;q^2\right] x^{n-1} y^m,\quad
{\del\over\del y} x^ny^m=q^{n}x^n\left[m;q^2\right] y^{m-1}\]
where $\left[m;q^2\right]={q^{2m}-1\over q^2-1}$. The derivatives obey the
relations
\[ {\del\over\del y}{\del\over\del x}=q^{-1}{\del\over\del x}{\del\over\del
y}\]
\end{example}
\proof We know from the theory above that the relations of the $\del^i$ are
necessarily the braided vector ones for $\C_{q^{-1}}^{2|0}$ in Example~3.5. The
action of the generators is $\del^ix_j=\delta^i{}_j$ which we then extend using
the braided-Leibniz rule (\ref{braleibb}) with the standard $R_{gl_2}$. This
gives the above results by an easy induction. Note that the $q$-derivatives act
like in the 1-dimensional case in each variable except that $\del\over\del y$
picks up a factor $q$ when it passes $x$ due to some braid statistics. This
example recovers partial differentiation on the quantum plane as deduced by
another approach (from the exterior algebra of forms) in
\cite{PusWor:twi}\cite{WesZum:cov}.\endproof

At another extreme, which works in any dimension, we can let $R$ be any
invertible solution of the QYBE and $R'=P$ the permutation matrix. Then
$\Vhaj(P,R)$ is the free algebra $\<x_i\>$ with no relations. This could be
called the {\em  free braided plane}\cite{Ma:fre}  associated to an
R-matrix. It is in a certain sense universal, with the others as quotients. The
vector algebra is the free algebra $\<v^i\>$ with no relations and is realised
by $\del^i$ acting as in (\ref{dxxx}). The R-matrix is still used, in the
braiding.

\subsection{Braided binomial theorem}

Next we would like to understand much better the braided coaddition on our
braided covectors from Section~3.1, namely how it looks on products of the
generators. We need $q$-binomial coefficients as in the 1-dimensional case in
Section~2, which we have to generalise now to our higher-dimensional R-matrix
setting. This was done (by the author)  following the inductive
way that binomial coefficients are usually defined. We require\cite{Ma:fre}
\ceqn{brabinoma}{ \left[{m\atop r};R\right]_{1\cdots m}\nquad\ =(PR)_{r,
r+1}\cdots (PR)_{m-1,
m}\left[{m-1\atop r-1};R\right]_{1\cdots m-1}\nquad\ +\left[{m-1\atop
r};R\right]_{1\cdots m-1}\\
  \left[{m\atop 0};R\right]=1, \qquad \left[{m\atop r};R\right]=0\quad {\rm
if}\ r>m}
where the suffices refer as usual to the matrix position in tensor powers of
$M_n$. This defines in particular
\ceqn{brabinomb}{\left[{m\atop m};R\right]=\left[{m-1\atop
m-1};R\right]=\cdots=\left[{1\atop 1};R\right]=1\\
 \left[{m\atop 1};R\right]=(PR)_{12}\cdots (PR)_{m-1,
m}+\left[{m-1\atop 1};R\right]=\cdots=\left[m;R\right].}
A similar recursion defines $\left[{m\atop 2};R\right]_{1\cdots m}$ in terms of
$ \left[{m\atop 1};R\right]$ (which is known) and $\left[{m-1\atop
2};R\right]$, and similarly (in succession) up to $r=m$.

The main result from a technical point of view is to compute
these braided-binomial coefficients. This is the
{\em braided binomial theorem}\cite{Ma:fre}
\eqn{binomthma}{ \left[r;R\right]_{1\cdots r}\left[{m\atop r};R\right]_{1\cdots
m}=\left[{m-1\atop r-1};R\right]_{2\cdots m}\left[m;R\right]_{1\cdots m}}
or formally
\alignn{binomthmb}{ \left[{m\atop r};R\right]_{1\cdots m}\equad&&
=\left[r;R\right]_{1\cdots
r}^{-1}\cdots \left[2;R\right]^{-1}_{r-1,
r}\left[m-r+1;R\right]_{r\cdots m}\cdots \left[m;R\right]_{1\cdots
m}\nonumber\\
&&=[r;R]!{}^{-1}_{1\cdots r}[m-r;R]!{}^{-1}_{r+1\cdots m}[m;R]!{}_{1\cdots m}}
where
\eqn{brafact}{ [m;R]!\equiv [2;R]_{m-1\, m}[3;R]_{m-2\, m}\cdots [m;R]_{1\cdots
m}}
is the {\em braided factorial}\cite{Ma:fre}. The proof is by induction and a
series of
lemmas, see \cite{Ma:fre} for details. Note that one does not really need the
braided-integers or factorials here to be invertible
(just as for the usual binomial coefficients). For example, the recursion
relation in the theorem implies that
\alignn{mm-1}{\left[{m\atop m-1};R\right]_{1\cdots
m}\equad&&=1+(PR)_{m-1,m}+(PR)_{m-1,m}(PR)_{m-2,m-1}+\cdots\nonumber\\
&&\qquad\quad\cdots+(PR)_{m-1,m}\cdots (PR)_{12}=\left[m;R_{21}\right]_{m\cdots
1}}
One can prove numerous other identities of this type in analogy with usual
combinatoric
identities.

The braided binomial theorem demonstrates the beginning of some kind of
braided-number-theory or braided-combinatorics. Because it holds for any
invertible solution of the QYBE, it corresponds to a novel  identity in the
group algebra of the braid group. Physically, it corresponds to `counting' the
`partitions' of a box of braid-statistical particles. Mathematically, it tells
us the coaddition on
products\cite{Ma:fre}
\ceqn{bracoadda}{ (a_{i_1}+x_{i_1})\cdots
(a_{i_m}+x_{i_m})=\sum_{r=0}^{r=m}a_{j_1}\cdots a_{j_r}x_{j_{r+1}}\cdots
x_{j_m}\left[{m\atop r};R\right]^{j_1\cdots j_m}_{i_1\cdots i_m}\\
{\it i.e.},\quad (\veca_1+\vecx_1)\cdots
(\veca_m+\vecx_m)=\sum_{r=0}^{r=m}\veca_1\cdots\veca_r\vecx_{r+1}
\cdots\vecx_m\left[{m\atop r};R\right]_{1\cdots m}}
or in more formal terms
\ceqn{coprodxxx}{\Delta(x_{i_1}\cdots x_{i_m})=\sum_{r=0}^{r=m}x_{j_1}\cdots
x_{j_r}\tens x_{j_{r+1}}\cdots x_{j_m}\left[{m\atop r};R\right]^{j_1\cdots
j_m}_{i_1\cdots i_m}\\
{\it i.e.},\quad
\Delta(\vecx_1\vecx_2\cdots\vecx_m)=\sum_{r=0}^{r=m}\vecx_1\cdots\vecx_r\tens
\vecx_{r+1}
\cdots\vecx_m\left[{m\atop r};R\right]_{1\cdots m}}
The proof is once again by induction. Suppose it is true for $m-1$, then
\align{&&\equad(\veca_1+\vecx_1)\cdots
(\veca_m+\vecx_m)\\
&&=(\veca_1+\vecx_1)\cdots
(\veca_{m-1}+\vecx_{m-1})\veca_m+(\veca_1+\vecx_1)\cdots
(\veca_{m-1}+\vecx_{m-1})\vecx_m\\
&&=\sum_{r=0}^{m-1}\veca_1\cdots\veca_r\vecx_{r+1}\cdots
\vecx_{m-1}\veca_m\left[{m-1\atop r};R\right]_{1\cdots
m-1}+(\veca_1+\vecx_1)\cdots (\veca_{m-1}+\vecx_{m-1})\vecx_m\\
&&=\sum_{r=1}^{m}\veca_1\cdots\veca_{r-1}\vecx_{r}\cdots\vecx_{m-1}
\veca_m\left[{m-1\atop r-1};R\right]_{1\cdots m-1}
\\
&&\qquad\qquad+\sum_{r=0}^{m-1}\veca_1\cdots\veca_r\vecx_{r+1}
\cdots\vecx_m\left[{m-1\atop r};R\right]_{1\cdots m-1}\\
&&=\sum_{r=1}^{m}\veca_1\cdots\veca_{r-1}\veca_r\vecx_{r+1}
\cdots\vecx_{m}(PR)_{r,r+1}\cdots (PR)_{m-1,m}
\left[{m-1\atop r-1};R\right]_{1\cdots m-1}\\
&&\qquad\qquad+\sum_{r=0}^{m-1}\veca_1\cdots\veca_r\vecx_{r+1}
\cdots\vecx_m\left[{m-1\atop r};R\right]_{1\cdots m-1}\\
&&=\sum_{r=1}^{m}\veca_1\cdots\veca_r\vecx_{r+1}\cdots\vecx_{m}\left[{m\atop
r};R\right]_{1\cdots m}
+\vecx_{1}\cdots\vecx_{m}\left[{m-1\atop 0};R\right]_{1\cdots m-1}}
using the induction hypothesis and (\ref{brabinoma}). The last term is also the
$r=0$ term in the desired sum, proving the result for $m$.

\subsection{Duality of braided vectors and covectors}

Next we apply the braided differentiation operators and braided binomial
theorem to establish a duality pairing between vectors and covectors. This is
needed to round-off our concepts of linear algebra from Sections~3 and~4. We
will need it also in the next section to define the braided-exponential. Two
braided groups $B,C$ say are said to be {\em in duality} if there is a map $\<\
,\ \>: B\tens C\to \C$ such that the product in one determines the coproduct or
coaddition $\Delta$ in the other, etc., according to
\eqn{bradual}{ \<ab,c\>=\<a,\<b,c\Bo\> c\Bt\>,\quad \<a,cd\>=\<a\Bo,\<a\Bt,c\>
d\>,\quad
\<Sa,c\>=\<a,S c\> }
where $\Delta c=c\Bo\tens c\Bt$ is the braided-coproduct and $S$ the
braided-antipode. We also require the pairing with 1 to be the counit. This
(\ref{bradual}) is
not the usual pairing because
we do not move $b$ past $c\Bo$ to evaluate on $c\Bt$ etc, as one would
usually do. It {\em is} possible to define such a more usual pairing by using
the
braiding $\Psi$
to make the transposition but the result would be equivalent to (\ref{bradual})
via the braided antipode, so we avoid such an unnecessary complication.

There is such a duality pairing of the braided vectors and covectors given
by\cite{Ma:fre}
\ceqn{vxdual}{\<v^{i_m}\cdots v^{i_2}v^{i_1},x_{j_1}x_{j_2}\cdots
x_{j_r}\>=\delta_{m,r}([m;R]!)_{j_1j_2\cdots j_m}^{i_1i_2\cdots i_m}\\
{\it i.e.},\quad \<\vecv_m\cdots\vecv_2\vecv_1,\vecx_1\vecx_2\cdots\vecx_r\>
=\delta_{m,r}[m;R]!,\quad  \<f(\vecv),g(\vecx)\>=\eps\circ f(\del)g(\vecx)}
This linear map is manifestly well-defined because the differentiation
operators $\del^i$ are well-defined on products of the $x_i$ by their very
construction, so the relations are respected on this side of the pairing. The
relations of the vector algebra on the other side of the pairing are also
respected by the result in Section~5.1 that $\del^i$ indeed realise the vector
algebra. We have still to check that the product on one side is the braided
coproduct of the other. This is
\align{&&\equad \<\vecv_m\cdots \vecv_{r+1}\cdot \vecv_r\cdots
\vecv_1,\vecx_1\cdots \vecx_m\>=[m;R]!\\
&&\equad \<\vecv_m\cdots \vecv_{r+1}\tens \vecv_r\cdots \vecv_1,\Delta
\vecx_1\cdots \vecx_m\>=[r,R]!_{1\cdots r}[m-r,R]!_{r+1\cdots m}\left[{m\atop
r};R\right]}
where we evaluate the inner $V,\Vhaj$ first and then the remaining outer two as
required in (\ref{bradual}). The coproduct is from (\ref{coprodxxx}) in
Section~5.2 and our two expressions coincide just by the braided-binomial
theorem (\ref{brabinoma}). Similarly for the coproduct of products of the $v^i$
by an analogous computation. The duality pairing with respect to the
unit/counit and antipode are clear from the form on the generators. Note that
one could
also turn this around and recover
\eqn{dev}{\del^i=(\<v^i,\ \>\tens\id)\circ\Delta}
if we start from the knowledge that our vectors and covectors are dual.

The second half of this proof is clear enough by the evident symmetry between
the braided vector and braided covector constructions. To give this in detail
one has to redevelop the various steps above under this symmetry. Thus we have
differentiation operators $\overleftarrow{\del}_i={\del\over\del v^i}$ acting
from the right and
defined by
\eqn{delva}{ f(\vecv)\overleftarrow{\del}_i=f(\vecv+\vecw)|_{{\rm coeff\ of}
w^i};\quad \vecw_1\vecv_2=R\vecv_2\vecw_1}
\ceqn{delvb}{\vecv_m\cdots\vecv_2\vecv_1\overleftarrow{\del}_i=[m;R]^{\rm
op}\vecv_m\cdots\vecv_3\vecv_2({\bf f_i})_1\\
{}[m;R]^{\rm op}=1+(PR)_{12}+(PR)_{23}(PR)_{12}+\cdots+(PR)_{m-1\, m}\cdots
(PR)_{12}}
where $\bf f_i$ is a basis vector  $(f_i)^j=\delta_i{}^j$. The matrices here
are like (\ref{branum}) but with the opposite product of matrices. We can
equally well write the pairing (\ref{vxdual}) as
\[ \<\vecv_m\cdots\vecv_2\vecv_1,\vecx_1\vecx_2\cdots\vecx_r\>
=\delta_{m,r}[m;R]^{\rm op}!,\quad \<f(\vecv),g(\vecx)\>
=\eps\circ f(\vecv)g(\overleftarrow{\del})\]
where
\eqn{opfact}{ [m;R]^{\rm op}!\equiv [m;R]_{1\cdots m}^{\rm op}\cdots
[3;R]_{m-2\, m}^{\rm op}[2;R]_{m-1\, m}^{\rm op}=[m;R]!}
by matrix computations similar to those in Section~5.2. The symmetry between
the $\del$
and the $\overleftarrow{\del}$ points of view expresses the symmetry (with
left-right reversal) in the axioms of a pairing.

This pairing between braided vectors and covectors is typically nondegenerate.
This is true for $q=1$ where it becomes the usual pairing between the functions
on $\R^n$ and the enveloping algebra of $\R^n$, which is by usual
differentiation. Non-degeneracy corresponds to the fact that the only functions
which have zero differentials are constant. But this is also a feature of the
braided-differentiation for standard $R$-matrices at generic $q$ and other
R-matrices near to the identity.

Finally, we mention briefly a more abstract or categorical way of thinking
about our pairing $\<\ ,\ \>$. This is as an {\em evaluation map}
\eqn{VVev}{\ev=\<\ ,\ \>=\epsfbox{cup.eps}}
making $\Vhaj$ the categorical predual of $V$. This is the line taken in
\cite{Ma:introp}, to which we refer the interested reader for further details.
The point is that in the finite-dimensional case, as soon as we have a proper
duality or evaluation map, we also have a {\em coevaluation} as the canonical
element $e_a\tens f^a\in \Vhaj\tens V$, where $e_a$ is a basis and $f^a$ a dual
basis. We next give a direct definition of the braided-exponential which, from
an abstract point of view, is nothing other than $\exp=\coev=\epsfbox{cap.eps}$
but developed as a formal powerseries rather than an element of the algebraic
tensor product.  This is the abstract reason that $\exp$ exists on a general
braided linear space as a formal powerseries.

\subsection{Braided exponentials}

The direct approach to the braided exponential is of course to seek
eigenfunctions of the operators $\del^i$. For the moment we
seek these among formal powerseries in the $x_i$ coordinates, but in our
application to braided Taylor's theorem only finitely many terms will be
nonvanishing. The only difference is that we consider both $\del$ and
$\overleftarrow{\del}$. Thus we define the {\em  braided exponential}
$\exp(\vecx|\vecv)$ as a formal powerseries such that
\ceqn{diffexp}{\del^i\exp(\vecx|\vecv)=\exp(\vecx|\vecv)v^i,\quad
(\eps\tens\id)\exp(\vecx|\vecv)=1\\
\exp(\vecx|\vecv)\overleftarrow{\del}_i=x_i\exp(\vecx|\vecv),\quad
(\id\tens\eps)\exp(\vecx|\vecv)=1.}
We usually also require it to be covariant under our background quantum group,
just as the constructions above were all covariant. This definition corresponds
to an infinitesimal version of the rigidity axioms for $\ev,\coev$ in the
diagrammatic language of \cite{Ma:introp}. It is also reasonable when we
consider
the `integrability condition' for the solution of these braided-differential
equations. Classically it means to
ask for the constraint imposed by commutativity of partial derivatives. Here we
have
\align{(\del_1\del_2-R'\del_2\del_1)\exp(\vecx|\vecv)\equad&&=\del_1
\exp(\vecx|\vecv)\vecv_2-R'\del_2\exp(\vecx|\vecv) \vecv_1\\
&&=\exp(\vecx|\vecv)(\vecv_1\vecv_2-R'\vecv_2\vecv_1)=0}
since the $v^i$ obey the vector algebra. Similarly from the other side in terms
of $\overleftarrow{\del}$. While not a proof, this tells us that, at least
generically, an $\exp$ should exist which is an eigenfunction with respect to
each input.

\begin{example}\cite{Ma:fre} If the braided-integers $[m;R]$ are all
invertible, then $\exp$ is given by
\[ \exp(\vecx|\vecv)=\sum_{m=0}^{\infty}
\vecx_1\cdots\vecx_m[m;R]!_{1\cdots m}^{-1}\vecv_m\cdots\vecv_1.\]
\end{example}
\proof We see this at once from (\ref{dxxx}). Differentiation from the left
brings down $[m;R]$ which reduces $[m;R]!^{-1}$ to $[m-1;R]!^{-1}$.  Similarly
for $\exp(\vecx|\vecv)\overleftarrow{\del}$
where we bring down $[m;R]^{\rm op}$ on the right which reduces $([m;R]!^{\rm
op})^{-1}$. We use the identity (\ref{opfact}). This example includes the
braided-line $\C_q$ and higher-dimensional free quantum planes where $R'=P$ and
$R$ is generic. \endproof

We have to work harder in the more common case where the $[m;R]$ are not all
invertible, but the strategy is just the same. We write as ansatz
\ceqn{Fexp}{ \exp(\vecx|\vecv)=\sum_{m=0}^{\infty}
\vecx_1\cdots\vecx_mF(m;R)\vecv_m\cdots\vecv_1\\
\vecx_1\cdots\vecx_mF(m;R)[m;R]!=\vecx_1\cdots\vecx_m,\quad
[m;R]!F(m;R)\vecv_m\cdots\vecv_1=\vecv_m\cdots\vecv_1}
and solve for $F$. To solve (\ref{diffexp}) in Example~5.3 we did not really
need full invertibility of the $[m;R]!$ but only the weaker form contained in
this ansatz (it is weaker because $\vecx_1\cdots\vecx_2$ and
$\vecv_m\cdots\vecv_1$ involve products of which some linear combinations could
be zero). We have
\align{&&\equad \del^i\exp(\vecx|\vecv)=\sum_{m=0}^\infty {\bf
e}^i_1\vecx_2\cdots\vecx_m[m;R]F(m;R)\vecv_m\cdots\vecv_1\\
&&=\sum_{m=0}^\infty {\bf e}^i_1\vecx_2\cdots\vecx_mF(m-1;R)_{2\cdots
m}([m-1;R]!)_{2\cdots m}[m;R]F(m;R)\vecv_m\cdots\vecv_1\\
&&=\sum_{m=0}^\infty {\bf e}^i_1\vecx_2\cdots\vecx_mF(m-1;R)_{2\cdots
m}[m;R]!F(m;R)\vecv_m\cdots\vecv_1\\
&&=\sum_{m=0}^\infty {\bf e}^i_1\vecx_2\cdots\vecx_mF(m-1;R)_{2\cdots
m}\vecv_m\cdots\vecv_1=\exp(\vecx|\vecv)v^i}
and a similar computation for $\overleftarrow{\del}$ using (\ref{opfact}). We
can also impose  further conditions (that $F(m;R)$ commutes with products
$\vect_1\cdots\vect_m$ of quantum group generators or $R$-matrix conditions to
this effect) if we want to ensure covariance. This is how the diagrammatic
definition $\exp=\coev$ translates in matrix terms. We have developed this
point of view in \cite{KemMa:alg} where it leads to a useful and general
Fourier theory.

\begin{example} If $R$ is Hecke and $R'=q^{-2}R$ then
\[ \exp(\vecx|\vecv)=\sum_{m=0}^\infty
{\vecx_1\cdots\vecx_m\vecv_m\cdots\vecv_1\over
[m;q^2]!}=\sum_{m=0}^\infty
{(\vecx\cdot\vecv)^m\over
[m;q^{-2}]!}\equiv e_{q^{-2}}^{\vecx\cdot\vecv}\]
\end{example}
\proof Since $\vecx_1\vecx_2=\vecx_1\vecx_2 PR'$ and
$PR'\vecv_2\vecv_1=\vecv_2\vecv_1$ we know without any calculation that
$\vecx_1\cdots\vecx_m[m;R]!=\vecx_1\cdots\vecx_m[m;q^2]!$ and
$[m;R]!\vecv_m\cdots\vecv_1=[m;q^2]!\vecv_m\cdots\vecv_1$ making it obvious
that
$F(m;R)=[m;q^2]!^{-1}$ solves (\ref{Fexp}). The second form\cite{ChrZum:tra}
follows at once
from $x_i
(\vecx\cdot\vecv)=q^2(\vecx\cdot\vecv)x_i$, which is valid in the Hecke
case as an easy consequence of the braid-statistics relations
(\ref{xvustatb})
between vectors and covectors. This example includes the quantum planes
$\C_q^{2|0}$ and $\C_q^{1|1}$ in Examples~3.1 and~3.2 and their
higher-dimensional analogues. There is clearly a similar result for the other
`fermionic-type' quantum planes where $R'=q^2R$.
\endproof

We assume then that we have these eigenfunctions $\exp(\vecx|\vecv)$. Since the
$\del^i$
themselves are a realisation of the vector algebra, we are now able to
formulate a {\em braided-Taylor's theorem} as c.f.\cite{Ma:fre}
\eqn{brataylor}{ \exp(\veca|\del) f(\vecx)=f(\veca+\vecx)=\Delta f(\vecx)}
where we use braided addition with $\vecx_1\veca_2=\veca_2\vecx_1R$ as usual.
This follows at once
from the braided-binomial theorem in Section~5.2 as
\align{ && \equad \exp(\veca|\del)
\vecx_1\cdots\vecx_m\\
&&=\sum_{r=0}^{r=m}\veca_{1'}\cdots
\veca_{r'}F(r;R)_{1'\cdots r'}\del_{r'}\cdots\del_{1'}
\vecx_{1}\cdots\vecx_{m} \\
&&=
\sum_{r=0}^{r=m}\veca_{1}\veca_{2'}\cdots\veca_{r'}F(r;R)_{1\cdots
r'}\del_{r'}\cdots\del_{2'}
\vecx_{2}\cdots\vecx_{m} \\
&&=\sum_{r=0}^{r=m}\veca_1\cdots\veca_r F(r;R)_{1\cdots r} \vecx_{r+1}\cdots
\vecx_m
\left[m-r+1;R\right]_{r\cdots m}\cdots \left[m;R\right]_{1\cdots m}\\
&&=\sum_{r=0}^{r=m}\veca_1\cdots\veca_r \vecx_{r+1}\cdots \vecx_m
F(r;R)[r;R]!\left[{m\atop r};R\right]=(\veca_1+\vecx_1)\cdots
(\veca_m+\vecx_m)=\Delta(\vecx_1\cdots\vecx_m).}
Here the $1',2'$ etc refer to copies of $M_n$ distinct from the copies labelled
by $1\cdots m$, but they are successively identified by the ${\bf e}^i$ (which
are Kronecker delta-functions) brought down by the application of $\del^i$.
The $ \vecx_{r+1}\cdots\vecx_m$ commute to the left and (\ref{binomthma}),
(\ref{coprodxxx})  give the result. There is clearly a corresponding form of
Taylor's theorem for $\overleftarrow{\del}$ recovering the coaddition of
braided vectors.

Finally, we can apply the braided Taylor's theorem to $\exp$ itself and deduce
its usual
bicharacter properties
\ceqn{bicharexp}{ \exp(\veca+\vecx|\vecv)=\exp(\vecx|\vecv)\exp(\veca(\ |\
)\vecv)\\
\exp(\vecx|\vecv+\vecw)=\exp(\vecx(\ |\
)\vecw)\exp(\vecx|\vecv)}
where  $(\ |\ )$ denotes a space for $\exp(\vecx|\vecv)$ to be inserted in each
term of the
exponentials. The additions here are braided ones, i.e. we use the braid
statistics  $\vecx_1\veca_2=\veca_2\vecx_1R $ and
$\vecw_1\vecv_2=R\vecv_2\vecw_1$ from (\ref{xvustatb}).
In the usual covariant case $\exp$ is bosonic and we do not need to write the
$(\ |\ )$ when we work in the appropriate braided tensor product algebra.

These formulae are rather important in physics, where they are key properties
of addition of plane waves in position and momentum space. On the other hand in
categorical terms they just say that the product in the vector algebra
corresponds to the additive coproduct in the covector algebra and vice-versa
via the exponential. I.e. they are just the statement that our braided-Hopf
algebras are {\em  copaired}. We expressed duality in Section~5.3 in terms of
the evaluation map $\ev=\<\ ,\ \>$. Now we see how it looks equivalently in
terms of the
coevaluation map $\coev=\exp$. This is the line taken in \cite{KemMa:alg}.

\subsection{Braided Gaussians}

Next we turn to the Gaussian. We proceed in the same direct way by writing down
a differential equation that characterises it as a formal powerseries. The
simplest (but not the only) case for which this strategy works is when there is
given a covariant metric in the sense explained in Section~4.3. We concentrate
on this case for simplicity.  Assuming such a metric, we  define the
corresponding Gaussian $g_\eta$ to be a formal powerseries in $x_i$ such
that\cite{KemMa:alg}
\eqn{diffg}{ \del^i g_\eta=-x_a \eta^{ai}g_\eta,\quad \eps(g_\eta)=1.}
As before, we check that this is a good definition by checking that this
equation is integrable. We have
\align{&&\equad
(\del^i\del^j-R'{}^i{}_a{}^j{}_b\del^b\del^a)g_\eta=-\del^i(x_a\eta^{aj}
g_\eta)+R'{}^i{}_a{}^j{}_b\del^b(x_c\eta^{ca}g_\eta)\\
&&=(R'{}^i{}_a{}^j{}_b\eta^{ba}-\eta^{ij})g_\eta+ R'{}^i{}_a{}^j{}_b x_d
R^d{}_c{}^b{}_e \eta^{ca}\del^e g_\eta-x_cR^c{}_a{}^i{}_d\del^d
\eta^{aj}g_\eta\\
&&=(x_c R^c{}_a{}^i{}_d x_e \eta^{ed}\eta^{aj}-R'{}^i{}_a{}^j{}_b x_d
R^d{}_c{}^b{}_e x_f \eta^{fe}\eta^{ca})g_\eta\\
&&=(x_c R^c{}_a{}^i{}_d x_e \eta^{ed}\eta^{aj}-R'{}^i{}_a{}^j{}_b x_d
\lambda^{-2}R^{-1}{}^d{}_c{}^f{}_e \eta^{eb}\eta^{ca}x_f)g_\eta\\
&&=(x_c R^c{}_a{}^i{}_d x_e \eta^{ed}\eta^{aj}-\eta^{ai}\eta^{bj}
R'{}^c{}_a{}^e{}_b  x_d x_f \lambda^{-2}R^{-1}{}^d{}_c{}^f{}_e )g_\eta\\
&&=(x_c R^c{}_a{}^i{}_d x_e \eta^{ed}\eta^{aj}-\eta^{ai}\eta^{bj}\lambda^{-2}
R^{-1}{}^e{}_b{}^c{}_a  x_d x_f R'^f{}_e{}^d{}_c )g_\eta\\
&&=(x_c R^c{}_a{}^i{}_d x_e \eta^{ed}\eta^{aj}-\eta^{ai}\eta^{bj}\lambda^{-2}
R^{-1}{}^e{}_b{}^c{}_a  x_e x_c )g_\eta=0}
where we use the Gaussian equation, braided-Leibniz rule (\ref{braleibb}),
(\ref{etainvRd}), the
Gaussian equation again, (\ref{etainvRb}), (\ref{etainvRa}) and (\ref{covecc}).
After
that we use the relations among the $x_i$ and (\ref{etainvRb}) again to obtain
zero. While not a proof, this computation suggests that $g_\eta$ exists at
least as a formal powerseries.

The Gaussian can be found explicitly if we suppose that $\eta$ obeys some
additional conditions.
For example\cite{KemMa:alg}
\eqn{etaRe}{ \eta^{ka}R'^l{}_j{}^i{}_a=R'^{-1}{}^l{}_j{}^k{}_a\eta^{ai}}
ensures that $\vecx\cdot\vecx=x_ax_b\eta^{ba}$ is central in the braided tensor
product algebra. This is just $x_i x_a x_b\eta^{ba}=x_d x_c x_b
R'{}^c{}_i{}^d{}_a\eta^{ba}=x_d x_c x_b \eta^{ad}R'{}^{-1}{}^c{}_i{}^b{}_a=x_d
x_a\eta^{ad} x_i$ using our assumption and the relations in the braided
covector algebra. On the other hand, the condition \cite{KemMa:alg}
\eqn{etaRf}{R^i{}_a{}^j{}_b \eta^{ba} = q^{-2} \eta^{ij}}
ensures the operator identity $\del^i\hat{\vecx}\cdot
\hat{\vecx}=(1+q^{-2})\hat x_a\eta^{ai}+ \lambda^{-2} \hat\vecx\cdot \hat\vecx
\del^i$ as a consequence of the braided Leibniz rule (\ref{braleibb}) and the
quantum metric identity (\ref{etainvRc}). With these constraints on the
metric, we deduce at once that
\eqn{gauss}{ g_\eta= e_{\lambda^{-2}}^{-[2;q^{-2}]^{-1}\vecx\cdot \vecx}}
solves our differential equation. There is a similar conclusion when there are
factors in (\ref{etaRe}) and (\ref{etaRf}). Concrete examples include
$q$-Euclidean space Example~3.8 and $q$-Minkowski space Example~3.10. The
Euclidean case also makes contact with the treatment of Gaussians on
$SO_q(N)$-covariant quantum planes in \cite{CWSW:cov}\cite{Fio:sym}.

\subsection{Braided integration}

Next we use the braided-Gaussian to define translation-invariant integration.
This should be a map $\int$ from suitable functions of $\vecx$ to $\C$. One
might think that integration over $\R^n$ is one thing that cannot be done
algebraically since polynomials are not integrable. However, we can reduce it
to algebra if we only want integrals of the form $\int f(\vecx)g_\eta$ where
$f$ is a polynomial in the braided co-vector coordinates. Indeed, such
integrals are done by parts under the boundary assumption
\eqn{intparts}{ \int\del^i(f(\vecx)g_\eta)=0,\quad \forall f}
which expresses translation-invariance on our class of functions assuming that
the Gaussian `vanishes at infinity'.  As is well-known in physics, the ratios
of such integrals with $\int g_\eta$ are  well-defined and
algebraically-computable objects.

There is an R-matrix algorithm to do just this in \cite{KemMa:alg}. It can also
be used without a quantum metric, but we concentrate on the nicer case where
there is one, obeying the conditions in Section~4.3. If  $g_\eta$ and $\int$
exist obeying (\ref{diffg}) and (\ref{intparts}), then the {\em Gaussian
weighted integral}\cite{KemMa:alg}
\eqn{Zinta}{ \CZ[\vecx_1\cdots\vecx_m]=\left(\int \vecx_1\cdots\vecx_m g_\eta
\right)
\left(\int g_\eta\right)^{-1}}
is a well-defined linear functional on the braided covector algebra and can be
computed inductively by\cite{KemMa:alg}
\ceqn{Zintb}{\CZ[1]=1,\quad  \CZ[x_i]=0, \quad
\CZ[x_ix_j]=\eta_{ba}R^a{}_i{}^b{}_j\lambda^2\\
 \CZ[\vecx_1\cdots \vecx_m]=\sum_{i=0}^{m-2} \CZ[\vecx_1\cdots
\vecx_i\vecx_{i+3}\cdots\vecx_{m}]\CZ[\vecx_{i+1}\vecx_{i+2}](PR)_{i+2\,
i+3}\cdots(PR)_{m-1\, m}\lambda^{2(m-2-i)}.}
Indeed, assuming $\int,g_\eta$ exist we compute
\align{&&\equad \int x_{i_1}\cdots x_{i_m} g_\eta=-\int x_{i_1}\cdots
x_{i_{m-1}}\eta_{i_m a}\del^a g_\eta\\
&&=-\int x_{i_1}\cdots x_{i_{m-2}}\eta_{i_m
a}\left(-\widetilde{R}^b{}_{i_{m-1}}{}^a{}_b g_\eta
+\widetilde{R}^c{}_{i_{m-1}}{}^a{}_b\del^b (x_c g_\eta)\right)\\
&&=-\int x_{i_1}\cdots x_{i_{m-2}}\left(-\lambda^2
R^b{}_{i_{m-1}}{}^a{}_{i_m}\eta_{ab}g_\eta+\lambda^2R^c{}_{i_{m-1}}{}^a
{}_{i_m}\eta_{ab}\del^b (x_c g_\eta)\right)\\
&&=\left(\int x_{i_1}\cdots x_{i_{m-2}}g_\eta\right)\CZ[x_{i_{m-1}}x_{i_m}]
-\left(\int x_{i_1}\cdots x_{i_{m-2}}\eta_{a b}\del^b (x_c
g_\eta)\right)\lambda^2 R^c{}_{i_{m-1}}{}^a{}_{i_m}}
where we used the braided-Leibniz rule (in reverse) for the second equality and
(\ref{etaRa}) for the third. When $m=2$ there is no second term here as it is a
total derivative, and we obtain $\CZ[x_ix_j]$ as stated. When $m>2$ we
recognise the form shown in terms of $\CZ[x_{i_{m-1}}x_{i_m}]$ and an
expression similar to our second expression but with a two-lower power of
$\vecx$ to the left of $\del$. We now repeat the above process, each time
lowering the degree in the residual term by 2, until we reach
$\int\del=0$. This gives the iterative formula for $\CZ$. One can just take it
as a definition even when the Gaussian and $\int$ themselves are not known.

It is worth noting that the quantum metric needed in these constructions can be
recovered if one knows the norm element  $\vecx\cdot\vecx$ in the braided
covector algebra, by partial differentiation. For example, in the setting
(\ref{etaRe})--(\ref{etaRf}) we have the identity
\eqn{etadel}{ \eta^{ij}=\del^i\del^j (1+q^{-2})^{-1}\vecx\cdot\vecx.}
The same idea applies more generally to generate $R'$-symmetric tensors with
more indices\cite{Ma:eps}.

It is clear that we have the ingredients now to do most of the constructions of
classical scalar field theory in our general braided setting. Using the
integral and $\exp$ from Section~5.4 we can write down
a {\em braided Fourier transform}\cite{KemMa:alg}
\eqn{Fou}{ \CF(f)(\vecv)=\int f(\vecx)\exp(\vecx|\vecv)}
and prove reasonable theorems such as a convolution theorem and the theorem
that turns a differential operation in position space $x_i$ into algebraic
multiplication in momentum space $v^i$. See \cite{KemMa:alg} for details. The
inverse Fourier transform takes a similar form but integrating over momentum
space. Our constructions above have been symmetric under position and momentum
interchange modulo a left-right reversal. So in principle we can now write down
such things as the braided-Green function
\eqn{green}{  G(\vecx)=\CF^{-1}((\vecv\cdot\vecv-m^2)^{-1});\qquad
\vecv\cdot\vecv=v^av^b\eta_{ba}}
using the quantum metric. In practice we have to
introduce a Gaussian regulator before we can apply $\int$ in the form $\CZ$
constructed above to compute $\CF$ and $\CF^{-1}$. We also have to expand the
propagator here and any exponentials as powerseries in monomials. A closed
formula for the Green function remains to be computed, but in principle it is
now constructed by the above.

\subsection{Braided electromagnetism}

One of the nice things about braided geometry is that it handles both
$q$-deformed bosonic
constructions and $q$-deformed fermionic ones equally well. We have seen this
right
from the start in Examples~3.1--3.6. So we can apply our braided geometrical
constructions above to the $q$-deformed Grassmann co-ordinates $\theta_i=\extd
x_i$ just as well as to the $q$-deformed bosonic $x_i$ on which we have
concentrated so far. In order to do this, we now impose the further conditions
\eqn{covecd}{ R_{12}R'_{13}R'_{23}=R'_{23}R'_{13}R_{12},\quad
R_{23}R'_{13}R'_{12}=R'_{12}R'_{13}R_{23} }
\eqn{covece}{ R'_{12}R'_{13}R'_{23}=R'_{23}R'_{13}R'_{12}}
in addition to (\ref{QYBE})--(\ref{covecc}) in Section~3.1. This ensures that
there is a symmetry\cite{Ma:eps}
\eqn{RR'sym}{ R \leftrightarrow -R'}
in this combined system of equations and means that we are free to reverse
their roles. It means in particular that we can define the {\em braided group
of forms} $\Lambda(R',R)\equiv \Vhaj(-R,-R')$ with generators $1,\theta_i$ say,
relations
\eqn{formsa}{  \theta_i\theta_j=-\theta_b\theta_aR{}^a{}_i{}^b{}_j,\quad {\it
i.e.},\quad \theta_1\theta_2=-\theta_2\theta_1R}
and braided addition law whereby $\theta_i''=\theta_i+\theta_i'$ obey the same
relations provided we have the braid statistics
\eqn{formsb}{  \theta'_i\theta_j=-\theta_b\theta'_aR'{}^a{}_i{}^b{}_j,\quad
{\it i.e.},\quad \theta'_1\theta_2=-\theta_2\theta'_1R'.}
The roles of relations and statistics have been reversed relative to the
co-ordinates $x_i$ in Section~3.1, and there is also a minus sign in each case.
Likewise, we have the {\em braided group of coforms} $\Lambda^*(R',R)\equiv
V(-R,-R')$ with generators $1,\phi^i$ and relations corresponding to the vector
case with upper indices.
When it comes to questions of covariance, we assume that the quantum group
obtained from $A(R')$ coincides with that obtained from $A(R)$.

This line has been taken by the author in \cite{Ma:eps} and we outline now some
of its
results. Namely, differentiating $\del\over\del\theta_i$ in form-space gives
tensors which must be manifestly $-R$-symmetric i.e., $R$-antisymmetric. For
example, in nice cases the algebra of forms will have an element of top degree
given by $\theta_1\cdots\theta_n$.
We then define\cite{Ma:eps}
\eqn{epsu}{ \eps^{i_1 i_2\cdots i_n}={\del\over\del \theta_{i_1}}\cdots
{\del\over\del
\theta_{i_n}}\theta_1\cdots\theta_n=([n;-R']!)^{i_n \cdots i_1}_{12\cdots n}}
and by the reasoning above, it will be $R$-antisymmetric. Likewise, an
$R$-antisymmetric tensor with lower indices
can be obtained by applying any $m$-th order operator built from
$\del\over\del\theta_i$ to
$\theta_{i_1}\cdots\theta_{i_m}$. For example, we define\cite{Ma:eps}
\eqn{epsd}{\eps_{i_1\cdots i_n}={\del\over\del \theta_n}\cdots {\del\over\del
\theta_1} \theta_{i_1}\cdots\theta_{i_n}=([n;-R']!)_{i_1 \cdots i_n}^{12\cdots
n}.}
Its total $R$-antisymmetry is inherited this time from antisymmetry of the
$\theta_i$ coordinates in form-space.

\begin{example} The $q$-epsilon tensors on $\C_q^{2|0}$ are the well known
`spinor metric'
\[ \eps^{ij}=\pmatrix{0&-q^{-1}\cr 1&0},\quad \eps_{ij}=\pmatrix{0&1\cr
-q^{-1}&0 }\]
\end{example}
\proof We have
$\eps^{ij}=(1-PR')^j{}_1{}^i{}_2=\delta^j{}_1\delta^i{}_2-R'{}^i{}_1{}^j{}_2$
from (\ref{epsu}) and  $\eps_{ij}=\delta^1{}_i\delta^2{}_j-R'{}^2{}_i{}^1{}_j$
from (\ref{epsd}). We just put in $R'$ from Example~3.1. The epsilon tensor on
 $\C_q^{2|0}$ is of course well-known by hand\cite{FRT:lie}. On the other hand,
putting in a different $R'$ gives the braided epsilon tensor for any other
2-dimensional
example such as $\C_q^{1|1}$ etc., just as well. \endproof

Such epsilon tensors form the next layer of braided geometry. One can use them
in conjunction with a metric to define the Hodge $*$-operator
$\Lambda\to\Lambda$ along the usual lines\cite{Ma:eps}\cite{Mey:wav}
\eqn{hodge}{(\theta_{i_1}\cdots\theta_{i_m})^*=\eps^{a_1\cdots a_mb_n\cdots
b_{m+1}}\eta_{a_1i_1}\cdots\eta_{a_mi_m}\theta_{b_{m+1}}\cdots\theta_{b_n}.}
There are also concrete applications to quantum group theory, such as
a general R-matrix formula
\eqn{detR}{ \det(\vect)\propto \eps_{i_1\cdots i_n}t^{i_1}{}_{j_1}\cdots
t^{i_n}{}_{j_n}\eps^{j_n\cdots j_1}}
for a $q$-determinant in our background quantum group $A(R)$. Usually such
objects are introduced by hand on a case by case basis, see e.g.
\cite{FRT:lie}\cite{Fio:det}\cite{Mey:wav} among others; the
general formulae  need the above braided geometry.

Finally, we consider both the space coordinates
$x_i$ and the forms $\theta_i$ together with the interpretation $\theta_i=\extd
x_i$. Thus we
consider the {\em right-handed exterior algebra}
\eqn{Omega}{\Omega_R=\Lambda\und\tens
\Vhaj,\quad\vecx_1\vecx_2=R'\vecx_2\vecx_1,\quad
\theta_1\theta_2=-R\theta_2\theta_1,\quad \vecx_1\theta_2=\theta_2\vecx_1R}
where $\und\tens$ is the braided tensor product with braid statistics as shown.
We
then define the components $\Omega_R^p$ of form-degree $p$ and the exterior
derivative
\eqn{extd}{ \overleftarrow\extd: \Omega_R^p\to\Omega_R^{p+1},\quad
(\theta_{i_1}\cdots\theta_{i_p}
f(\vecx))\overleftarrow\extd=\theta_{i_1}\cdots\theta_{i_p} \theta_a
{\del\over\del x_a}
f(\vecx).}
In our conventions it obeys a right-handed $\Z_2$-graded-Leibniz
rule\cite{Ma:eps}
\eqn{extdleib}{ (fg)\overleftarrow\extd=f(g\overleftarrow\extd)+
(-1)^pf\overleftarrow\extd g,\quad\forall f\in\Omega,\ g\in\Omega^p.}
and gives us a differential complex when
\eqn{d2}{\overleftarrow\extd^2=0,\quad {\it i.e.}\quad
\theta_1\theta_2\del_2\del_1=0.}
The latter holds quite generally because $\del^i$ are symmetric and $\theta_i$
are antisymmetric. It is immediate when  $PR'=f(PR)$ for some function $f$ with
$f(-1)\ne 1$ and can be verified in other cases too according to the form of
$R$ and $R'$, see \cite{Ma:eps} for such examples. We can equally well define a
{\em left-handed exterior algebra}  $\Omega_L=\Vhaj\und\tens \Lambda$ as the
braided tensor product in the other order. It has exterior
derivative
\eqn{leftextd}{ \extd
\left(f(\vecx)\theta_{i_1}\cdots\theta_{i_m}\right)=f(\vecx)\overleftarrow
{\del}_i\theta_i\theta_{i_1}\cdots
\theta_{i_m},\quad \theta_1\vecx_2=\vecx_2\theta_1R}
which obeys the usual left-handed Leibniz rule for exterior differentials. The
proof of these properties for the right-handed exterior algebra is in
\cite{Ma:eps}. For the left-handed theory we just reflect the diagram-proofs
there
and reverse crossings. One can also verify the Leibniz properties using
properties of
the
braided integers along the lines in Section~5.2. Note that
$x_i\overleftarrow\extd=\theta_i=\extd x_i$ so the two exterior algebras
coincide on basic forms. The general $\Omega_L$ reduces to the rectangular
quantum matrix (\ref{ROmega}) in the Hecke case.

This is the construction of the exterior differential calculus on a quantum or
braided vector space coming out of braided geometry. The resulting R-matrix
formulae are just
\ceqn{dxdx}{\Omega_R:\quad \vecx_1\vecx_2=\vecx_2\vecx_1R',\quad
\extd\vecx_1\,\extd\vecx_2=-\extd\vecx_2\,\extd\vecx_1R,\quad
\vecx_1\extd\vecx_2=\extd\vecx_2\,\vecx_1 R\\
\Omega_L:\quad \vecx_1\vecx_2=\vecx_2\vecx_1R',\quad
\extd\vecx_1\,\extd\vecx_2=-\extd\vecx_2\,\extd\vecx_1R,\quad
\extd\vecx_1\,\vecx_2=\vecx_2\extd\vecx_1 R}
which are the  relations usually deduced by consistency
arguments\cite{WesZum:cov} within the axiomatic framework of
Woronowicz\cite{Wor:dif}. In the braided approach $\extd$ is explicitly
constructed as an operator from partial differentials, which in turn come from
the braided coaddition.

We have already covered the exterior algebra on the quantum plane in
Example~4.4. If we want the exterior algebra $\Omega(A(R))$ on a quantum
matrix with $R$ Hecke, we just use the braided covector form (\ref{tRR'})
explained in
Section~3.2, computing (\ref{dxdx}) with the multi-index $\bf R',R$. The cross
relations are just the same braid statistics as for coaddition so we can just
as
well write the exterior algebra (say the right-handed one) in the original
matrix form
\eqn{tOmega}{R\vect_1\vect_2=\vect_2\vect_1R,\quad \vect_1\extd
\vect_2=R_{21}\extd \vect_2\, \vect_1 R,\quad \extd
\vect_1\, \extd\vect_2=-R_{21}\extd\vect_2\, \extd\vect_1 R}
due to Sudbery\cite{Sud:alg}. The 1-form relations are again of the same
structure with a minus sign. This is the general story and we
see that our braided covector point of view recovers known formulae. The same
applies when we look at $\Omega(\bar A(R))$ and $\Omega(B(R))$ in the
braided-covector form
(\ref{xRR'}) and (\ref{uRR'b}) respectively and use the formulae (\ref{dxdx})
with the multi-index $\bf R',R$,
\eqn{dxdxR}{ \vecx_1\vecx_2=\vecx_2\vecx_1{\bf R'},\quad \extd\vecx_1\, \extd
\vecx_2=-\extd\vecx_2\, \extd\vecx_1{\bf R},\quad \vecx_1\extd \vecx_2=\extd
\vecx_2\, \vecx_1{\bf R}}
\eqn{duduR}{\vecu_1\vecu_2=\vecu_2\vecu_1{\bf R'},\quad \extd\vecu_1\, \extd
\vecu_2=-\extd\vecu_2\, \extd\vecu_1{\bf R},\quad \vecu_1\extd \vecu_2=\extd
\vecu_2\, \vecu_1{\bf R}}
These may obviously be rearranged as in (\ref{xRR'}) and (\ref{uRR'b}) to the
matrix form
\eqn{xOmega}{R_{21}\vecx_1\vecx_2=\vecx_2\vecx_1R,\quad
\extd\vecx_1\extd \vecx_2=-R\extd\vecx_2\extd\vecx_1 R,\quad \vecx_1\extd
\vecx_2=R\extd \vecx_2\, \vecx_1 R}
\eqn{uOmega}{\nquad\ R_{21}\vecu_1R\vecu_2=\vecu_2R_{21}\vecu_1R,\ \,
R^{-1}\extd\vecu_1R\extd\vecu_2=-\extd\vecu_2 R_{21}\extd\vecu_1 R,\ \,
R^{-1}\vecu_1R\extd\vecu_2=\extd\vecu_2 R_{21}\vecu_1 R.}
Some authors\cite{AzcRod:dif} prefer the second form for $\Omega(B(R))$ but let
us stress that
they are just the Wess-Zumino construction (\ref{dxdx}) in the matrix notation
and not mathematically new once the
required matrices $\bf R',R$ in (\ref{uRR'a}) had been introduced by the author
and U. Meyer in \cite{Ma:exa},\cite{Mey:new}. Some interesting new results
about $\Omega(B(R))$ from the point of view of braided groups are in
\cite{Vla:coa}\cite{Isa:int}.

\begin{example} The (right handed) quantum exterior algebra $\Omega(M_q(2))$ of
$M_q(2)$ as a 4-dimensional braided covector space is given by $\pmatrix{a&b\cr
c&d}$ as in Example~3.7 and $\pmatrix{\extd a&\extd b\cr \extd c&\extd d}$ with
 relations\cite{Mal:str}
\cmath{\extd a\extd a=0,\quad \extd b\extd b=0,\quad \extd c\extd c=0,\quad
\extd d\extd d=0\\
\extd a\extd b=-q\extd b\extd a,\quad \extd a\extd c=-q\extd c\extd a,\quad
\extd a\extd d=-\extd d\extd a,\quad \extd b\extd d=-q\extd b\extd b,\quad
\extd c\extd d=-q\extd d\extd c\\
\extd b\extd c=-\extd c\extd b+(q-q^{-1})\extd a\extd d}
and cross relations given by the braid statistics in Example~3.7. So
\[ a\extd a=q^2 \extd a\, a,\quad b\extd b=q^2\extd b\, b,\quad a\extd b=q\extd
b\,
a,\quad b\extd a=q\extd a\, b+(q^2-1) \extd b\, a,\quad {\it etc.}\]
\end{example}
\proof We do not need to make any fresh computations: the cross relations
between the 1-forms and co-ordinates are read off from the braid statistics for
coaddition displayed in Example~3.7. The relations among the 1-forms are of
just the same form with an extra minus sign, forcing a finite-dimensional
algebra as shown. So we recover a known example computed in \cite{Mal:str},
cf.\cite{Wor:dif}, now from our general braided approach. \endproof

We also have the $q$-epsilon tensor and the other constructions above on
$M_q(2)$ regarded as a braided covector space. Now we can put all our
constructions and ideas together to see that we have the basic formulae for the
theory of electromagnetism in our setting of general braided vector and
covector spaces. Actions such as
\eqn{YM}{ {\rm YM}=\int F(A)^*F(A),\quad {\rm CS}=\int A\extd A}
are now defined by the above as elements of $\C$.  Here $A=A^i(\vecx) \extd
x_i$ is the gauge potential treated as a 1-form and $F=\extd A$ is its
curvature. We use the Hodge $*$ operator from (\ref{hodge}) and the integral
from Section~5.6, understood  on an $n$-form to be  $\int$ computed on the
coefficient of the top form $\theta_1\cdots\theta_n$. We can of course also
write down the Maxwell equations $(\extd (F^*))^*=J$ etc, without trying to
compute the action itself. The case of these equations on $q$-Minkowski space
has
been studied recently in \cite{Mey:wav}. The latter also studied $q$-scaler
electrodynamics with some interesting results. A different (spinorial) approach
to wave equations in the $q$-Minkowski example is in \cite{Pil:def} and may
relate
to the above for spin 1. Braided geometry provides however, the only systematic
R-matrix
approach that works quite generally.

\section{Covariance}

We have mentioned from time to time that all our constructions are covariant
under a `background quantum group'. This is an automatic or inherent feature of
the whole braided approach, as we shall see in this section. To explain it
requires rather more familiarity with advanced aspects of quantum group
theory\cite[Sec. 3]{Ma:qua} which is why we have left it till the end.
Covariance is the reason that we can write the constructions of braided
geometry as braid diagrams (see Section~2) and is therefore rather a deep
feature. It ensures for example that the product map $\epsfbox{prodfrag.eps}$
can be pulled through a braid crossing, as explained in Section~2. Likewise
invariance of $\exp=\epsfbox{cap.eps}$ makes it bosonic so that we can treat it
like a free node:
\eqn{functex}{\epsfbox{functex.eps}\quad {\it etc.}}
all hold as a consequence of this hidden quantum group symmetry in our
constructions. This is the role of quantum groups in braided geometry.

The conditions (\ref{coveca}) in Section~3.1 essentially ensure such a
covariance for our braided covector and vector algebras $x_i$ and $v^i$.
The covariance is  like the transformation properties in Section~4.1 but
without a braiding between the
coacting quantum group and the vector or covector: the quantum group is treated
bosonically,
\eqn{xvcov}{x_i\mapsto x_a\tens t^a{}_i,\quad v^i\mapsto v^a\tens
St^i{}_a,\quad {\it i.e.},\quad
\vecx\to\vecx\vect,\quad[\vecx_1,\vect_2]=0,\quad \vecv\to\vect^{-1}\vecv,\quad
[\vecv_1,\vect_2]=0.}
Here $t^i{}_j$ is the quantum group generator obtained from the quantum matrix
$A(R)$. The transformation of $\vecv$ is a right coaction but written on the
left as a matrix action. (There are also left-covariant vectors which do not
need the antipode or `inversion' operation
and corresponding left-covariant covectors using it).  This formulation works
fine in practice although strictly speaking, to be sure of covariance from only
the conditions (\ref{coveca}) requires us to work with the full symmetry
quantum group of the system as obtained by Tannaka-Krein
reconstruction\cite{Ma:poi} from the R-matrix. In practice we usually already
know the quantum group which we are going to get by the reconstruction and the
generators $t^i{}_j$ by which we are going to describe it. In this case it is
easier to check the covariance directly
 by adopting the covariance condition
\eqn{R'tt}{ R'\vect_1\vect_2=\vect_1\vect_2R'}
in place of (\ref{coveca}). Then clearly we have
$\vecx_1\vect_1\vecx_2\vect_2=\vecx_1\vecx_2\vect_1\vect_2=\vecx_2\vecx_1
R' \vect_1\vect_2=\vecx_2\vecx_1\vect_2\vect_1R' =\vecx_2\vect_2\vecx_1
\vect_1R'$ so that the transformed $x_i$ obey the same relations. Similarly for
the $v^i$. The conditions (\ref{coveca}) are recovered automatically by
applying
the fundamental and conjugate-fundamental representations $\rho^\pm$ to
(\ref{R'tt}).

Once we have ensured covariance of our vector and covector algebras, the
covariance of the additive braid statistics and the braided coaddition is
ensured by the QYBE or braid relations (\ref{QYBE}). So all the structure maps
of our braided covector and vector spaces are covariant (intertwiners for the
quantum group action). Next, all our constructions based on diagrams will
remain covariant because we work in a braided category\cite{Ma:introp}. The
category is the category of objects covariant under our chosen quantum group,
so it is really inherent. Sometimes an algebra may be in more than one braided
category at the same time. Finally, when we add special data, such as
the quantum metric $\eta$ in Section~4.3, we will have to add conditions on it
to keep in our categorical setting. Again, the easiest way is to demand it
directly in
terms of $\vect$ as
\eqn{etacov}{\eta_{ab}t^a{}_it^b{}_j=\eta_{ij} ,\quad {\it or},\quad
t^i{}_at^j{}_b\eta^{ba}=\eta^{ji}}
which indeed imply (\ref{etaRb})--(\ref{etaRc}) etc. We will see this below.
Likewise, the differentiation and duality pairing in Section~5 are
automatically
covariant, as will be the integration and Hodge $*$ operator etc., when the
metric is covariant. If the exponential is defined abstractly as coevaluation,
it will also be covariant. But if we try to construct it by means of an ansatz
solving (\ref{diffexp}) we need also to ensure covariance by demanding
\eqn{expcov}{ \vect_1\cdots\vect_m F(m;R)=F(m;R)\vect_1\cdots\vect_m.}

The algebras $A(R)$, $\bar A(R)$ and $B(R)$ in Section~3 are examples of
braided covector algebras so they are covariant by the above under a quantum
group $t^I{}_J$ obtained form $A({\bf R})$. As usual, we can also write their
covariances in matrix terms. For $\bar A(R)$ this is the right
coaction\cite{Ma:euc}
\eqn{xcov}{ \vecx \mapsto \vecs^{-1}\vecx\vect,\quad
R\vecs_1\vecs_2=\vecs_2\vecs_1R,\ R\vect_1\vect_2=\vect_2\vect_1R,\
[\vecs_1,\vect_2]=[\vecs_1,\vecx_2]=[\vect_1,\vecx_2]=0}
where the quantum group is obtained from $A(R)\tens A(R)$ with generators
$\vecs,\vect$ respectively. This is easily checked\cite{Ma:euc} and is
connected with the multi-index braided-covector point of view by
\eqn{Lst}{t^I{}_J\mapsto S s^{j_0}{}_{i_0}\tens t^{i_1}{}_{j_1}}
and $\bf R$ from (\ref{xRR'}). The matrix form of covariance of $A(R)$ is
similar but more natural in a bicovariant (i.e. left- and right-covariant)
setting as in Section~4.2. That is why we prefer $\bar A(R)$ for a simpler
right-covariant (or left-covariant) theory. This
quantum symmetry becomes the $q$-orthogonal group in our $\bar M_q(2)$ example
in Section~7.1.

For $B(R)$ the covariance is under a more complicated quantum group obtained
from the double cross product $A(R)\dcross A(R)$ with generators $\vecs,\vect$
and transformation as before but the non-trivial cross
relations\cite{Ma:mor}\cite{CWSSW:lor}
\eqn{ucova}{ \vecu \mapsto \vecs^{-1}\vecu\vect,\quad
R\vecs_1\vecs_2=\vecs_2\vecs_1R,\ R\vect_1\vect_2=\vect_2\vect_1R,\
R\vect_1\vecs_2=\vecs_2\vect_1R,\ [\vecs_1,\vecu_2]=[\vect_1,\vecu_2]=0.}
The same formula (\ref{Lst}) connects it with the braided covector description
$\vect^I{}_J$ but with $\bf R$ now from (\ref{uRR'a}). This bigger symmetry
becomes the $q$-Lorentz group in the case of $BM_q(2)$ in Section~7.2. There is
also a homomorphism $A(R)\dcross A(R)\to A(R)$ given by multiplication, with
the result that $B(R)$ is covariant under \cite{Ma:exa}\cite{Ma:lin}
\eqn{ucovb}{u^i{}_j\mapsto u^a{}_b\tens (St^i{}_a)t^b{}_j,\quad{\it i.e.},\quad
\vecu \to \vect^{-1}\vecu\vect}
which puts our braided matrices $u^i{}_j$ into the same category as the
$x_i,v^i$ above. This is useful if we want to do braided linear algebra
involving all three objects as in Section~4.1. The additive coproduct
(\ref{uhopff}) and its braiding have the larger symmetry (\ref{ucova}) while
the diagonal case (\ref{ucovb}) is the symmetry of the multiplicative coproduct
(\ref{uhopfb}) and its braiding as introduced by the author in
\cite{Ma:exa}\cite{Ma:lin}.
Note that because we have covariance built in from the start in the concept of
braided groups, it is surprising to see the covariance of some of these
algebras presented sometimes subsequently as new results.

\subsection{Induced braiding}

Here we explain how the background quantum group symmetry mentioned above leads
to
the braid statistics and functoriality properties such as (\ref{functex}). We
will be brief because this is well covered in the author's original papers on
this topic in which braided groups
were introduced\cite{Ma:bra}\cite{Ma:bg}\cite{Ma:exa}\cite{Ma:lin}\cite{Ma:poi}
and also reviewed in \cite{Ma:introp}.

The main concept needed, introduced by the author in
\cite{Ma:pro}\cite{Ma:eul}\cite{Ma:bg}\cite{Ma:lin} and in an earlier form in
\cite[Sec.
3]{Ma:qua} is
that of a {\em dual-quasitriangular structure} or `universal R-matrix
functional' $\CR:A\tens A\to \C$ on a Hopf algebra or quantum
group $A$. It is characterised by axioms which are the dual of those of
Drinfeld\cite{Dri}, namely
\ceqn{dqua}{\CR(a\tens bc)=\CR(a\o\tens c)\CR(a\t\tens b),\quad \CR(ab\tens
c)=\CR(a\tens c\o)\CR(b\tens c\t)\\
b\o a\o\CR(a\t\tens b\t)=\CR(a\o\tens b\o)a\t b\t}
for all $a,b,c\in A$ and $\Delta a=a\o\tens a\t$ the coproduct. The main
theorem we need is the result due to the author in \cite[Sec. 3]{Ma:qua} that
the quantum matrices $A(R)$ indeed have such a universal R-matrix functional
such that
\eqn{R(tt)a}{\CR(t^i{}_j\tens t^k{}_l)=R^i{}_j{}^k{}_l,\quad {\it i.e.},\quad
\CR(\vect_1\tens\vect_2)=R}
on the matrix generators. This is where we need that $R$ obeys the QYBE, not to
ensure that $A(R)$ is a bialgebra as often mistakenly
written! The proof is in \cite[Sec. 3]{Ma:qua} as well as in the more modern
form
in \cite{Ma:introm} and Chapter~4 of my forthcoming book.

The reason we need this $\CR$ is that in this case any two vector spaces $V,W$
say on which the quantum group $A$ coacts have an intrinsic braiding
\cite{Ma:pro}\cite{Ma:eul}\cite{Ma:bg}\cite{Ma:lin}
\eqn{coactPsi}{\Psi_{V,W}(v\tens w)=w\uo\tens v\uo\,\CR(v\ut\tens w\ut)}
where $v\mapsto v\uo\tens v\ut$ denotes the coaction $V\to V\tens A$, etc. We
have not explained the term coaction very formally here, but
(\ref{xvcov})--(\ref{ucovb}) are typical examples; see \cite{Ma:qua} for more
details. Moreover, which is the {\em fundamental lemma for the theory of
braided groups}, if $B,C$ are two algebras on which the quantum group $A$
coacts covariantl (the coaction is an algebra homomorphism) then the algebra
$B\und\tens C$ defined with $\Psi$ as in (\ref{brahom}) is again an algebra on
which the quantum group coacts
covariantly\cite{Ma:bra}\cite{Ma:exa}\cite{Ma:bg}. This is the reason that the
braided tensor product construction $\und\tens$ was introduced (by the author).

We can apply these ideas directly to our braided covectors if we are content to
work with the quantum symmetry as $A(R)$ without necessarily an
antipode. Then (\ref{R(tt)a}) and (\ref{coactPsi}) immediately gives the
braiding in Section~3.1 and~4.1 between braided covectors. If we have an
inverse $\vect^{-1}$ then we can also do the braided vectors, braided matrices
and indeed all the braidings between them using the relevant coaction put into
(\ref{coactPsi}). For example,
\align{&&\equad\Psi(x_i\tens v^j)=v^b\tens x_a\CR(t^a{}_i\tens
St^j{}_b)=v^b\tens x_a \tilde{R}^a{}_i{}^j{}_b\\
&&\equad\Psi(u^i{}_j\tens x_k)=x_b\tens u^m{}_n\CR((St^i{}_m)t^n{}_j\tens
t^b{}_k)=x_b\tens u^m{}_n\CR^{-1}(t^i{}_m\tens t^b{}_c)\CR(t^n{}_j\tens
t^c{}_k)\\
&&\equad \Psi(u^i{}_j\tens u^k{}_l)=u^c{}_d\tens
u^a{}_b\CR((St^i{}_a)t^b{}_j\tens (St^k{}_c)t^d{}_l)\\
&&\qquad=u^c{}_d\tens u^a{}_b\CR(St^i{}_a\tens
(St^e{}_c)t^d{}_f)\CR(t^b{}_j\tens
(St^k{}_e)t^f{}_l)\\
&&\qquad=u^c{}_d\tens u^a{}_b\CR(St^i{}_m\tens St^e{}_c)\CR(St^m{}_a\tens
t^d{}_f)
\CR(t^n{}_j\tens (St^k{}_e)\CR(t^b{}_n\tens t^f{}_l)}
and so on. We use (\ref{dqua})--(\ref{R(tt)a}). The first of these ensures that
when there is an antipode, it obeys
\eqn{R(tt)b}{ \CR(S\vect_1\tens\vect_2)=R^{-1},\quad \CR(\vect_1\tens
S\vect_2)=\tilde R,}
which we also use. This is how the various mutual braidings (\ref{xvustata})
and their corresponding braid statistics (\ref{xvustatb}) were obtained in
\cite{Ma:lin}. We see that {\em quantum group covariance induces braid
statistics on whatever the quantum group acts on}.
This is the fundamental reason that braided geometry and not more conventional
 non-commutative geometry is the correct concept
of geometry for which quantum groups  are the generalised
symmetries\cite{Ma:bra}\cite{Ma:exa}. Note also that the same algebra could be
covariant under two different quantum groups, resulting possibly in two
different braid statistics: these are not therefore intrinsic
properties of the algebra (like fermions or bosons) but depend on which quantum
group covariance we are interested in. Some constructions may
be covariant under one quantum group using its induced statistics, and other
constructions under another. The addition and multiplication of braided
matrices $B(R)$ are a case in point.

There is one technicality which we have glossed over until now. Namely, while
(\ref{R(tt)a}) works in any normalisation, i.e. on any $A(R)$ with $R$
invertible, our later assumption that we can adjoin $\vect^{-1}$ may not. This
is because one typically obtains an antipode by quotienting by determinant-like
and other relations (such as cutting  $M_q(2)$ down to $SL_q(2)$ by the
constraint $ad-q^{-1}bc=1$). Such relations may well not be compatible with
$\CR$. For most R-matrices there is no problem provided $R$ is normalised
correctly (the {\em quantum group normalisation}) when defining $\CR$ in
(\ref{R(tt)a}). These are called {\em regular}\cite{Ma:lin}. All the R-matrices
the reader is likely to encounter are regular in this sense, although not
usually presented in the quantum group normalisation. In this paper we have
used either the braided covector normalisation needed for (\ref{covecb}) in
Section~3.1 or the Hecke normalisation needed for (\ref{q-Hecke}) in
Sections~3.2, 3.3 and~4.2. Therefore we suppose that $\lambda R$ is the quantum
group normalisation, where $\lambda$ is called the {\em quantum group
normalisation constant}\cite{Ma:lin}, and use this $\lambda R$ in
place of $R$ in (\ref{R(tt)a}), (\ref{R(tt)b}).

Keeping this in mind we can derive correctly the covariance identities for the
quantum metric, such as (\ref{etaRb})--(\ref{etaRc}) etc. in Section~4.3. For
example,
\[ \eta_{ab}\CR(t^i{}_c\tens t^b{}_l)\CR(t^c{}_j\tens
t^a{}_k)=\eta_{ab}\CR(t^i{}_j\tens t^a{}_kt^b{}_l)=\CR(t^i{}_j\tens
1)\eta_{kl}=\delta^i{}_j\eta_{kl}\]
derives (\ref{etaRb}) from the covariance condition (\ref{etacov}). Similarly
for the other quantum metric covariance conditions.

On the other hand, we really do not want a $\lambda$ in the braiding
$\Psi(x_i\tens x_j)$ between braided covectors in Section~3.1. Unless the
quantum group normalisation coincides with the braided covector normalisation
(which requires that $PR$ in the quantum group normalisation has an eigenvalue
-1) the covariance of our algebras does not induce the correctly normalised
braid statistics for coaddition. To induce the correctly normalised braid
statistics we have to extend our quantum group $A$ to the {\em  dilatonic
extension} $A[\varsigma]$
where we adjoin a new generator $\varsigma$ (the dilaton or dilatation element)
with\cite{Ma:poi}
\eqn{dila}{\Delta\varsigma=\varsigma\tens\varsigma,\quad \eps\varsigma=1,\quad
S\varsigma=\varsigma^{-1},\quad [A,\varsigma]=0,\quad \CR(\dila ^m\tens \dila
^n)=\lambda^{-mn}}
where we extend the $\CR$ as shown. For its value on mixed products we use
(\ref{dqua}). The need for a central element like $\dila$ has been realised by
hand in specific constructions (see below), but we see the need now at a
fundamental level to do with the braiding\cite{Ma:poi}. The dual
quasitriangular structure
$\CR$ for the dilaton is the new and crucial ingredient coming out of braided
geometry. We then let $\dila$ coact by scale
transformations. For example, we extend the coactions (\ref{xvcov}) of
$A=\<\vect\>$ to coactions of $A[\dila]=\<\vect,\dila\>$ acting by
\eqn{xvcovdila}{\vecx\to\vecx\vect\dila,\quad
\vecv\to\dila^{-1}\vect^{-1}\vecv}
where we include this scale transformation. This time the covariance under the
new extended quantum group induces the correctly normalised braid statistics
$\Psi(\vecx_1\tens\vecx_2)=\vecx_1\tens\vecx_2\CR(\vect_1 \dila \tens \vect_2
\dila )=\vecx_1\tens\vecx_2\CR(\dila \tens\vect_2\dila
)\CR(\vect_1\tens\vect_2\dila )=\vecx_1\tens\vecx_2\CR(\dila
\tens \dila )\CR(\dila \tens
\vect_1)\CR(\vect_1\tens \dila )$
$\CR(\vect_1\tens\vect_2)=\vecx_1\tens\vecx_2\lambda^{-1}\lambda
R =\vecx_1\tens\vecx_2R$ as explained in \cite{Ma:poi}. Likewise for the
vector-vector and mixed vector-covector braid
statistics. This is the correct derivation of
(\ref{xvustata})--(\ref{xvustatb}). The braid statistics involving $\vecu$ in
this group are not affected; one can consider that it has zero scale dimension
and transforms as before in (\ref{ucovb}) with zero power of $\dila$.

\subsection{Induced Poincar\'e group}

Most of the time in braided geometry we do not need to worry about the
background quantum group symmetry: it is there and makes sure that our
constructions are coherent in the way explained above. Sometimes however, the
covariance of the system gets into the algebra and we see $A[\dila]$ directly.
One of these situations is the construction of the $q$-Poincar\'e quantum group
in the approach of \cite{Ma:poi}. The idea is very simple: if we consider one
of our covector spaces $x_i$ as `space' then the symmetry (\ref{xvcovdila}) is
obviously some kind of `extended rotation'. But because all the braided
covector constructions are fully covariant under this, it is natural to make a
semidirect product by this coaction. The theorem in \cite{Ma:bos}\cite{Ma:poi}
is
that this semidirect product or `bosonisation' is always an ordinary (bosonic)
Hopf algebra and that its corepresentations are equivalent to the covariant
representations of the original $x_i$. The semidirect product $
A[\dila]\rbiprod\Vhaj$ is then an `extended Poincar\'e or Euclidean group'. We
call it the induced Poincar\'e group if the braided-covector algebra is
regarded as spacetime. The best proof is diagrammatic\cite{Ma:bos} and works
for {\em any} braided group $B$ where the braiding is induced by some quantum
group covariance: just make a semidirect product by that quantum symmetry. So
this is a very general phenomenon and not an ad-hoc construction.

The idea for the construction comes from the Jordan-Wigner transformation in
physics where the
super-representations of a fermionic or graded system are equivalent to the
ordinary representations of a new algebra obtained by adjoining the degree
operator. The grading corresponds in our language to a background quantum group
$\Z_2'$ under
which everything is covariant. It has a generator $\dila$ with $\dila^2=1$ and
a non-trivial dual quasitriangular structure $\CR$. This `hidden quantum group
of supersymmetry' is what generates the $\pm1$ Bose-Fermi statistics we
encounter, using exactly the formalism above. We explained this in detail in
\cite{Ma:bos}\cite{MacMa:spe} in a form where $\Z_2'$ was quasitriangular
rather than dual-quasitriangular, the two being equivalent. See also the
Appendix below.

The same mathematical ideas apply just as well in our geometrical example where
the `hidden quantum group' is much more like a $q$-deformation of a rotation
group than a discrete one: the mathematical formalism of Hopf algebras {\em
unifies} two quite different concepts in physics. One is the concept of
supersymmetry where everything is $\Z_2$-graded and the other is the concept of
Lorentz transformation or rotational invariance where everything is covariant
under a
background linear co-ordinate transformation. One needs the concept of quantum
groups to do this (ordinary groups will not do) but once we have that concept,
these two physically different ideas are just extremes of the same phenomenon.
{\em This unification of the concept of Lorentz transformation or rotation (or
other quantum group) covariance and supersymmetry is one of the deepest reasons
to be
interested in quantum groups in physics}.

We content ourselves here with the formulae from \cite{Ma:poi} for how these
constructions look in practice in R-matrix form. We use the braided covector
algebra
of Section~3.1 but call the generators momentum $p_i$. It lives in the category
of $A[\dila]$-covariant systems as explained in Section~6.1, where $A[\dila]$
has generators $t^i{}_j,\dila$. The constant $\lambda$ is such that $\lambda R$
is in the quantum group normalisation. Then the induced Poincar\'e quantum
group has cross relations and (ordinary) Hopf algebra structure\cite{Ma:poi}
\ceqn{poinc}{ \vecp \dila =\lambda^{-1}\dila \vecp,\quad
\vecp_1\vect_2=\lambda\vect_2\vecp_1R ,\quad \Delta \vect=\vect\tens\vect, \
\Delta \dila =\dila \tens \dila \\
 \Delta \vecp=\vecp\tens \vect \dila +1\tens\vecp,\quad \eps\vect=\id,\quad\eps
\dila =1,\quad \eps\vecp=0\\
S\vect=\vect^{-1},\quad S\dila =\dila ^{-1},\quad S\vecp=-\vecp \dila
^{-1}\vect^{-1}. }
Some examples of such type were first considered by hand in \cite{SWW:inh}, but
the braided approach in \cite{Ma:poi} provided the first systematic
construction for general $R,R'$. Moreover, which was entirely missing before
\cite{Ma:poi}, it provided for a covariant action of this R-Poincar\'e group on
another copy $x_i$ of the braided covectors, regarded this time as space,
by\cite{Ma:poi}
\eqn{poinccov}{x_i\mapsto x_a\tens t^a{}_i\dila+p_i,\quad {\it i.e.},\quad
\vecx\mapsto \vecx\vect \dila  + \vecp.}
This is ensured by the general braided group theory of bosonisation in
\cite{Ma:bos} which says that braided-covariant systems under any braided group
become automatically ordinary covariant systems under its bosonisation. We have
seen in Section~5.1 that the braided covectors $x_i$ indeed coact covariantly
on
themselves by the braided coaddition or finite braided translation. Hence by
this theorem the induced Poincar\'e coacts too. In keeping with the theme of
this paper, we can give a direct proof too:
\align{(\vecx_1\vect_1 \dila +\vecp_1)(\vecx_2\vect_2
\dila +\vecp_2)\equad&&=\vecx_1\vecx_2\vect_1\vect_2
\dila ^2+\vecp_1\vecp_2+\vecp_1\vecx_2\vect_2 \dila +\vecx_1\vect_1 \dila
\vecp_2\\
(\vecx_2\vect_2 \dila +\vecp_2)(\vecx_1\vect_1
\dila +\vecp_1)R'\equad&&=\vecx_2\vecx_1\vect_2\vect_1R'
\dila ^2+\vecp_2\vecp_1R'+\vecx_2\vect_2 \dila \vecp_1 R'
+\vecp_2\vecx_1\vect_1
\dila R'\\
&&=\vecx_2\vecx_1R'\vect_1\vect_2 \dila ^2+\vecp_2\vecp_1R'+ \vecx_1\vect_1
\dila \vecp_2+\vecx_2\vect_2 \dila \vecp_1R}
where we use covariance (\ref{R'tt}) for the first term and the cross-relations
(\ref{poinc}), the condition
(\ref{covecc}) and the action of the usual permutation $P$ for the last two
terms. Using again the
cross-relations (\ref{poinc}) and the algebra relations of the $\vecx,\vecp$ we
see that the two expressions coincide, i.e.
the transformed $\vecx\vect \dila +\vecp$ obey the same braided covector
algebra.

\begin{example} The bosonisation of $\C_q^{2|0}=\{x,y\}$ is the ordinary Hopf
algebra $GL_q(2) \rbiprod \C_q^{2|0}$ with relations
\[ \dila ^2=ad-q^{-1}bc,\quad x\dila =q^{3\over 2} \dila x,\quad y\dila
=q^{3\over 2} \dila y\]
\[x\pmatrix{a&b\cr c&d}=\pmatrix{q^2a&qb\cr q^2c&qd}x,\quad y\pmatrix{a&b\cr
c&d}=\pmatrix{qay+(q^2-1)bx&q^2by\cr qcy+(q^2-1)dx&q^2dy}\]
where $\pmatrix{a&b\cr c&d}$ have the relations of $M_q(2)$ with
$(ad-q^{-1}bc)^{\pm \h}$
adjoined. The coproduct is
\[ \Delta x=x\tens a+y\tens c+1\tens x,\quad \Delta y=x\tens b+ y\tens d+1\tens
y\]
and the matrix coproduct for the $GL_q(2)$ generators.
\end{example}
\proof We consider $\C_q^{2|0}$ a braided covector algebra with the standard
$R$ as in Example~3.1. Then $\lambda=q^{-{3\over 2}}$ takes us to the quantum
group normalisation. The quantum group $SL_q(2)[\dila]$ consists of $M_q(2)$
generated by $t^i{}_j$ as in Example~3.7 with a $q$-determinant relation added
(so $SL_q(2)$ as usual) and the  element $\dila $ adjoined as in (\ref{dila}).
It is convenient to the redefine generators $\pmatrix{a&b\cr c&d}=\vect \dila
$. They have the same commutation relations as the quantum matrices $M_q(2)$
but instead of $\det_q(\vect)=1$ we have the relation $ad-q^{-1}bc=\dila ^2$.
So we can identify $SL_q(2)[\dila]$ as a version of
$GL_q(2)$ consisting of $M_q(2)$ and $(ad-q^{-1}bc)^{\pm \h}$ adjoined. The
commutation relations are then obtained from (\ref{poinc}) using the explicit
form of $R$. The coaction and corresponding coproduct are given by
$\pmatrix{a&b\cr c&d}$ acting on $(x,y)$ from the right as a matrix. \endproof

This is a more or less well-known example and in the general family in
\cite{SWW:inh} and elsewhere,
but demonstrates the kind of structure one gets. From \cite{Ma:poi} we know
that it necessarily
coacts covariantly on $\C_q^{2|0}$ by matrix multiplication and translation
(\ref{poinccov}).
Moreover, the
braided approach allows us to bosonise all the other braided covector algebras
in
this paper just as well and obtain their natural induced `Poincar\'e' quantum
groups and their coactions.

For example, when $R$ is Hecke, we can apply the bosonisation to the additive
braided groups $A(R),\bar A(R)$ and $B(R)$ just as well. We just use $\bf R,R'$
from (\ref{xRR'}), (\ref{tRR'}), (\ref{uRR'a}) respectively. The Poincare
groups consist
of adjoining the Lorentz generators $t^I{}_J$ obtained from $A(\bf R)$. We can
also give a
spinorial or matrix version using the quantum symmetry in the matrix form, as
given for
$\bar A(R)$ and $B(R)$ in (\ref{xcov}) and (\ref{ucova}) respectively. In these
cases we chose the
quantum group normalisation constant $\lambda$ such that $A(R)$ has a
quotient dual-quasitriangular Hopf algebra $A$ and from two copies of this we
build $A\tens A$ as the quantum covariance of $\bar A(R)$ and $A\dcross A$ as
the
quantum covariance of $B(R)$. We use for the latter the $\dcross$ construction
introduced by the
author and making use of the dual universal R-matrix $\CR$ of $A$, see
\cite[Sec. 4]{Ma:poi}. In both cases we then adjoin $\dila$.

The R-matrix form of the induced Poincar\'e quantum group $(A\tens
A)[\dila]\rbiprod \bar A(R)$ then has
cross relations and coproduct \cite{Ma:euc}
\ceqn{eucpoi}{ \vecp_1\vect_2=\lambda \vect_2\vecp_1R,\quad
\vecp_1\vecs_2=\vecs_2\lambda^{-1} R^{-1}\vecp_1,\quad \vecp\dila
=\lambda^{-2}\dila \vecp\\
\Delta p^i{}_j=p^a{}_b\tens (Ss^i{}_a)t^b{}_j\dila+1\tens p^i{}_j,\quad {\it
i.e.},\quad \Delta\vecp=\vecp\tens \vecs^{-1}(\ )\vect\dila +1\tens\vecp}
where the indices of $\vecp$ have to be inserted into the space. There is also
a counit $\eps \vecp=0$ and antipode. The standard R-matrix put into these
formulae provides the 4-dimensional $q$-Euclidean space Poincar\'e group
appropriate
to the $q$-Euclidean space algebra in Section~7.1 below.

The R-matrix form of the induced Poincar\'e group $(A\dcross A)[\dila]\rbiprod
B(R)$
has the cross-relations and coproduct\cite{MaMey:bra}
\ceqn{minkpoi}{\vecp_1\vect_2=\lambda^2\vect_2R_{21}\vecp_1 R,\quad
\vecp_1\vecs_2=\vecs_2 R^{-1}\vecp_1 R,\quad \vecp \dila =\lambda^{-2}\dila
\vecp\\
\Delta \vecp=\vecp\tens\vecs^{-1}(\ )\vect \dila +1\tens\vecp}
in a similar way. There is also a counit $\eps \vecp=0$ and antipode. The
standard R-matrix inserted into these formulae gives us the $q$-Minkowski space
Poincar\'e group appropriate to the $q$-Minkowski algebra in Section~7.2. Of
course, our
constructions (\ref{eucpoi})--(\ref{minkpoi}) are quite general and not limited
to the standard R-matrix.

We conclude here by mentioning that also in \cite{Ma:poi} was given the
corresponding Poincar\'e enveloping algebra-type quantum group for the general
construction (\ref{poinc}). We give it in terms of the $\vecl^\pm$ generators
of $U_q(\cg)$ in \cite{FRT:lie}. One can also use it for non-standard quantum
enveloping algebras when the universal R-matrix is known. This enveloping
algebra is our `Lorentz or rotation' algebra and all our constructions are
covariant under its
action (rather than coaction as before). First we extend $U_q(\cg)$ by
adjoining an infinitesimal scaling central generator $\xi$ dual to the finite
dilaton $\dila$,
\eqn{envdila}{ \Delta\xi=\xi\tens 1+1\tens\xi,\quad
\CR_\xi=\lambda^{-\xi\tens\xi},\quad  \<\dila ,\xi\>=1,\quad{\rm or}\quad
\<\dila ,\lambda^\xi\>=\lambda}
We use its {\em quantum line} universal R-matrix\cite{Ma:csta}. The universal
R-matrix of the extension of $U_q(\cg)$ is its  usual
one\cite{Dri}\cite{Jim:dif} times this $\CR_\xi$. We focus on braided vectors
$p^i$ (say) as momentum generators covariant under this extension. Bosonisation
proceeds using the general theorem in \cite{Ma:bos} and gives the cross
relations and Hopf structure \cite{Ma:poi}
\ceqn{envpoi}{ \lambda^\xi\vecp\lambda^{-\xi}=\lambda^{-1}\vecp,\quad \vecl_1^+
\vecp_2=\lambda^{-1}R_{21}^{-1}\vecp_2\vecl^+_1,\quad \vecl_1^-\vecp_2=\lambda
R\vecp_2\vecl^-_1\\
 \Delta\vecp=\vecp\tens 1+\vecl^-\lambda^\xi\tens
\vecp,\quad \eps \vecp=0,\quad S\vecp=-\lambda^{-\xi} S\vecl^-\vecp}
and a covariant action on braided covectors $x_i$ by
\eqn{envpoiact}{ \vecl^+_2\la\vecx_1=\vecx_1\lambda R,\quad
\vecl^-_2\la\vecx_1=\vecx_1\lambda^{-1}R_{21}^{-1},\quad
\lambda^\xi\la\vecx=\lambda^{-1}\vecx,\quad p^i\la x_j=\delta^i{}_j.}
The action of $p^i$ on products is according to the coproduct and is equivalent
to a version (in some other conventions) of the braided-Leibniz rule
(\ref{braleiba}) for
braided derivatives. We can just as easily give the enveloping algebra form of
the matrix Poincar\'e groups (\ref{eucpoi})--(\ref{minkpoi}) by the same
techniques. See \cite{Ma:euc} for the action of the $\vecl^\pm$ in these cases.
We note that
there are many other
applications too of the bosonisation theory. See
\cite{SchZum:bra}\cite{Dra:bra}
for recent applications in connection with the differential calculus and Koszul
complex of $U_q(\cg)$.

\section{$q$-Deformed spacetime}

At various points in our tour of braided geometry we have mentioned that the
examples $\bar M_q(2)$ and $BM_q(2)$ make natural $q$-Euclidean and
$q$-Minkowski spaces when $q$ is real. We just have to specialise our general
$\bar A(R)$ and
$B(R)$ constructions to the case of $R$ given by the standard quantum plane or
Jones polynomial
R-matrix in the Hecke normalisation. We have seen above many results about
these general
algebras, all of which constitutes the {\em braided approach} to $q$-spacetime
due to the
author\cite{Ma:exa}\cite{Ma:lin}\cite{Ma:mec}\cite{Ma:poi}\cite{Ma:euc} and U.
Meyer\cite{Mey:new}\cite{MaMey:bra}. Among other general results, there is a
theory of
{\em quantum Wick rotation} introduced in \cite{Ma:euc} which strictly connects
$\bar A(R)$
and $B(R)$ by an algebraic twisting construction. There is also a purely
quantum isomorphism $B(R)\isom U(\CL)$ where
$\CL$ is a braided-Lie algebra and $U(\CL)$ its braided enveloping
algebra\cite{Ma:lie}. These are among the more advanced results of braided
geometry which were not able to cover here.

Many authors have considered what algebras should be taken as the
$q$-deformation of Euclidean and
Minkowski spaces. In the Euclidean case the braided geometry approach based on
$\bar M_q(2)$ is compatible with
the $N=4$ case of the theory of $SO_q(N)$-covariant quantum planes in
\cite{FRT:lie}\cite{CWSW:cov}\cite{Fio:sym}\cite{HebWei:fre}\cite{Wei:mec} and
elsewhere. It has been taken quite far by Fiore, Weich and others. In the
Minkowski case our approach based on the braided matrices $BM_q(2)$ is
compatible with the approach of Carow-Watamura, Schlieker, Scholl, Watamura,
Wess, Zumino and others, based on the idea of $q$-Minkowski space as a tensor
product of two quantum planes, see
\cite{CWSSW:lor}\cite{CWSSW:ten}\cite{OSWZ:def} and elsewhere. The mathematical
basis for this overlap is the homomorphism (\ref{rank1})--(\ref{bm2dec})
introduced by the author in \cite{Ma:lin}, though its significance was not
appreciated until somewhat later when the two approaches began to
converge \cite{Ma:poi}. The final details for this convergence were worked out
in \cite[Sec. 3]{MaMey:bra} using the dual-universal
R-matrix functional $\CR$ on the $q$-Lorentz group.

In the following two sections we explain cleanly and simply how some of our
various braided group constructions  look
in the case of $\bar M_q(2)$ and $BM_q(2)$. Some comments about which were
known before for these particular algebras and which were  obtained as far as I
know for the first time as examples of the braided approach are collected at
the end of each section.

\subsection{$q$-Euclidean space}

The algebra $\bar M_q(2)$
\ceqn{eucalg}{ba=qab,\quad ca=q^{-1}ac,\quad da=ad,\quad db=q^{-1}bd\quad
dc=qcd\\
bc=cb+(q-q^{-1})ad}
was already computed in Example~3.8. The covariance (\ref{xcov}) is
$SU_q(2)\tens SU_q(2)$  regarded as the spinorial version of $SO_q(4)$.
We can also work with $O_q(4)$ obtained from $\bf R$ but here we concentrate on
the spinorial setting.

Next, the element $ad-qcb$ is central in the algebra and invariant under
$SU_q(2)\tens SU_q(2)$. Hence we take it as a natural radius function. The
corresponding metric by differentiation (\ref{etadel}) is
\eqn{eucmetric}{\eta^{IJ}=\pmatrix{0&0&0&1\cr 0&0&-q&0\cr 0&-q^{-1}&0&0\cr
1&0&0&0}={\del\over\del x_I}{\del\over\del x_J}(ad-qcb).}
It obeys all the conditions in Section~4.3 with $\lambda=q^{-\h}$, and the
conditions (\ref{etaRe})--(\ref{etaRf}) as well. The Gaussian has the simple
form (\ref{gauss}).

Next, there is a natural $*$-structure
\eqn{euc*}{\pmatrix{a^*&b^*\cr c^*&d^*}=\pmatrix{d&-q^{-1}c\cr -qb&a}}
on this algebra, making it into a $*$-braided group as explained in
Section~4.4. It corresponds to the metric under the multi-index version of
(\ref{x*}). We can therefore chose new variables
\eqn{euctxyz}{t={a-d\over 2\imath},\quad x={c-qb\over 2}, \quad y={c+qb\over
2\imath},\quad z={a+d\over 2}}
which are self-adjoint in the sense $t^*=t,x^*=x,y^*=y,z^*=z$. They are the
physical spacetime
coordinates and
\eqn{euclmetric}{ ad-qcb=({1+q^2\over
2})t^2+x^2+y^2+({1+q^2\over 2})z^2}
so that the signature of the metric is the Euclidean one. This justifies our
defining $\R_q^4\equiv \bar M_q(2)$, the algebra of `co-ordinate functions' on
$q$-Euclidean
space.

After this, we can routinely apply the rest of the $\bar A(R)$ constructions of
braided
geometry with this interpretation. The braided vector algebra of derivatives
from (\ref{barA(R)vec}) is
\cmath{ {\del\over\del d} {\del\over\del b}=q^{-1}  {\del\over\del b}
{\del\over\del
d}, \quad {\del\over\del d} {\del\over\del c} = {\del\over\del c}
{\del\over\del d} q,\quad
{\del\over\del d} {\del\over\del a} = {\del\over\del a} {\del\over\del d}\\
{\del\over\del b} {\del\over\del c}= {\del\over\del c} {\del\over\del b} +
(q-q^{-1} ) {\del\over\del a} {\del\over\del d},\quad
{\del\over\del b} {\del\over\del a} = {\del\over\del a} {\del\over\del b}
q,\quad {\del\over\del c} {\del\over\del a} =q^{-1}   {\del\over\del a}
{\del\over\del c}.}
Their Leibniz rule (\ref{braleibb}) is read off from the braiding listed in
Example~3.8.

The algebra of forms from (\ref{dxdxR}) is
\cmath{ \extd a \extd a=0,\quad \extd b\extd b=0,\quad \extd c\extd c=0,\quad
\extd d\extd d=0\\
 \extd b  \extd a =-q^{-1}   \extd a  \extd b ,\quad \extd c  \extd a  =- \extd
a  \extd c  q,\quad
\extd d   \extd b  =- \extd b  \extd d   q\\
 \extd c  \extd b  =- \extd b  \extd c ,\quad \extd d   \extd c =-q^{-1}
\extd c  \extd d  ,\quad
\extd d   \extd a =- (q-q^{-1} ) \extd b  \extd c  - \extd a  \extd d }
or deduced immediately from the braid statistics in Example~3.8 with an extra -
sign. These relations generate a finite-dimensional algebra with the dimension
at each degree being the same as in the classical case. The exterior algebra
$\Omega(\R_q^4)$ is generated by these forms and the co-ordinates, with
cross-relations again being read off from the additive braid statistics in
Example~3.8 as explained in Section~5.7. So we do not list it again.

The $q$-epsilon tensor from (\ref{epsd}) is
\cmath{\eps_{abcd}=-\eps_{acbd}=\eps_{adbc}=-\eps_{adcb}
=\eps_{bcad}=-\eps_{bcda}=1\\
-\eps_{cbad}=\eps_{cbda}=-\eps_{dabc}=\eps_{dacb}
=-\eps_{dbca}=\eps_{dcba}=1\\
 \eps_{acdb}=-\eps_{cdba}=-\eps_{dcab}=q,\quad
-\eps_{abdc}=-\eps_{bacd}=\eps_{bdca}=\eps_{dbac}=q^{-1}\\
-\eps_{cadb}=\eps_{cdab}=q^2,\quad  \eps_{badc}=-\eps_{bdac}=q^{-2},\quad
-\eps_{adad}=\eps_{dada}=q-q^{-1}.}
We see that there are a few unexpected entries which are zero at $q=1$.

Finally, the Gaussian-weighted integral on degree two from Section~5.6 is
$\CZ[x_Ix_J]=q^{-4}\eta_{IJ}$. Its nonzero values on degree four are
\cmath{ \CZ[abcd]=-q^{-11},\quad \CZ[acbd]=-q^{-7}\\
\CZ[a^2d^2]=q^2\CZ[b^2c^2]=q^{-2}\CZ[c^2b^2]=\CZ[bc^2b]=\CZ[cb^2c]
=q^{-10}(q^{2}+1)}
along with the other cases implied by the relations in $\R_q^4$. In terms of
the spacetime coordinates we have
\cmath{ \CZ[t^2]=\CZ[z^2]=\h q^{-4},\quad
\CZ[x^2]=\CZ[y^2]={[2;q^2]\over 4}q^{-4}\\
\CZ[t^4]=\CZ[z^4]={3[2;q^2]\over8} q^{-10},\quad
\CZ[x^4]=\CZ[y^4]={[3;q^2]!\over 8} q^{-10}}
which shows a degree of spherical symmetry even in the noncommutative case.

Our approach to $q$-Euclidean space is a useful R-matrix one introduced by the
author in \cite{Ma:euc} based on the formulae
$R_{21}\vecx_1\vecx_2=\vecx_2\vecx_1R$ etc., and making contact in the above
specific example with the $N=4$ case of the approach based on
$SO_q(N)$-covariance.
The general $N$ could also be studied in our braided formalism as
braided covector spaces. The differentials, epsilon tensor, Gaussian and
integration in this family were known already
\cite{CWSW:cov}\cite{Fio:det}\cite{Fio:sym}\cite{HebWei:fre} while our braided
approach adds such things as the braided addition law, the R-matrix form
(\ref{eucpoi}) of the Poincar\'e group\cite{Ma:euc}, its covariant action on
the $\vecx$, existence of the
braided-exponential and several other results, including the quantum wick
rotation by twisting\cite{Ma:euc} which systematically turns the above into the
following results for $q$-Minkowski space. The specific relations
(\ref{eucalg}) are in fact
isomorphic in this example with usual quantum matrices $M_q(2)$ which were
already anticipated as $q$-Euclidean space in \cite[Appendix]{CWSSW:ten}, with
$SU_q(2)\tens SU_q(2)$ as rotations.

\subsection{$q$-Minkowski space}

Finally we follow the same constructions as in the preceding section but now
for the braided matrices $B(R)$. It is ironic that the simplest example
$BM_q(2)$ was the very first braided group known\cite{Ma:exa} but also turns
out to be the most intersting for $q$-deformed physics.
We have already given the algebra in Example~3.10 as
\ceqn{minkalg}{ba=q^2ab,\quad ca=q^{-2}ac,\quad d a=ad,\qquad
bc=cb+(1-q^{-2})a(d-a)\\
d b=bd+(1-q^{-2})ab,\quad cd=d c+(1-q^{-2})ca.}
It has the covariance $SU_q(2)\dcross SU_q(2)$ from (\ref{ucova}). Its natural
$*$-structure is not the usual tensor product one but involves also reversing
the two $SU_q(2)$ factors. So these two $SU_q(2)$ are like the $SU_q(2)$ which
physicists use when they work with $SO_q(1,3)$
at the level of a complex Lie algebra.

This algebra has a nice braided-coproduct too corresponding to matrix
multiplication as explained in Section~3.3 and this at once picks out
the braided determinant $\und\det(\vecu)=ad-q^2cb$ as a natural radius
function. It is group-like, central and bosonic with respect to the
multiplicative braid statistics as explained in  (\ref{bdeta})--(\ref{bdetc}).
We obtain the quantum metric from (\ref{etadel}) as
\eqn{minkmetric}{\eta^{IJ}={\del\over\del u_I}{\del\over\del
u_J}(ad-q^2cb)=\pmatrix{q^{-2}-1&0&0&1\cr 0&0&-q^2&0\cr 0&-1&0&0\cr
1&0&0&0}.}
The conditions in Section~4.3 hold with $\lambda=q^{-\h}$, as well as
(\ref{etaRe})--(\ref{etaRf}) so the Gaussian has the simple form (\ref{gauss}).

There is a natural $*$-structure for real $q$ which is the Hermitian one
\[ \pmatrix{a^*&b^*\cr c^*&d^*}=\pmatrix{a&c\cr b&d}\]
giving a $*$-braided group under both braided addition and multiplication. This
is from the general construction \cite{Ma:mec}\cite{Ma:star} recalled in
Section~4.4. This leads to a change of variables
\eqn{minktxyz}{t={qd+q^{-1}a\over 2},\quad x={b+c\over 2},\quad y={b-c\over
2i},\quad
z={d-a\over 2}}
which are self-adjoint in the sense $t^*=t$ etc, and the metric becomes
\eqn{minkdet}{ ad-q^2cb={4q^2\over(q^2+1)^2}t^2-{q^2}x^2-{q^2} y^2-
{2(q^4+1)q^2\over (q^2+1)^2}z^2+{2q}\left({q^2-1\over q^2+1}\right)^2
tz.}
This has the Minkowski signature when $q=1$ and justifies the definition
$\R_q^{1,3}\equiv BM_q(2)$. The natural time direction
$t$ is the quantum trace element in Example~3.10 and is central and bosonic
with respect to the multiplicative braid statistics.

The rest of the structure is routinely computed from the R-matrix formulae for
$B(R)$.
The braided vector algebra of differentiation operators from (\ref{B(R)vec}) is
\cmath{ {\del\over\del d}{\del\over\del b}=q^{-2}{\del\over\del
b}{\del\over\del
d},\quad
{\del\over\del d}{\del\over\del c}= {\del\over\del c}{\del\over\del d}q^2,\quad
{\del\over\del d}{\del\over\del a}= {\del\over\del a}{\del\over\del d}\\
{\del\over\del c}{\del\over\del a}={\del\over\del a}{\del\over\del
c}+{\del\over\del c}{\del\over\del d}(q^2-1),\quad
{\del\over\del b}{\del\over\del a}={\del\over\del a}{\del\over\del
b}+{\del\over\del b}{\del\over\del d}(q^{-2}-1)\\
{\del\over\del b}{\del\over\del c}={\del\over\del c}{\del\over\del
b}+{\del\over\del d}{\del\over\del d}(q^{-2}-1)+{\del\over\del a}{\del\over\del
d}(q^2-1)}
Their Leibniz rule comes from the additive braid statistics listed in
Example~3.10.

The algebra of forms from (\ref{duduR}) is
\cmath{ \extd c \extd c=0,\quad \extd a \extd a=0,\quad \extd b \extd b=0,\quad
\extd b \extd a=-\extd a \extd b\\
 \extd c \extd a=-\extd a \extd c,\quad \extd c \extd b=-\extd b \extd
c,\quad\extd d\extd d=\extd b \d
c(1-q^{-2})\\
\extd d \extd c=-\extd c\extd d q^{-2}+\extd a \extd c(1-q^{-2}),\quad\extd d
\extd b=- \extd b\extd d q^2
- \extd a \extd b(q^2 -1)\\
\extd d \extd a=- \extd b \extd c (q^2 -1)- \extd a\extd d}
also obtained from the additive braid statistics (with an extra minus sign).
These relations generate a finite-dimensional algebra with the dimension at
each degree being the same as in the classical case. The braided exterior
algebra $\Omega(\R_q^{1,3})$ is generated by the space-time co-ordinates and
these forms, with relations again read off at once from the additive braid
statistics in Example~3.10.

The $q$-epsilon tensor from (\ref{epsd}) is
\cmath{\eps_{addd}=-\eps_{bdcd}=-\eps_{dadd}
=\eps_{dbdc}=\eps_{ddad}=-\eps_{ddda}=1-q^{-2}\\
-\eps_{adad}=-\eps_{cdbd}=\eps_{dada}=\eps_{dcdb}=q^2-1\\
\eps_{abcd}=-\eps_{acbd}=\eps_{adbc}=-\eps_{adcb}
=-\eps_{bacd}=\eps_{bcad}=-\eps_{bcda}=\eps_{cabd}=1\\
-\eps_{cbad}=\eps_{cbda}=-\eps_{dabc}=\eps_{dacb}
=\eps_{dbac}=-\eps_{dbca}=-\eps_{dcab}=\eps_{dcba}=1\\
\eps_{acdb}=-\eps_{cadb}=\eps_{cdab}=-\eps_{cdba}=q^2,\
-\eps_{abdc}=\eps_{badc}=-\eps_{bdac}=\eps_{bdca}=q^{-2}}
and has even more nonzero elements than in the Euclidean case.

Finally, the Gaussian-weighted integral on degree two from Section~5.6 is
$\CZ[u_Iu_J]=q^{-4}\eta_{IJ}$. Its nonzero values on degree four are
\cmath{ \CZ[abcd]=-q^{-12},\quad \CZ[acbd]=-q^{-8},\quad
\CZ[bcd^2]=-q^{-10}(1-q^{-4})\\
\CZ[cbd^2]=-\h\CZ[d^4]=-q^{-8}(1-q^{-2})^2,\quad
\CZ[ad^3]=q^{-10}(1+2q^2)(1-q^{-2})\\
\CZ[a^2d^2]=q^4\CZ[b^2c^2]=\CZ[c^2b^2]=q^2\CZ[bc^2b]
=q^2\CZ[cb^2c]=q^{-10}(q^2+1)}
along with various other cases implied by the relations in $\R_q^{1,3}$. In
terms of spacetime
coordinates (\ref{minktxyz}) we have
\cmath{ \CZ[t^2]={[2;q^2]\over 4}q^{-4},\quad
\CZ[x^2]=\CZ[y^2]=\CZ[z^2]=-{[2;q^2]\over 4}q^{-6}\\
\CZ[z^4]={3[2;q^2]\over 8}q^{-12},\quad \CZ[t^4]=q^4\CZ[x^4]=q^4
\CZ[y^4]={[3;q^2]!\over 8}q^{-10}.}
We see that the Gaussian-weighted integral $\CZ$ is quite similar to the
Euclidean one in its values, except for the sign in the spacelike directions.
We also
note that the negative sign in the space-like directions agrees with the sign
obtained in
physics by Wick rotation to make sense of Gaussian integrals in Minkowski
space.

We have developed $q$-Minkowski space cleanly and simply as a braided matrix
$R_{21}\vecu_1R\vecu_2=\vecu_2R_{21}\vecu_1 R$ in the braided approach started
in \cite{Ma:exa}\cite{Ma:skl}\cite{Ma:lin}\cite{Ma:mec}. Because of the
mapping (\ref{bm2dec}) our results must necessarily
recover in the standard $2\times 2$ case the pioneering work in the spinorial
approach of
\cite{CWSSW:lor}\cite{CWSSW:ten}\cite{OSWZ:def} when one looks at the explicit
algebra. On the other hand, our braided formulae are
always in a general R-matrix form; for example the spinorial form
(\ref{minkpoi}) for the
$q$-Poincar\'e group is a new result in \cite{MaMey:bra} and can be contrasted
with the explicit algebra relations found in \cite{OSWZ:def} by other means.
The role of $SU_q(2)\dcross SU_q(2)$ as $q$-Lorentz group is due to
\cite{PodWor:def}\cite{CWSSW:lor} although its abstract structure is due to
\cite{Ma:mor}\cite[Sec. 4]{Ma:poi}, in the second of which the link with
twisting of the Euclidean rotation group was made precise. The algebra of
differential forms\cite{OSWZ:def} is not new but developed now in a
constructive way from the braided coaddition\cite{Ma:eps}. The braided approach
added principally such things as the coaddition itself\cite{Mey:new}, the
comultiplication\cite{Ma:exa}, the duality with braided vectors, the covariant
action of the $q$-Poincar\'e group on the spacetime generators\cite{Ma:poi},
the existence of the braided exponential (though no very good formula for it),
the braided Gaussian and associated integral\cite{KemMa:alg} even for this
particular example. Further results in the braided approach are in
\cite{Mey:new} where a braided covector picture of both spacetime and the
Lorentz group is developed, and \cite{Mey:wav}.

The braided approach also added the canonical
isomorphism\cite{Ma:lie}\cite{Ma:mex}
\eqn{minklieisom}{ \R_q^{1,3}\equiv BM_q(2)\isom U(gl_{q,2})}
which is a `purely quantum' phenomenon since at $q=1$ the left hand side
becomes the commutative
co-ordinates on Minkowski space while the right hand side becomes the
non-commutative enveloping algebra
of $su_2\oplus u(1)$. This is obviously very interesting for particle physics
since two fundamental ingredients
of electroweak theory are unified by $q$-deformation. Under this
quantum-geometry transformation the
mass shell $ad-q^2cb=1$ in $q$-Minkowski space (the braided group $BSL_q(2)$ in
\cite{Ma:exa}) is isomorphic to the
algebra of $U_q(sl_2)$\cite{Ma:skl}. This also provided of course plenty of
representations of the $q$-Minkowski space algebra by pulling back the usual
finite-dimensional, $q$-boson or other favourite representations of the latter.
A further application of braided-Lie algebras is in \cite{Ma:sol}.

\appendix

\section{Transmutation}

Braided groups were first constructed (by the author) using a process of {\em
transmutation} which turns any quantum group containing a strict subquantum
group
(with universal R-matrix) into a braided
one\cite{Ma:bra}\cite{Ma:tra}\cite{Ma:bg}. We give here an introduction to this
theory.
As in Section~6, we require the reader to be more familiar with advanced
aspects of quantum group theory\cite{Ma:qua}. Of course, it is not
necessary to know this construction if one just wants to work with braided
groups already given to you, which is the line we have
taken above.

The idea of transmutation is that the type of an algebraic object (the kind of
object it is) is to some extent a
matter of choice. As an example, we could start with a bosonic object (an
ordinary quantum group) and consider the collection or category of things on
which it (co)acts. This category might well be equivalent to some other
category, such as the category of braided-coactions of a braided group. By
applying braided-Tannaka-Krein reconstruction we could reconstruct that
braided-group\cite{Ma:bg}. The same principle applies quite generally, whenever
we have a concept of `representation' powerful enough to reconstruct the object
being represented. We start with one type of object, take its representations,
identify that category as equivalent to another category of representations of
some other type of object, and reconstruct it. I call this principle
`transmutation' because it changes the flavour of the object. It is a kind of
Fourier transform technique for mathematical concepts.

In our example, this idea becomes the following
theorem\cite{Ma:bg}\cite{Ma:euc}. If $f:A\to A_1$ is a Hopf algebra
homomorphism between two quantum groups, where $A_1$ is dual-quasitriangular in
the sense that there is a universal R-matrix functional $\CZ:A_1\tens A_1\to
\C$ obeying (\ref{dqua}) as in Section~6, then $A$ can be transmuted by $f$
into a braided group $B$ as follows. It has the same linear space and coproduct
as $A$ but a new product, antipode and braiding\cite{Ma:bg}\cite{Ma:euc}
\ceqn{trans}{a\und\cdot b=a\t b\t \CZ(f((Sa\o)a\th)\tens f(Sb\o)),\quad
\und{S}a= S  a\t\CZ(f((S ^2a\th)S  a\o)\tens f(a_{(4)}))\\
\Psi(a\tens b)=b\t\tens a\t\CZ(f((Sa\o)a\th)\tens f((Sb\o)b\th)).}
We are underlining the braided product and antipode here to distinguish them
from the quantum group ones for $A$ in terms of which they are given. This
braided group lives in the braided category of objects covariant under $A_1$,
by the coaction
\eqn{transcov}{b\mapsto b\t\tens f((Sb\o)b\th)}
which induces the braiding $\Psi$ via (\ref{coactPsi}).

If $A$ has its own dual-quasitriangular structure $\CR:A\tens A\to \C$ obeying
(\ref{dqua}) then the product of $B$ is braided-quantum-commutative in the
sense
\alignn{transcom}{&&\equad\CZ(f(b\o)\tens f( a\o))\CR(a\t\tens
b\t)\CZ(f(Sa\th)\tens
f(b\th))a\fo \CZ(f(a\fiv)\tens f( b\fo))\und\cdot b\fiv\nonumber\\
&&\qquad=b\o\CZ(f(b\t)\tens f( a\o))\und\cdot a\t\CR(a\th\tens b\th)}
as one verifies easily from the definition of its product in (\ref{trans}) and
the axioms (\ref{dqua}) for $\CZ$ and $\CR$. This expresses the theorem in
\cite{Ma:bg} that $B$ in this case is a braided-quantum group with
braided-universal R-matrix functional, the above being the braided version of
the second of (\ref{dqua}). Finally, all of this applies just as well at the
bialgebra level, if $A$ does not have an antipode. Just replace
$f\circ S$ by $f^{-1}$, the
convolution-inverse of $f$ which we can suppose instead\cite{Ma:bg}.

This is the general theory of transmutation in function algebra form. We also
gave it in enveloping algebra form\cite{Ma:tra}. As an immediate example,
suppose there is a quantum group homomorphism $f:A(R)\to A(Z)$ sending
$t^i{}_j$ to the matrix generator $f^i{}_j$ say of $A(Z)$. The
condition for this is $R\vecf_1\vecf_2=\vecf_2\vecf_1R$ which is just the same
as our covariance assumption (\ref{R'tt}). It implies and is essentially
equivalent\cite{Ma:poi} to the matrix condition
\eqn{RZZ}{R_{12}Z_{13}Z_{23}= Z_{23}Z_{13}R_{12},\quad
R_{23}Z_{13}Z_{12}=Z_{12}Z_{13}R_{23}}
as explained in Section~6 when deriving (\ref{coveca}). From the transmutation
formula (\ref{trans}) and (\ref{R(tt)a}) we obtain a braided group $B(R,Z)$
with product cf.\cite{Ma:lin}
\eqn{transRZ}{u^i{}_j=t^i{}_j,\  u^i{}_ju^k{}_l=t^a{}_bt^d{}_lZ^i{}_a{}^c{}_d
\widetilde Z^b{}_j{}^k{}_c,\quad
{\it i.e.},\quad \vecu=\vect,\ \vecu_1Z\vecu_2=Z\vect_1\vect_2,\quad{\it etc.}}
where we write the generators of $B(R,Z)$ as $\vecu$ and use their product
$\und\cdot$, while on the right we use the original product of $A(R)$. The
braiding comes out as in (\ref{uhopfb}) with $Z$ in place of $R$ and
corresponding braid-statistics $Z^{-1}\vecu'_1Z\vecu_2=\vecu_2 Z^{-1}\vecu'_1
Z$ for braided matrix comultiplication. The braided-quantum-commutativity
(\ref{transcom}) becomes the relations
\eqn{transcomRZ}{ Z_{21}RZ^{-1}\vecu_1 Z\vecu_2=\vecu_2 Z_{21}\vecu_1 R}
which is also immediate by transmutation (\ref{transRZ}) of the FRT relations
of $A(R)$ as $RZ^{-1}\vecu_1Z\vecu_2=R\vect_1\vect_2=\vect_2\vect_1R
=Z_{21}^{-1}\vecu_2Z_{21}\vecu_1 R$. We assume in the derivation that $Z$
is regular so that in principle one has a quantum group with antipode
$\vecf^{-1}$, but this vanishes from the final formulae using (\ref{R(tt)b}).
We gave these explicit formulae in \cite{Ma:exa}\cite{Ma:lin} for the case
$B(R)\equiv B(R,R)$, where we checked everything directly from the matrix
data as a self-contained example of a braided group. This case was emphasised
because the abstract braided-universal R-matrix functional is the ratio of the
universal R-matrix functionals of $A(R),A(Z)$ and therefore trivial when
$R=Z$, i.e. $B(R)$ has  been made totally `braided commutative'. In this
extreme  the quantum non-commutativity of $A(R)$ is fully traded for
braid-statistics of $B(R)$ as a `classical' but braided matrix. Another extreme
is
$B(R,\id)=A(R)$. Other cases
of  $B(R,Z)$ are intermediate between these extremes but their derivation
and the direct check that they form a braided group follows in the same way
as for $B(R)$. \cite{Hla:bra},\cite{Lu:bra} seem to be the first to
explicitly study such algebras $B(R,Z)$ and check them directly from
(\ref{transcomRZ}), (\ref{RZZ}). More importantly, Hlavaty
obtained some interesting examples, see \cite{Hla:bra}. One can obtain further
examples in
the context of quantum principle bundles\cite{BrzMa:gau} and quantum
homogeneous spaces, where homomorphisms between two quantum groups, which are
all we need, abound.

The transmutation theory tells of course not only that $B(R,Z)$ is a braided
group, but also that the representations of $A(R)$ become automatically
$Z$-braided representations of $B(R,Z)$. This is the essence of transmutation,
as explained above. To see how it works in the matrix setting, transmutation
tells us that whenever our braided covectors $x_i$ from Section~3.1 are
$A(R)$-covariant under (\ref{xcov}) the same linear map $\vecx\to\vecx\vecu$
becomes a $Z$-braided coaction of $B(R,Z)$. The $A(R)$-covariance becomes
through the map $f$ an $A(Z)$-covariance, and the $Z$-braiding of the $x_i$
with anything else is induced by this, e.g.,
$\Psi(\vecx_1\tens\vecx_2)=\vecx_2\tens\vecx_1\CZ(f(\vect_1)\tens
f(\vect_2))=\vecx_2\tens\vecx_1 Z$ and likewise
$\Psi(\vecu_1\tens\vecx_2)=\vecx_2Z^{-1}\tens \vecu_1 Z$, being derived in the
same was as we derived (\ref{xvustata})--(\ref{xvustatb}) in Section~6.1. The
braided-covariance under $B(R,Z)$ has to work by the general theory above, but
one can check it directly as
$\vecx_1\vecu'_1\vecx_2\vecu'_2=\vecx_1\vecx_2Z^{-1}\vecu'_1
Z\vecu'_2=\vecx_2\vecx_1R'Z^{-1}\vecu'_1
Z\vecu'_2=\vecx_2\vecx_1Z_{21}^{-1}\vecu'_2 Z_{21}\vecu'_1R'=\vecx_2\vecu'_2
\vecx_1\vecu'_1R'$ where the third equality is the covariance assumption
(\ref{R'tt}) in terms of $\vecu$. Recall that this in turn implies and is
essentially implied by (\ref{coveca}). The diagonal case $Z=R$ of this theory
is how we obtained the braided coactions of $B(R)$ in \cite{Ma:lin} as
described in Section~4.1 above.
Also, if $\vect$ has an antipode then so does $\vecu$, which is how to obtain
the braided antipode on $BSL_q(2)$ in \cite{Ma:exa} from that of $SL_q(2)$.
Indeed, all constructions for quantum groups have braided parallels because of
transmutation. Details of the above and some further results about $B(R,Z)$
from the point of view of transmutation will appear in \cite{MaPla:mat}.

To give some other simple examples of transmutation, we let $G$ be an Abelian
group, $\C G$ its group algebra and $\beta$ a bicharacter on $G$, i.e. a
function on $G\times G$ such that $\beta(g,\ )$ and $\beta(\ ,g)$ are
multiplicative for each $g\in G$.  We know from \cite{Ma:csta} that any
bicharacter on $G$ defines a (dual)-quasitriangular structure
$\CZ(g,h)=\beta(g,h)$ extended linearly. So we have a dual quantum group. The
idea to use bicharacters in this was introduced in \cite{Ma:any}\cite{Ma:csta}
as a quasitriangular structure on the Hopf algebra $\C(G)$, which is
equivalent. Transmutation was also covered there and we are just repeating it
in the dual form for convenience of the reader. So any quantum group $A$
mapping onto $\C G$ gets transmuted to a quantum braided group $B$ with
structure from (\ref{trans}).

\begin{example}cf\cite{MaPla:uni} Let $R$ be a matrix solution of the QYBE and
let its indices $i=1,\cdots, n$ be assigned a degree $|i|\in G$, where $G$ is
an Abelian group and such that $R$ has $G$-{\em degree zero} in the sense
\[ R^i{}_j{}^k{}_l=0\quad {\it for\ all}\quad |i|\cdot|k|\ne |j|\cdot|l|.\]
Then for any bicharacter $\beta$ the quantum matrices $A(R)$ have a
corresponding transmutation $B(R,\beta)$ with braided group structure
\cmath{\beta(i,k)\beta(j,b)\und R^i{}_a{}^k{}_b u^a{}_j
u^b{}_l=\beta(j,l)\beta(b,i)u^k{}_b u^i{}_a \und R^a{}_j{}^b{}_l;\qquad \und
R^i{}_j{}^k{}_l\equiv R^i{}_j{}^k{}_l\beta(j,l)^{-1}\\
\beta(i,j)\equiv\beta(|i|,|j|),\quad
\und\Delta \vecu=\vecu\tens\vecu,\quad \und\eps\vecu=\id,\quad
\Psi(u^i{}_j\tens u^k{}_l)=u^k{}_l\tens
u^i{}_j{\beta(i,k)\beta(j,l)\over\beta(j,k)\beta(i,l)}.}
\end{example}
\proof This is immediate as an application of transmutation. The choice of
$G$-degree $|\ |$ and the degree-zero condition give us precisely a
homomorphism $A(R)\to \C G$ by $f(t^i{}_j)=\delta^i{}_j |i|$. It clearly maps
the coproducts in the sense $(f\tens f)\circ\Delta=\Delta\circ f$ where $\Delta
g=g\tens g$ and $\Delta \vect=\vect\tens\vect$, as required for a bialgebra
homomorphism. The convolution-inverse is
$f^{-1}(t^i{}_j)=Sf(t^i{}_j)=\delta^i{}_j|i|^{-1}$
using the group-inverse. The new product from (\ref{trans}) is just
\eqn{transbeta}{ u^i{}_j u^k{}_l=t^a{}_b
t^c{}_l\CZ(\delta^i{}_a|i|^{-1}\delta^b{}_j
|j|\tens\delta^k{}_c|k|^{-1})=t^i{}_j
t^k{}_l\beta(|i|^{-1}|j|,|k|^{-1})=t^i{}_jt^k{}_l{\beta(i,k)\over\beta(j,k)}.}
The braided-quantum commutation relations (\ref{transcom}) reduce to the ones
stated, where we have written them in terms of a matrix $\und R$ obeying a
$\beta$-version of the QYBE. The braided group lives in the category of $\C
G$-covariant algebras with coaction from (\ref{transcov})
\eqn{transcovbeta}{ u^i{}_j\mapsto u^a{}_b \tens
f^{-1}(t^i{}_a)f(t^b{}_j)=u^i{}_j\tens |i|^{-1}\cdot |j|}
which then induces the braiding  $\Psi(u^i{}_j\tens u^k{}_l)=u^k{}_l\tens
u^i{}_j\beta(|i|^{-1}|j|,|k|^{-1}|l|)$ as shown. We also have all the theory of
comodules too, automatically. Thus our braided covectors $x_i$ and vectors
$v^i$ from Section~3.1 have the coaction
\eqn{xcovbeta}{x_i\to x_a\tens f(t^a{}_i)= x_i\tens |i|,\quad v^i\mapsto
v^a\tens f^{-1}(t^i{}_a)=v^i\tens |i|^{-1}}
and with the induced braiding $\Psi(u^i{}_j\tens x_k)=x_k\tens
u^i{}_j\beta(|i|^{-1}|j|,|k|)=x_k\tens u^i{}_j{\beta(j,k)\over\beta(i,k)}$ etc.
Their usual $A(R)$-covariance becomes a braided-covariance under
$\vecx\to\vecx\vecu$ etc., provided we remember these induced braid statistics.
 Note also that if $\vect$ has an antipode (if we work with a Hopf algebra
obtained from $A(R)$) then the corresponding $\vecu$ has a braided antipode
(\ref{trans}) computed in the present setting.
\endproof

These simplest braided-quantum groups where the braiding is given by a
$\C$-number bicharacter $\beta$ are called {\em $\C$-statistical braided
groups} in \cite{Ma:csta}. Their construction is an elementary application of
the author's transmutation theory. On the other hand, to find interesting
examples and applications is rather more challenging. The very simplest example
is the super case where $G=\Z_2$ and $\beta=\pm 1$ (this is the quantum group
$\Z_2'$ which is the hidden covariance of supersymmetry\cite{Ma:tra} as
mentioned in Section~6.1). The above transmutation in this case becomes the
process of {\em superisation} studied by the author and M.J. Rodriguez-Plaza
in \cite{MaPla:uni} (in an equivalent form). The next simplest is $G=\Z_N$ and
$\beta(g,h)=e^{2\pi\imath gh\over N}$ in an additive notation
\cite{Ma:any}\cite{MaPla:any}\cite{MaPla:mat}. Here \cite{MaPla:uni} and
\cite{Ma:any} use transmutation in the enveloping form while \cite{MaPla:mat}
uses the dual form as above. Further details of Example~A.1, further results
about $B(R,Z)$ and some concrete examples will appear in \cite{MaPla:mat}.
Another choice is $G=\Z^n$ or $\C^n$ and
$\beta(\vec\phi,\vec\psi)=e^{\imath(\vec\phi,\vec\psi)}$ induced by an
arbitrary non-degenerate bilinear form, see \cite[Sec. 3]{Ma:csta} where
braided groups of such $\C$-statistics were introduced and studied from a
bosonisation point of view as a novel approach to quantisation of free fields.
Following exactly our transmutation ideas from \cite{Ma:exa}\cite{Ma:lin} there
appeared recently in \cite{CouLev:gen} some explicit examples $M_{q,\mu}(n)$ of
the $G=\Z^n$ type obtained by transmutation of the usual quantum matrices
$M_q(n)$. The underlying mathematics is not new (in view of the above
transmutation theory) but these examples are nevertheless interesting because
of a connection with statistical mechanical models in physics\cite{CouLev:gen}.
Finally, we note that braided-quantum groups have as rich a theory as usual
quantum groups on account of their braided-universal R-matrix, as developed
in \cite{Ma:tra} by diagrammatic means. Recent results using such means are
in \cite{Bes:cro}.

\itemsep 0pt
%\bibliographystyle{unsrt}
%\bibliography{biblio}
%
%\end{document}

\end{document}